\documentclass[a4paper,10pt,twocolumn,showpacs,notitlepage,floatfix,nofootinbib,superscriptaddress,prd]{revtex4-2}
\usepackage{graphicx}
\usepackage{amsfonts}
\usepackage{amsmath}
\usepackage{hyperref}
\usepackage{float}
\usepackage{amssymb,amsbsy}
\usepackage{epstopdf}
\usepackage{color}
\usepackage{dsfont}
\usepackage{combelow}
\usepackage{mathtools}
\usepackage[normalem]{ulem}
\usepackage{tikz}
\usepackage{pgfplots}

\renewcommand{\k}{\mathbf{k}}
\newcommand{\p}{\mathbf{p}}
\renewcommand{\d}{\mathrm{d}}

\def\bk{{\mathbf{k}}}

\def\nora#1{{\mathfrak{#1}}}

\allowdisplaybreaks[2]

\begin{document}

\title{Analytical structure of the binary collision integral and the ultrarelativistic limit of transport coefficients of an ideal gas}

\author{David Wagner}
	\email{dwagner@itp.uni-frankfurt.de}
\affiliation{Institut f\"ur Theoretische Physik, Johann Wolfgang Goethe-Universit\"at, Max-von-Laue-Straße 1, D-60438 Frankfurt am Main, Germany}%
\affiliation{Department of Physics, West University of Timi\cb{s}oara, \\
Bd.~Vasile P\^arvan 4, Timi\cb{s}oara 300223, Romania}

\author{Victor E. Ambru\cb{s}}
\email[Corresponding author: ]{victor.ambrus@e-uvt.ro}
\affiliation{Department of Physics, West University of Timi\cb{s}oara, \\
Bd.~Vasile P\^arvan 4, Timi\cb{s}oara 300223, Romania}

\author{Etele Moln\'ar}
\affiliation{Institut f\"ur Theoretische Physik, Johann Wolfgang Goethe-Universit\"at, Max-von-Laue-Straße 1, D-60438 Frankfurt am Main, Germany}%
\affiliation{Department of Physics, West University of Timi\cb{s}oara, \\
Bd.~Vasile P\^arvan 4, Timi\cb{s}oara 300223, Romania}
\affiliation{Incubator of Scientific Excellence–Centre for Simulations of Superdense Fluids,
University of Wroclaw, pl. M. Borna 9, PL-50204 Wroclaw, Poland}

\begin{abstract}
	In this paper we discuss the analytical properties of the binary collision integral
	for a gas of ultrarelativistic particles interacting via a constant cross-section. 
	Starting from a near-equilibrium expansion over a complete basis of irreducible tensors in momentum space we compute the linearized collision matrices analytically.
	Using these results we then numerically compute all transport-coefficients of relativistic fluid dynamics
	with various power-counting schemes that are second-order in Knudsen and/or inverse Reynolds numbers.
	Furthermore, we also exactly compute the leading-order contribution 
	with respect to the particle mass to the coefficient of bulk viscosity, 
	the relaxation time, and other second-order transport coefficients of the bulk viscous pressure. 
\end{abstract}

\date{\today }
\maketitle

%%% 
\section{Introduction}

The kinetic theory of rarefied gases contains a collision term which describes the interaction among constituents through collisions. The well known collision term defined by Boltzmann's \textit{Stoßzahlansatz}, or the assumption of \textit{molecular chaos}, defines the number of binary collisions through a product of two single-particle distribution functions.
The resulting integro-differential equation, the Boltzmann transport equation, describes the space-time
evolution of the single-particle distribution function~\cite{Groot.1980,Cercignani.2002,Denicol.2021}
\begin{equation}
	k^{\mu }\partial _{\mu }f_{\mathbf{k}}=C\left[ f\right] \; ,  
	\label{BTE}
\end{equation}%
where $C\left[ f\right] $ is the collision term. In the case of binary
elastic collisions, the collision term reads, 
\begin{align} \nonumber
	C\left[ f\right] &\equiv  \frac{1}{2}\int \d K^{\prime }\d P\d P' \, 
	\left( W_{\mathbf{pp}' \rightarrow \mathbf{kk}' } 
	f_{\mathbf{p}}f_{\mathbf{p}'}\tilde{f}_{\mathbf{k}}\tilde{f}_{\mathbf{k}'} \right.  \\
	&- \left. W_{\mathbf{kk}' \rightarrow \mathbf{pp}' } 
	f_{\mathbf{k}}f_{\mathbf{k}'}\tilde{f}_{\mathbf{p}}\tilde{f}_{\mathbf{p}'} \right) \; , 
	\label{COLL_INT}
\end{align}
where $f_{\mathbf{k}} \equiv  f_{\mathbf{k}}(x^\mu,k^\mu)$ denotes the Lorentz-invariant single-particle distribution function, while $\tilde{f}_{\mathbf{k}}\equiv 1-af_{\mathbf{k}}$, with $a= \pm 1$ for 
fermions/bosons and $a=0$ for classical particles.  
The Lorentz-invariant differential element is
$\d K\equiv g\,\d^{3}\mathbf{k/}\left[ (2\pi)^{3}k^{0}\right]$, while $g$ denotes the number of internal degrees of freedom. The $1/2$ factor removes the double counting from the integrations with respect to $\d P$ and $\d P'$. 
The four-momentum of particles $k^{\mu}=(k^0,\mathbf{k})$  is normalized to their rest mass
squared, $k^{\mu }k_{\mu }=m_{0}^{2}$, where $k^{0}=\sqrt{\mathbf{k}^{2} + m_{0}^{2}}$ is the 
on-shell energy. In this paper we use natural units $\hbar=c=k_{B}=1$.

The binary transition rate is defined as 
\begin{equation}
	W_{\mathbf{kk}' \rightarrow \mathbf{pp}'} \equiv \frac{s}{g^2} (2\pi)^6 \frac{\d\sigma(\sqrt{s},\Omega)}{\d\Omega} \delta(k^\mu + k'^\mu - p^\mu -p'^\mu) \;,
	\label{eq:W_general}
\end{equation}
where the factor $(2\pi)^6 / g^2$ appears due to our convention for the momentum-space integration measure. For simplicity, in the remainder of this paper we set $g = 1$. 
 The delta function ensures the conservation of the energy and momentum in binary collision.
The transition rate depends on the total center-of-momentum (CM) energy squared 
$s \equiv (k^\mu + k'^\mu)^2 = (p^\mu+ p'^\mu)^2$, while the total cross section integrated over the solid angle $\Omega$ is defined as~\cite{Fotakis:2022usk}
\begin{equation}
	\sigma_{T}(\sqrt{s}) = \frac{1}{2} \int \d\Omega 
	\frac{\d\sigma(\sqrt{s},\Omega)}{\d\Omega}\;.
\end{equation}
In this paper we employ the so-called \textit{hard-sphere} 
approximation which assumes that the transport cross section is isotropic and independent of the total CM energy,
\begin{equation}
	\sigma_{T} \equiv 2\pi \frac{\d\sigma(\sqrt{s},\Omega)}{\d\Omega} = \frac{1}{n_0 \lambda_{\rm mfp}} \; ,
\end{equation}
where $n_0$ is the particle density and
$\lambda_{\textrm{mfp}}$ is the mean free path between collisions.

The relativistic Boltzmann equation provides a framework for studying various properties of 
matter in and out of equilibrium, as well as for deriving the macroscopic conservation laws, 
i.e., fluid dynamics, based on the microscopic properties of the system.

A vanishing collision term, $C\left[ f_{0\mathbf{k}}\right] = 0$, due to detailed balance,
defines the local equilibrium distribution, the J\"uttner distribution function~\cite{Juttner,Cercignani.2002,Denicol.2021}, 
\begin{equation}
	f_{0\mathbf{k}}= \left[ \exp \left( \beta k^\mu u_\mu - \alpha\right) + a\right]^{-1} \;,  \label{f_0k}
\end{equation}
where $u^{\mu}=\gamma (1,\mathbf{v})$  is the timelike fluid-flow four-velocity normalized to $u^{\mu}u_{\mu}= 1$, while $\gamma=(1-\mathbf{v}^{2})^{-1/2}$. 
Furthermore, $\beta = 1/T$ is the inverse temperature and $\alpha = \beta\mu$, with $\mu$ the chemical potential.
Out of equilibrium, the distribution function is separated as
\begin{equation}
	f_\k \equiv f_{0\k} + \delta f_\k \;.
	\label{eq:f=f0+deltaf}
\end{equation}

In this paper we apply a relativistic version of Grad's method of moments \cite{Grad:1949}, as formulated
by Denicol, Niemi, Moln\'ar and Rischke, (referred to as DNMR)~\cite{Denicol:2012cn} to obtain the transport coefficients for a classical gas of massless particles interacting via an isotropic constant cross-section. 
Therein the irreducible moments of tensor-rank $\ell$ of $\delta f_\k$ are defined as
\begin{equation}
	\rho_{r}^{\mu_{1}\cdots \mu_{\ell}} \equiv   \int \d K E_\bk^r 
	k^{\langle \mu_1}\cdots k^{\mu_\ell\rangle}\delta f_\bk \; .
	\label{eq:rho_def}
\end{equation}
Here, $r$ denotes the power of energy $E_{\mathbf{k}}\equiv k^{\mu }u_{\mu }$, 
while $k^{\left\langle \mu _{1}\right.}\cdots k^{\left. \mu _{\ell}\right\rangle }
=\Delta_{\nu_{1}\cdots \nu_{\ell }}^{\mu_{1}\cdots \mu_{\ell }}k^{\nu_{1}}\cdots k^{\nu_{\ell }}$
are the irreducible tensors forming a complete orthogonal basis \cite{Denicol:2012cn,Groot.1980}. 
The four-momentum is decomposed as $k^{\mu} = E_{\mathbf{k}} u^{\mu} + k^{\langle \mu \rangle}$, where 
$k^{\left\langle \mu \right\rangle } \equiv \Delta _{\nu }^{\mu }k^{\nu }$ is defined using the
elementary projection operator $\Delta^{\mu \nu} \equiv g^{\mu \nu} - u^\mu u^\nu$, with $g^{\mu \nu}=\text{diag}(+,-,-,-)$ being the metric tensor. 
The symmetric, traceless, and orthogonal
projection tensors of rank $2\ell$, 
$\Delta_{\nu _{1}\cdots \nu_{\ell }}^{\mu_{1}\cdots\mu _{\ell }}$, are constructed 
using the $\Delta ^{\mu \nu }$ projectors. 

Expressing the comoving derivative of irreducible moments, 
$\dot{\rho}_{r}^{\left\langle \mu _{1}\cdots \mu _{\ell }\right\rangle }
\equiv \Delta_{\nu _{1}\cdots \nu _{\ell }}^{\mu _{1}\cdots \mu _{\ell }}
u^\alpha \partial_\alpha \rho _{r}^{\nu _{1}\cdots \nu_{\ell }}$, 
the equations of motion for these moments follow from the Boltzmann equation~(\ref{BTE}). 
For the sake of concision, we do not list the complete equations of motion, since they 
can be found in Eqs.~(35)--(46) of Ref.~\cite{Denicol:2012cn}, and instead quote just the following terms:
\begin{align}
	\dot{\rho}_{r}-C_{r-1}& =\alpha _{r}^{\left( 0\right) }\theta + (\textrm{higher-order terms}) \;, \label{D_rho} \\ 
	\dot{\rho}_{r}^{\left\langle \mu \right\rangle } - C_{r-1}^{\left\langle \mu \right\rangle } 
	& = \alpha _{r}^{\left( 1\right) }\nabla ^{\mu }\alpha
	+  (\textrm{higher-order terms}) \; , \label{D_rho_mu} \\
	\dot{\rho}_{r}^{\left\langle \mu \nu \right\rangle }-C_{r-1}^{\left\langle\mu \nu \right\rangle }
	&= 2\alpha _{r}^{\left( 2\right) }\sigma ^{\mu \nu } +  (\textrm{higher-order terms}) \;,
	\label{D_rho_munu}
\end{align}
where the irreducible moments of the collision term~\eqref{COLL_INT} are defined similarly 
to Eq.~\eqref{eq:rho_def} as
\begin{equation}
	C_r^{\langle \mu_1 \cdots \mu_\ell \rangle} \equiv   
	\int \d K\, E_\bk^r k^{\langle \mu_1}\cdots 
	k^{\mu_\ell \rangle} C[f]\;.
\end{equation}
The coefficients $\alpha^{(\ell)}_r \equiv \alpha^{(\ell)}_r(\alpha,\beta)$ are thermodynamic functions.
Furthermore, $\nabla^\mu = \Delta^{\mu\nu} \partial_\nu$ is the gradient operator, $\theta\equiv \nabla_{\mu}u^{\mu }$ is the expansion scalar and
$\sigma^{\mu \nu }\equiv \frac{1}{2}\left(\nabla^{\mu}u^{\nu} + \nabla^{\nu}u^{\mu}\right) - \frac{1}{3}\theta \Delta^{\mu \nu}$ is the shear tensor.

The conservation of particle number as well as of energy and momentum in binary collisions require that the corresponding moments of the collision term vanish equivalently, 
i.e., $C_0 = 0$, $C_1 = 0 $, and $C_1^{\langle \mu \rangle} = 0$. 
The resulting equations of motion are the conservation laws of fluid dynamics,
\begin{equation}
	\partial_\mu N^{\mu} = 0 \;, \quad \partial_\mu T^{\mu \nu} = 0 \;, 
	\label{cons_eqs}
\end{equation}
where the particle four-current and energy-momentum tensor are given by
\begin{align}\label{N_mu}
	N^{\mu} &= n_0 u^{\mu} + V^{\mu} \;, \\ \label{T_munu}
	T^{\mu \nu} &= e_0 u^{\mu}u^{\nu} - \left(P_0 + \Pi \right)\Delta^{\mu\nu} + \pi^{\mu \nu}\;.
\end{align}
Here, $n_0$, $e_0$, and $P_0$, are the particle density, energy density, and the isotropic pressure, in equilibrium.
The bulk viscous pressure, the particle diffusion four-current and the shear-stress tensor are defined by
\begin{align} \label{Pi}
	\Pi &\equiv -\frac{1}{3} T^{\mu \nu} \Delta_{\mu \nu} - P_0 = -\frac{m^2_0}{3} \rho_0 \;, \\ \label{V_mu}
	V^{\mu} &\equiv  \Delta^{\mu}_{\alpha} N^{\alpha} = \rho^{\mu}_0 \;,  \\  \label{pi_munu} 
	\pi^{\mu \nu} &\equiv \Delta^{\mu \nu}_{\alpha \beta}T^{\alpha \beta} = \rho^{\mu \nu}_0 \;.  
\end{align}
In the above decompositions the fluid-flow four-velocity is  
the timelike eigenvector of the energy-momentum tensor, $u^\mu = T^{\mu \nu} u_{\nu}/e_0$, as per Landau's definition \cite{Landau.2014}, hence
\begin{equation}
	\rho^{\mu}_1 \equiv \Delta^{\mu}_{\alpha} T^{\alpha \beta} u_{\beta} = 0 \;.
	\label{Landau_flow}
\end{equation}
The chemical potential and the temperature are determined through the 
Landau matching conditions,
\begin{equation}
	\rho_1 \equiv N^{\mu }u_{\mu } - n_0 = 0 \;, \quad 
	\rho_2 \equiv T^{\mu \nu } u_{\mu }u_{\nu } - e_0 = 0  \;. \label{delta_ne}
\end{equation}

The equations of motion for the primary dissipative quantities 
$\rho_{0} = -3 \Pi/m^2_0$,  $\rho_{0}^{\mu} = V^{\mu}$, and $\rho_{0}^{\mu \nu} = \pi^{\mu\nu}$ 
follow from Eqs.~(\ref{D_rho})--(\ref{D_rho_munu}).
The five conservation equations~\eqref{cons_eqs} couple to these
nine transport equations which contain various transport coefficients that 
explicitly depend on the underlying approximations and the influence of all non-dynamical moments 
included in this truncation.
These equations are truncated according to a power-counting scheme in Knudsen and inverse Reynolds numbers.  
The Knudsen number, $\textrm{Kn} \equiv \lambda_\textrm{mfp}/L$, is the ratio of the particle mean free
path $\lambda_{\text{mfp}}$ and a characteristic macroscopic scale $L$, while the inverse Reynolds number $\mathrm{Re}^{-1}$ is the ratio of an out-of-equilibrium and a local equilibrium macroscopic field.
The resulting equations of fluid dynamics are of second-order in Knudsen and/or inverse Reynolds numbers, and  are closed in terms of 14 dynamical moments contained in $N^{\mu}$ and $T^{\mu \nu}$.

Focusing on this second-order theory of relativistic fluid dynamics, 
we compute the moments of the linearized collision term for a gas of ultrarelativistic hard spheres with constant cross-section. 
Introducing a novel anisotropic decomposition of the collision integral in the center-of-momentum frame,
the calculation of the linearized collision matrices is done analytically in the ultrarelativistic limit. 
Using these exact results in the 14-dynamical moment approximation we collect and compute all transport coefficients, with five significant digits of precision, for truncation orders $N_0-2=N_1-1=N_2 = 1000$, corresponding to $N_0+3N_1+5N_2+9= 9014$ moments included in the basis.

We also compare the effect of three slightly different power-counting schemes for 
the non-dynamic moments introduced in Refs.~\cite{Denicol:2012cn}, \cite{Wagner:2022ayd}, and \cite{Ambrus:2022vif}, on all transport coefficients.
Specifically, we will consider and compare
\begin{enumerate}
    \item the DNMR approach with additional corrections to the $O({\rm Kn}^2)$ transport coefficients of  Refs.~\cite{Denicol:2012cn,Molnar.2014},
    \item the Inverse Reynolds Dominance (IReD) approach of Ref.~\cite{Wagner:2022ayd}, where all $O({\rm Kn}^2)$ terms are rewritten and absorbed into the $O({\rm Re}^{-1}{\rm Kn})$ terms,
    \item the corrected DNMR (cDNMR) approach of Ref.~\cite{Ambrus:2022vif}, where the $O({\rm Kn}^2)$ transport coefficients receive contributions only from the asymptotic matching of the moments $\rho^{\mu_1 \cdots \mu_\ell}_{r>0}$.
\end{enumerate}
All three schemes considered here fully account for all second-order terms with respect to ${\rm Kn}$ and ${\rm Re}^{-1}$ and are therefore equivalent (up to terms of third order) within this truncation scheme. 
At asymptotically long times, when the magnitudes of the Knudsen and 
the inverse Reynolds numbers are of the same order, i.e., $\textrm{Kn} \sim \textrm{Re}^{-1}$, 
also known as the order of magnitude approximation~\cite{Molnar.2014,Fotakis:2022usk,Struchtrup.2004},
there is freedom to re-arrange the transport coefficients. 
The IReD approximation of Ref.~\cite{Wagner:2022ayd} expresses the thermodynamic forces in terms 
of the primary dissipative quantities to replace $\textrm{Kn}^2 \rightarrow \textrm{Kn}\,\textrm{Re}^{-1}$ 
and hence removes terms that are of second order in the Knudsen number from the fluid-dynamical equations 
of motion.
In this paper, we focus solely on the linearized binary collision integral, which does not provide ${\rm Re}^{-2}$ terms. However, one may include nonlinear contributions to the collision integral as described in Ref.~\cite{Molnar.2014}.

The main results of this paper are the closed-form computation of the collision matrices for the scalar, 
vector and tensor moments in the case of massless ultrarelativistic particles interacting through a constant isotropic cross-section. 
This interaction model reduces in the non-relativistic limit to the well-studied hard-sphere interaction model, for which the first-order transport coefficients, i.e., the shear viscosity and heat conductivity, can be obtained in terms of the so-called Chapman-Cowling collision integrals~\cite{Chapman:1991,Liboff:2003} via a successive iterative refinement procedure. 
This method can be extended into the relativistic regime~\cite{Stewart:1971,Cercignani.2002}, where the exact expression requires a resummation over the entire hierarchy of moments~\cite{Denicol:2012cn}. 
We will demonstrate the truncation-order dependence with an analytical result only for the leading-order contribution with respect to the particle mass $m_0$ to the bulk viscosity coefficient $\zeta$ and to the relaxation time $\tau_\Pi$ for the bulk viscous pressure. For all other transport coefficients, we rely on numerical methods to obtain their values in the limit of infinite truncation order.

Another collision model for which the transport coefficients are obtained with similar accuracy as for the hard-sphere model is that of the so-called Maxwell molecules~\cite{Cercignani:1988,Liboff:2003}, interacting via a potential $V(r) \sim r^{-5}$, with $r$ being the distance between two interacting particles. 
Two relativistic generalizations of this model correspond to the Israel particles model~\cite{Israel.1963} 
and the Polak model~\cite{polak:1973}. 
More recently, the collision operator corresponding to the $\lambda \phi^4$ scalar field theory was studied in Refs.~\cite{Mullins:2022fbx,Denicol:2022bsq}. These results were used to compute transport coefficients in several fluid dynamical theories in Ref.~\cite{Rocha:2023hts}. 
The present work complements these studies  considering the analogous problem for hard spheres.

This paper is structured as follows. In Sect.~\ref{sec:DNMR}, we introduce the expansion of the distribution function and the linearized collision integral in terms of irreducible moments. 
Then we discuss the various power-counting methods and the transport coefficients of second-order fluid dynamics.
Section~\ref{sec:coll} clarifies the analytical structure of the collision integrals appearing in the moment equations up to tensor-rank two. These expressions are the main results of this work. 
In Sect.~\ref{sec:2nd_order_coeff} all first- and second-order transport coefficients are computed in the ultrarelativistic limit. 
The exact results for the coefficient of bulk viscosity and the relaxation time of the bulk viscous pressure are computed in Sect.~\ref{sec:bulk_exact}.
Finally,  Sect.~\ref{sec:conc} concludes this work. For reasons of brevity and clarity, various computations 
are relegated to the Appendices.

%%%
\section{Second-order fluid dynamics with 14 dynamical moments}
\label{sec:DNMR}

For reasons of completeness, we first summarize the derivation of second-order relativistic
fluid dynamics from the Boltzmann equation based on Ref.~\cite{Denicol:2012cn}.
The near-equilibrium expansion is summarized in Sect.~\ref{sec:DNMR:moms}. 
In Sect.~\ref{sec:DNMR:powerc}, we discuss the linearized collision integral and the various 
power-counting schemes in a unitary fashion.
Finally, Sect.~\ref{sec:DNMR:hydro} provides the relaxation equations of second-order fluid 
dynamics with 14 dynamical moments. 
The particle rest mass $m_0$ is considered arbitrary (non-vanishing) throughout this section.

%%%
\subsection{Near-equilibrium expansion over a complete basis of irreducible tensors}
\label{sec:DNMR:moms}

The expansion of $\delta f_\k$ is given by,
\begin{align}
 	\delta f_{\mathbf{k}} &\equiv f_{0\mathbf{k}} \tilde{f}_{0\mathbf{k}} \phi_{\mathbf{k}} \nonumber\\ 
	&= f_{0{\mathbf{k}}} \tilde{f}_{0\mathbf{k}} \sum_{\ell =0}^{\infty }\sum_{n=0}^{N_\ell}
	 \rho_{n}^{\mu_{1}\cdots \mu_{\ell }}k_{\left\langle \mu _{1}\right. }\cdots
 	k_{\left. \mu_{\ell }\right\rangle }\mathcal{H}_{\mathbf{k}n}^{(\ell)} \;,
 	\label{eq:delta_f_expansion}
\end{align}
where the coefficient $\mathcal{H}_{\bk n}^{(\ell )}$ is a polynomial in
energy of order $N_\ell \rightarrow \infty$ defined through another polynomial 
$P_{\mathbf{k} m}^{(\ell)}$ as
\begin{equation}
 	\mathcal{H}_{\bk n}^{(\ell)}\equiv  \frac{W^{(\ell)}}{\ell!}
 	\sum_{m=n}^{N_\ell}a_{mn}^{(\ell)} P_{\mathbf{k} m}^{(\ell)} \;,\quad 
 	P_{\mathbf{k} m}^{(\ell)} \equiv \sum_{r=0}^{m}a_{mr}^{(\ell)}E_{\mathbf{k}}^{r} \;.
 	\label{eq:Hfunctions}
\end{equation}

The negative-order moments $\rho_{r<0}^{\mu_1\cdots \mu_\ell}$ are not included in the expansion (\ref{eq:delta_f_expansion}) 
but they are expressed by a linear combination of positive-order moments through
\begin{equation}
	\rho_{-r}^{\mu_1\cdots \mu_\ell} = \sum_{n=0}^{N_\ell} \mathcal{F}_{rn}^{(\ell)} 
	\rho_n^{\mu_1\cdots \mu_\ell} \;,
	\label{eq:rhoneg}
\end{equation}
where we defined
\begin{equation}
	\mathcal{F}_{rn}^{(\ell)} \equiv \frac{\ell !}{(2\ell+1)!!} \int \d K E_\k^{-r} 
	\left(\Delta^{\alpha \beta} k_\alpha k_\beta \right)^\ell \mathcal{H}_{\k n}^{(\ell)} 
	f_{0\k} \tilde{f}_{0\k} \;,
	\label{F_rn}
\end{equation}
such that $\mathcal{F}_{-r,n}^{(\ell)} = \delta_{rn}$ by construction.

The coefficients $a_{mn}^{(\ell )}$ are obtained via the Gram-Schmidt orthogonalization procedure 
by imposing the following condition,
\begin{equation}
	\int \d K\, \omega^{(\ell)} P_{\mathbf{k}m}^{(\ell)} P_{\mathbf{k}n}^{(\ell)} = \delta_{mn} \;,
	\label{eq:P_ortho}
\end{equation}
where the weight $\omega^{(\ell)}$ is defined as
\begin{equation}
	\omega^{(\ell)} \equiv  \frac{W^{(\ell)}}{(2\ell + 1)!!} 
	\left(\Delta^{\alpha\beta} k_\alpha k_\beta \right)^\ell f_{0\mathbf{k}} \tilde{f}_{0\mathbf{k}}\;,
	\label{eq:omega_def}
\end{equation}
while the normalization parameter $W^{(\ell)}$ is fixed according to 
$P_{\mathbf{k}0}^{(\ell)} \equiv a_{00}^{(\ell)} = 1$, leading to 
\begin{equation}
	W^{(\ell)} = \frac{(-1)^\ell}{J_{2\ell,\ell}} \;. \label{W_ell}
\end{equation}
Here the thermodynamic integrals are defined as
\begin{align} \label{def:Inq}
	I_{nq}(\alpha, \beta) &\equiv \frac{(-1)^q}{(2q+1)!!} \int \d K E^{n-2q}_\k 
	\left(\Delta^{\alpha \beta} k_\alpha k_\beta \right)^q f_{0\k} \;, \\
	J_{nq} (\alpha, \beta) &\equiv 
	\left(\frac{\partial I_{nq}(\alpha, \beta)}{\partial \alpha}\right)_\beta \;, \label{def:Jnq}
\end{align}
where  $(2q+1)!! = (2q+1)!/(2^q q!)$ is the double factorial of odd integers.
For classical particles, $a = 0$ and $J_{nq}(\alpha, \beta) = I_{nq} (\alpha, \beta) $.
Using these integrals, the particle number density, the energy density, and the isotropic pressure are $n_0 = I_{10}$, $e_0 = I_{20}$, and $P_0 = I_{21}$. 

The coefficients $\alpha^{(\ell)}_r(\alpha,\beta)$ in Eqs.~(\ref{D_rho}--\ref{D_rho_munu}) are
\begin{align}
	\alpha_r^{(0)} &\equiv -\beta J_{r+1,1} - \frac{n_0}{D_{20}} 
	\left(h_0 G_{2r} - G_{3r} \right) \; ,\label{alpha0}\\
	\alpha_r^{(1)} &\equiv J_{r+1,1} - \frac{J_{r+2,1}}{h_0} \; ,\label{alpha1}\\
	\alpha_r^{(2)} &\equiv \beta J_{r+3,2} \;, \label{alpha2}
\end{align}
where $h_0 \equiv \left(e_0 + P_0\right)/n_0$ is the enthalpy per particle and 
\begin{align}
	G_{nm} &\equiv J_{n0} J_{m0} - J_{n-1,0} J_{m+1,0} \;, \label{eq:G_nm} \\
	D_{nq} &\equiv J_{n+1,q} J_{n-1,q} - J_{nq}^2 \;.
	\label{eq:D_nq}
\end{align}
Note that the thermodynamic integrals introduced in this subsection are computed explicitly for the case of ultrarelativistic particles in Appendix \ref{sec:matrix_el:polys}.

%%%
\subsection{The linearized collision integral and power counting  methods}
\label{sec:DNMR:powerc}

Substituting the near-equilibrium distribution function from Eq.~\eqref{eq:f=f0+deltaf} into the binary collision term \eqref{COLL_INT} and
using the identity $f_{0\mathbf{p}}f_{0\mathbf{p}^{\prime }}\tilde{f}_{0\mathbf{k}}\tilde{f}_{0\mathbf{k}^{\prime }} = f_{0\mathbf{k}}f_{0\mathbf{k}^{\prime }}\tilde{f}_{0\mathbf{p}}\tilde{f}_{0\mathbf{p}^{\prime}}$ while neglecting quadratic terms in $\delta f_\bk$, the irreducible moments of the linearized collision integral become
\begin{align} 
	\hspace{-7pt} C_{r}^{\left\langle \mu _{1}\cdots \mu _{\ell }\right\rangle } 
	&=\frac{1}{2}\int \d K \d K' \d P \d P' \,  W_{\mathbf{kk}' \rightarrow \mathbf{pp}'} \, 
	f_{0\k}f_{0\k'}\tilde{f}_{0 \p}\tilde{f}_{0\p'} \nonumber \\ 
	&\times E_{\k}^{r} k^{\left\langle \mu_{1}\right. }\cdots
	k^{\left. \mu_{\ell }\right\rangle}  \left( \phi_{\p}+\phi_{\p'}-\phi_{\k}-\phi_{\k'}\right) \; .
 \label{Lin_Coll_Int_aux}
\end{align}
Now, inserting the expansion from Eq.~\eqref{eq:delta_f_expansion} into the above formula, 
the moments of the collision integral can be expressed in terms of a linear 
combination of irreducible moments from Eq.~\eqref{eq:rho_def} as
\begin{equation}
	C_{r-1}^{\langle \mu_{1}\cdots \mu_{\ell }\rangle } = -\sum_{n=0}^{N_{\ell}} \mathcal{A}_{rn}^{\left( \ell \right)} \rho_{n}^{\mu_{1}\cdots \mu_{\ell }} \;.
	\label{Lin_Coll_Int}
\end{equation}
The matrix $\mathcal{A}_{rn}^{\left( \ell \right)}$ is defined as~\cite{Denicol:2012cn}
\begin{align}
	\mathcal{A}^{(\ell)}_{rn}
	&\equiv \frac{1}{2 (2\ell+1)}
	\int \d K \d K' \d P \d P' \, W_{\mathbf{kk}' \rightarrow \mathbf{pp}'}  \nonumber \\ 
	&\times  f_{0\k} f_{0\k'} \tilde{f}_{0\p} \tilde{f}_{0\p'}  E_{\k}^{r-1} 
	k^{\left\langle \mu_{1}\right. }\cdots k^{\left. \mu_{\ell }\right\rangle }  \nonumber \\
	&\times \left( \mathcal{H}_{\k n}^{(\ell)}
	k_{\left\langle \mu_{1}\right. } \cdots k_{\left. \mu_{\ell}\right\rangle}
	+\mathcal{H}_{\k' n}^{(\ell)} 
	k^{\prime}_{\left\langle \mu_{1}\right. }\cdots k^{\prime}_{\left. \mu_{\ell}\right\rangle } \right. \nonumber \\
    &\quad \left. 
	- \mathcal{H}_{\p n}^{(\ell)} p_{\left\langle \mu_{1}\right. } \cdots p_{\left. \mu_{\ell}\right\rangle } - \mathcal{H}_{\p' n}^{(\ell)} 
	p'_{\left\langle \mu_{1}\right.}\cdots p'_{\left. \mu_{\ell}\right\rangle } \right) \;,
	\label{eq:CollisionMatrix_Tensor}
\end{align}
and it is separated in loss and gain parts as
\begin{equation}
 \mathcal{A}_{rn}^{\left( \ell \right)} = \mathcal{A}_{rn}^{\left( \ell \right),{\rm l}} - \mathcal{A}_{rn}^{\left( \ell \right), {\rm g}} \;.
\end{equation}
Now, using Eq.~\eqref{eq:Hfunctions} to express $\mathcal{H}^{(\ell)}_{\bk n}$, we obtain
\begin{align}
\mathcal{A}_{rn}^{(\ell),{\rm l}}
&\equiv \frac{W^{(\ell)}}{\ell!}\sum_{m=n}^{N_\ell}\sum_{q=0}^m a^{(\ell)}_{mn}a^{(\ell)}_{mq}  \mathcal{L}^{(\ell)}_{r-1,q} \; , \label{eq:A_expansion_loss} \\
\mathcal{A}_{rn}^{(\ell),{\rm g}} &\equiv 
\frac{W^{(\ell)}}{\ell!}\sum_{m=n}^{N_\ell}\sum_{q=0}^m a^{(\ell)}_{mn}a^{(\ell)}_{mq} \mathcal{G}^{(\ell)}_{r-1,q} \;, \label{eq:A_expansion_gain}
\end{align}
where the corresponding summands $\mathcal{L}_{rn}^{(\ell)}$ and $\mathcal{G}_{rn}^{(\ell)}$ are given by
\begin{align} 
	\mathcal{L}_{rn}^{(\ell)} &\equiv
	\frac{1}{2(2\ell+1)} \int \d P \d P' \d K \d K' \, 
	W_{\mathbf{kk}' \rightarrow \mathbf{pp}'} \nonumber\\
	&\times f_{0\k} f_{0\k'} \tilde{f}_{0\p} \tilde{f}_{0\p'}  E_\k^{r}  k^{\langle \mu_1}\cdots k^{\mu_\ell\rangle} \nonumber\\
	&\times \left(E_\k^n k_{\langle\mu_1}\cdots k_{\mu_\ell\rangle} +  E_{\k'}^n k^{\prime}_{\langle\mu_1}\cdots k^{\prime}_{\mu_\ell\rangle} \right) , \label{eq:loss_aux} 
\end{align}
and
\begin{align}
    \mathcal{\mathcal{G}}_{rn}^{(\ell)} &\equiv \frac{1}{(2\ell+1)}
	\int \d P \d P' \d K \d K' \,  W_{\mathbf{kk}' \rightarrow \mathbf{pp}'} \nonumber\\
    & \times f_{0\k} f_{0\k'} \tilde{f}_{0\p} \tilde{f}_{0\p'}  E_\k^{r} E_\p^n  k^{\langle \mu_1}\cdots k^{\mu_\ell\rangle} p_{\langle\mu_1}\cdots p_{\mu_\ell\rangle} \;, 
	\label{eq:gain_aux}
\end{align}
respectively.
The computation of these summands and of the collision matrix in the ultrarelativistic limit 
$m_0 \beta \rightarrow 0$ is one of the main purposes of this paper and will be discussed in Sect.~\ref{sec:coll}.

The inverse of the collision matrix, the relaxation-time matrix, contains the microscopic 
time scales proportional to the mean free time between collisions $\tau_{\rm mfp} = \lambda_{\rm mfp}/c$,
\begin{equation}
	\tau^{(\ell)}_{rn} \equiv \left(\mathcal{A}^{(\ell)} \right)^{-1}_{rn} = 
	\sum_{m = 0}^{N_\ell} \Omega^{(\ell)}_{rm}
	\frac{1}{\chi^{(\ell)}_m} \left(\Omega^{(\ell)} \right)^{-1}_{mn}\;.
	\label{eq:DNMR_tau}
\end{equation}
Here the matrix $\Omega^{(\ell)}$ diagonalizes $\mathcal{A}^{(\ell)}$, 
leading to eigenvalues that are arranged in 
increasing order, $\chi^{(\ell)}_{r} \le \chi^{(\ell)}_{r+1}$,
\begin{equation}
	\left(\Omega^{(\ell)} \right)^{-1} \mathcal{A}^{(\ell)} \Omega^{(\ell)} = 
	{\rm diag} \left(\chi_0^{(\ell)}, \chi_1^{(\ell)}, \cdots \right) ,
	\label{eq:DNMR_diag}
\end{equation}
where without loss of generality we set $\Omega^{(\ell)}_{00} = 1$. 

The diagonalization of the collision term identifies the slowest microscopic 
time scale that dominates the evolution of the linearized Boltzmann equation~\cite{Denicol:2012cn}. 
However, as discussed in Ref.~\cite{Wagner:2022ayd}, the diagonalization procedure is not required for the computation of the inverse collision matrix $\tau^{(\ell)}$, since it can be obtained by directly inverting $\mathcal{A}^{(\ell)}$ as apparent in Eq.~\eqref{eq:DNMR_tau}.

Using the relaxation-time matrices~\eqref{eq:DNMR_tau} and the $\alpha^{(\ell)}_r$ coefficients from Eqs.~\eqref{alpha0}--\eqref{alpha2}, the first-order transport coefficients 
$\zeta_r$, $\kappa_r$, and $\eta_r$, are defined as
\begin{gather} 
	\zeta_r \equiv \frac{m_0^2}{3} \sum_{n = 0, \neq 1,2}^{N_0} \!
	\tau_{rn}^{(0)} \alpha^{(0)}_n , \nonumber \\ 
	\kappa_r \equiv \sum_{n = 0, \neq 1}^{N_1} \!
	\tau^{(1)}_{rn} \alpha^{(1)}_n , \label{NS_coeff}
	\quad 
	\eta_r \equiv \sum_{n = 0}^{N_2} 
	\tau^{(2)}_{rn} \alpha^{(2)}_n ,
\end{gather}
where the exclusions of $n\neq 1,2$ and $n \neq 1$ from the first and second summations are imposed by the conservation laws~\eqref{cons_eqs}.

The equations of motion for the primary dissipative quantities, $\Pi = -m_0^2 \rho_0/3$,  
$V^{\mu} = \rho^{\mu}_0$, and $\pi^{\mu\nu} = \rho^{\mu \nu}_0$, 
are obtained by performing the matrix multiplication of Eqs.~(\ref{D_rho})--(\ref{D_rho_munu}) with $\tau^{(\ell)}_{nr}$, followed by setting $n = 0$. 
In these equations, the terms of second order in Knudsen and/or inverse Reynolds numbers also contain irreducible moments $\rho_r^{\mu_1 \cdots \mu_\ell}$ with $r\neq0$, which need to be specified, 
while moments with tensor rank $\ell>2$ are omitted in what follows.

Following the DNMR method~\cite{Denicol:2012cn}, the irreducible moments for $0 < r \leq N_\ell$ are approximated by their asymptotic solutions as 
\begin{subequations}\label{eq:DNMR_matching}
\begin{align}
	\rho_{r> 0} &\simeq -\frac{3}{m_0^2}\Omega^{(0)}_{r0} \Pi 
	+\frac{3}{m_0^2} (\zeta_r - \Omega^{(0)}_{r0} \zeta_0) \theta \;, 
	\label{eq:DNMR_matching_rho} \\
	\rho^\mu_{r> 0} &\simeq \Omega^{(1)}_{r0} V^\mu
	+ (\kappa_r - \Omega^{(1)}_{r0} \kappa_0) \nabla^\mu \alpha \; , 
	\label{eq:DNMR_matching_rho_mu} \\
	\rho^{\mu\nu}_{r> 0} &\simeq \Omega^{(2)}_{r0} \pi^{\mu\nu} 
	+ 2(\eta_r - \Omega^{(2)}_{r0} \eta_0) \sigma^{\mu\nu} \; .
	\label{eq:DNMR_matching_rho_munu}
\end{align}
\end{subequations}
The remaining moments of negative order $\rho^{\mu_1 \cdots \mu_\ell}_{-r}$ are obtained 
by substituting only the first terms, 
and hence neglecting terms of order $O(\textrm{Kn})$, from the right-hand sides of Eqs.~\eqref{eq:DNMR_matching_rho}--\eqref{eq:DNMR_matching_rho_munu} 
into Eq.~\eqref{eq:rhoneg}, leading to 
\begin{subequations}\label{DNMR_rho_neg}
\begin{align}
\rho_{-r} &\simeq -\frac{3}{m_0^2}\gamma^{(0)}_{r0} \Pi + \frac{3}{m_0^2}(\Gamma^{(0)}_{r0} - \gamma^{(0)}_{r0}) \zeta_0 \theta\;, \label{DNMR_rho0_neg}\\
\rho_{-r}^\mu &\simeq \gamma^{(1)}_{r0} V^\mu + (\Gamma^{(1)}_{r0} - 
\gamma^{(1)}_{r0}) \kappa_0 \nabla^\mu \alpha\;, \label{DNMR_rho1_neg}\\
\rho_{-r}^{\mu\nu} &\simeq \gamma^{(2)}_{r0} \pi^{\mu\nu} + 2(\Gamma^{(2)}_{r0} 
- \gamma^{(2)}_{r0}) \eta_0 \sigma^{\mu\nu}\;.,\label{DNMR_rho2_neg}
\end{align}
\end{subequations}
where we displayed explicitly both the $O({\rm Re}^{-1})$ and the $O({\rm Kn})$ contributions.
The DNMR coefficients, $\gamma_{r0}^{(\ell)}$, are
\begin{gather}
	\gamma_{r0}^{(0)} \equiv \sum_{n = 0,\neq 1,2}^{N_0} \! \mathcal{F}^{(0)}_{rn}\Omega^{(0)}_{n0} \;,
	\nonumber\\
	\gamma_{r0}^{(1)} \equiv \sum_{n = 0,\neq 1}^{N_1} \!\mathcal{F}^{(1)}_{rn} \Omega^{(1)}_{n0} \;, \quad 
	\gamma_{r0}^{(2)} \equiv \sum_{n = 0}^{N_2} \mathcal{F}^{(2)}_{rn} \Omega^{(2)}_{n0} \;.
	\label{gamma}
\end{gather}
In addition, we introduced the so-called corrected DNMR coefficients, $\Gamma^{(\ell)}_{r0}$, defined as \cite{Wagner:2022ayd,Ambrus:2022vif}
\begin{gather}
	\Gamma^{(0)}_{r0} \equiv \sum_{n = 0, \neq 1, 2}^{N_0} \mathcal{F}^{(0)}_{rn} \frac{\zeta_n}{\zeta_0} \;,
	\nonumber\\ 
	\Gamma^{(1)}_{r0} \equiv \sum_{n = 0, \neq 1}^{N_1} \mathcal{F}^{(1)}_{rn} \frac{\kappa_n}{\kappa_0} \;,
	\quad
	\Gamma^{(2)}_{r0} \equiv \sum_{n = 0}^{N_2} \mathcal{F}^{(2)}_{rn} \frac{\eta_n}{\eta_0} \;.
	\label{Gamma}
\end{gather}

On the other hand, in the cDNMR approach, the thermodynamic forces can be replaced by the Navier-Stokes relations, 
$\theta = -\Pi / \zeta_0$, $\nabla^\mu \alpha = V^\mu / \kappa_0$, 
and $\sigma^{\mu\nu} = \pi^{\mu\nu} / (2 \eta_0)$. Therefore substituting the right-hand sides of Eqs.~\eqref{DNMR_rho_neg} eliminates the $O({\rm Kn})$ contributions to the negative-order moments $\rho^{\mu_1 \cdots \mu_\ell}_{r < 0}$ and yields~\cite{Wagner:2022ayd,Ambrus:2022vif}
\begin{equation}
	\rho_{-r} \simeq -\frac{3}{m_0^2}\Gamma^{(0)}_{r0} \Pi \;, \
	\rho_{-r}^\mu \simeq \Gamma^{(1)}_{r0} V^\mu \;, \
	\rho_{-r}^{\mu\nu} \simeq \Gamma^{(2)}_{r0} \pi^{\mu\nu} \;.
	\label{IReD_rhoneg}
\end{equation}

Finally, the so-called Inverse-Reynolds Dominance (IReD) approximation of Ref.~\cite{Wagner:2022ayd}
defines a power counting scheme without the diagonalization procedure, such that the irreducible moments are of order $O({\rm Re}^{-1})$. 
The non-dynamical positive-order moments are given by
\begin{equation}
	\rho_{r> 0} \simeq -\frac{3}{m_0^2} \mathcal{C}^{(0)}_{r0} \Pi \;, \
	\rho^\mu_{r> 0} \simeq \mathcal{C}^{(1)}_{r0} V^\mu\;, \
	\rho^{\mu\nu}_{r> 0}\simeq \mathcal{C}^{(2)}_{r0}\pi^{\mu\nu} \;,
	\label{IReD_matching}
\end{equation}
where the corresponding IReD coefficients, $\mathcal{C}^{(\ell)}_{r0}$, are 
\begin{equation}
    \mathcal{C}^{(0)}_{r0}\equiv \frac{\zeta_r}{\zeta_0}\;, \quad
    \mathcal{C}^{(1)}_{r0}\equiv \frac{\kappa_r}{\kappa_0} \;,\quad
    \mathcal{C}^{(2)}_{r0}\equiv \frac{\eta_r}{\eta_0}\;,
    \label{eq:C_def}
\end{equation}
while the negative-order moments are given by Eqs.~\eqref{IReD_rhoneg}. 

To simplify our notation we will introduce a common variable, $\xi^{(\ell)}_r$, for the transport coefficients~\eqref{NS_coeff} in what follows,
\begin{equation}
	\xi^{(0)}_r=\zeta_r \;, \quad \xi^{(1)}_r=\kappa_r \;, \quad \xi^{(2)}_r = \eta_r \;.
\end{equation}
To study the different power-counting schemes, we introduce the following notation for the non-dynamical moments encompassing the DNMR, the cDNMR, and the IReD approximations,
\begin{align} 
	\rho_{r} &= -\frac{3}{m^2_0}\mathcal{X}^{(0)}_{r0} \Pi 
	+ \frac{3}{m^2_0} \mathcal{Y}^{(0)}_{r0} \theta \;,  \label{rho_XY}\\
	\rho_{r}^\mu &= \mathcal{X}^{(1)}_{r0} V^\mu + \mathcal{Y}^{(1)}_{r0} \nabla^\mu \alpha \; , \label{rho_mu_XY} \\
	\rho_{r}^{\mu\nu} &= \mathcal{X}^{(2)}_{r0} \pi^{\mu\nu} +  2\mathcal{Y}^{(2)}_{r0} \sigma^{\mu\nu}\; .
	\label{rho_munu_XY}
\end{align}
Here, for $r=0$, in all cases, by definition,
\begin{equation}
	\mathcal{X}^{(\ell)}_{00} = \Omega^{(\ell)}_{00} = \mathcal{C}^{(\ell)}_{00} = 1 \;, \;\;\;\; \mathcal{Y}^{(\ell)}_{00} = 0 \;.
\end{equation}
For $r\neq 0$, the DNMR coefficients are
\begin{align} \label{DNMR_X}
	\mathcal{X}^{(\ell)}_{r0}  &= 
	\begin{cases}
	\Omega^{(\ell)}_{r0} \;, \;\; &r > 0\;,\\
	\gamma^{(\ell)}_{-r,0} \;, \;\; &r< 0\;,
	\end{cases}
\end{align}
and 
\begin{align} \label{DNMR_Y}
	\mathcal{Y}^{(\ell)}_{r0}  &= 
	\begin{cases}
		\xi^{(\ell)}_r - \Omega^{(\ell)}_{r0} \xi^{(\ell)}_0\;,  &r> 0\;, \\
		\left(\Gamma^{(\ell)}_{-r,0} -\gamma^{(\ell)}_{-r,0} \right) \xi^{(\ell)}_0 \;,  &r< 0\;,
	\end{cases}
\end{align}
as follows from Eqs.~\eqref{eq:DNMR_matching} and \eqref{DNMR_rho_neg}.

Similarly, the cDNMR coefficients are
\begin{align} \label{cDNMR_X}
	\mathcal{X}^{(\ell)}_{r0}  = 
	\begin{cases}
	\Omega^{(\ell)}_{r0} \;, \;\; &r > 0\;,\\
	\Gamma^{(\ell)}_{-r,0} \;, \;\; &r< 0\;,
	\end{cases}
\end{align}
and
\begin{align} \label{cDNMR_Y}
	\mathcal{Y}^{(\ell)}_{r0}  &= 
	\begin{cases}
		\xi^{(\ell)}_r - \Omega^{(\ell)}_{r0} \xi^{(\ell)}_0\;,  &r> 0\;, \\
		0  \;,  &r < 0\;,
	\end{cases}
\end{align}
as it is apparent from Eqs.~\eqref{eq:DNMR_matching} and \eqref{IReD_rhoneg}.

Finally, the IReD coefficients can be identified from Eqs.~\eqref{IReD_rhoneg} and \eqref{IReD_matching}:
\begin{align} \label{IRed_X}
	\mathcal{X}^{(\ell)}_{r0}  &= \begin{cases}
	\mathcal{C}^{(\ell)}_{r0} \;, \;\; &r > 0\;,\\
	\Gamma^{(\ell)}_{-r,0} \;, \;\; &r< 0\;,
	\end{cases}
\end{align}
while, by definition,
\begin{equation} \label{IRed_Y}
	\mathcal{Y}^{(\ell)}_{r0} = 0\;, \;\;\;\; r \neq 0\;.
\end{equation}

Note that the following relation holds for all of the approaches:
\begin{align}
	\xi^{(\ell)}_0 \mathcal{X}^{(\ell)}_{r0} + \mathcal{Y}^{(\ell)}_{r0}
	&= 
	\begin{cases}
		\xi^{(\ell)}_r \;, \;\; &r \geq 0\;,\\
		\xi^{(\ell)}_0 \Gamma^{(\ell)}_{-r,0} \;, \;\; &r < 0\;.
	\end{cases}\label{eq:relation_XY}
\end{align}

%%%
\subsection{Second-order fluid dynamical equations}
\label{sec:DNMR:hydro}

The relaxation equations for the irreducible moments 
are obtained by multiplying Eqs.~\eqref{D_rho}--\eqref{D_rho_munu} by $\tau^{(\ell)}_{nr}$ and then
summing over $r$. 
Employing the expression
\begin{equation}
	\sum^{N_\ell}_{r=0} \tau^{(\ell)}_{nr}  C_{r-1}^{\langle \mu_{1}\cdots \mu_{\ell }\rangle } 
	= -\rho_{n}^{\langle \mu_{1}\cdots \mu_{\ell }\rangle } \;,
\end{equation}
valid for the linearized collision model \eqref{Lin_Coll_Int_aux}--\eqref{Lin_Coll_Int} and derived using the property
\begin{equation}
	\sum_{r = 0}^{N_\ell}\tau^{(\ell)}_{nr} \mathcal{A}^{(\ell)}_{rm} = \delta_{nm} \;,
\end{equation}
the second-order transport equations with a linearized collision integral  for 
$\Pi$, $V^\mu$, and $\pi^{\mu\nu}$ from Ref.~\cite{Denicol:2012cn} read
\begin{align}
	\tau_\Pi \dot{\Pi} + \Pi &=  -\zeta \theta + \mathcal{J} + \mathcal{K} \;, \label{Pidot} \\
	\tau_V \dot{V}^{\langle \mu \rangle} + V^\mu &= \kappa \nabla^\mu \alpha + 
	\mathcal{J}^\mu + \mathcal{K}^\mu \;, \label{Vdot} \\
	\tau_\pi \dot{\pi}^{\langle \mu \nu \rangle} + \pi^{\mu\nu} &= 
	2 \eta \sigma^{\mu\nu} + \mathcal{J}^{\mu\nu} + \mathcal{K}^{\mu\nu} \; . \label{pidot}
\end{align}
Here, $\tau_\Pi$, $\tau_V$, and $\tau_\pi$ are the relaxation 
times, while $\zeta = \zeta_0$, $\kappa = \kappa_0$, and $\eta=\eta_0$ 
are the first-order transport coefficients, 
\begin{alignat}{3}
	\zeta &= \frac{m_0^2}{3} \!\sum_{r = 0,\neq 1,2}^{N_0} \!
	\tau_{0r}^{(0)} \alpha^{(0)}_{r} \;, \quad
	&&\tau_\Pi = \sum_{r = 0,\neq 1,2}^{N_0} \! \tau_{0r}^{(0)}  \mathcal{X}_{r0}^{(0)} \;, \\ 
	\kappa &= \sum_{r = 0,\neq 1}^{N_1} \! \tau_{0r}^{(1)} \alpha^{(1)}_{r} \;, \quad 
	&&\tau_V = \sum_{r = 0,\neq 1}^{N_1} \! \tau_{0r}^{(1)}  \mathcal{X}_{r0}^{(1)} \;, \\ 
	\eta &= \sum_{r = 0}^{N_2}  \tau_{0r}^{(2)} \alpha^{(2)}_{r} \;, \quad 
	&&\tau_\pi = \sum_{r = 0}^{N_2} \tau_{0r}^{(2)}  \mathcal{X}_{r0}^{(2)} \;. 
 \label{eq:toceffs_firstorder}
\end{alignat}
Furthermore, $\mathcal{J}, \mathcal{J}^{\mu}$, and $\mathcal{J}^{\mu\nu}$ 
collect the terms of order $O({\rm Re}^{-1} \rm{Kn})$,
\begin{align}
	\mathcal{J} &= -\ell_{\Pi V} \nabla_\mu V^\mu - 
	\tau_{\Pi V} V_\mu \dot{u}^\mu - \delta_{\Pi\Pi} \Pi \theta \nonumber\\
	& - \lambda_{\Pi V} V_\mu \nabla^\mu \alpha + 
	\lambda_{\Pi \pi} \pi^{\mu\nu} \sigma_{\mu\nu} \;,\label{J}\\
	\mathcal{J}^\mu &= -\tau_V V_\nu \omega^{\nu\mu} - \delta_{VV} V^\mu \theta 
	- \ell_{V\Pi} \nabla^\mu \Pi \nonumber\\
	& + \ell_{V\pi} \Delta^{\mu\nu} \nabla_\lambda \pi^\lambda{}_\nu + \tau_{V\Pi} \Pi \dot{u}^\mu - 
	\tau_{V\pi} \pi^{\mu\nu} \dot{u}_\nu \nonumber\\
	& -\lambda_{VV} V_\nu \sigma^{\mu\nu} + 
	\lambda_{V \Pi} \Pi \nabla^\mu \alpha - \lambda_{V\pi} \pi^{\mu\nu} \nabla_\nu \alpha \;,\label{J_mu}\\
	\mathcal{J}^{\mu\nu} &= 2\tau_\pi \pi^{\langle\mu}_\lambda \omega^{\nu\rangle \lambda} - 
	\delta_{\pi\pi} \pi^{\mu\nu} \theta - 
	\tau_{\pi\pi} \pi^{\lambda\langle \mu}\sigma^{\nu\rangle}_\lambda + 
	\lambda_{\pi \Pi} \Pi \sigma^{\mu\nu} \nonumber\\
	& - \tau_{\pi V} V^{\langle \mu} \dot{u}^{\nu \rangle} + \ell_{\pi V} \nabla^{\langle \mu} 
	V^{\nu \rangle} + \lambda_{\pi V} V^{\langle \mu} \nabla^{\nu \rangle} \alpha \;.
	\label{J_munu}
\end{align}
Finally, the tensors $\mathcal{K}$, $\mathcal{K}^\mu$, and 
$\mathcal{K}^{\mu\nu}$ contain all contributions of order $O({\rm Kn}^2)$, given by
\begin{align}
	\mathcal{K} &= \widetilde{\zeta}_1  \omega_{\mu\nu} \omega^{\mu\nu} 
	+ \widetilde{\zeta}_2 \sigma_{\mu\nu} \sigma^{\mu\nu} + \widetilde{\zeta}_3 \theta^2 
	+ \widetilde{\zeta}_4 I^\mu I_\mu \nonumber \\
	&+ \widetilde{\zeta}_5 \dot{u}^\mu \dot{u}_\mu 
	+ \widetilde{\zeta}_6 I^\mu \dot{u}_\mu  + \widetilde{\zeta}_7 \nabla^\mu I_\mu 
	+ \widetilde{\zeta}_8 \nabla^\mu \dot{u}_\mu  \;,\label{eq:K}\\
	\mathcal{K}^\mu &= \widetilde{\kappa}_1 \sigma^{\mu\nu} I_\nu + \widetilde{\kappa}_2 \sigma^{\mu\nu} \dot{u}_\nu 
	+ \widetilde{\kappa}_3 I^\mu \theta + \widetilde{\kappa}_4 \dot{u}^\mu \theta \nonumber \\
	&+ \widetilde{\kappa}_5 \omega^{\mu\nu} I_\nu 
	+ \widetilde{\kappa}_6 \Delta^\mu_\lambda \nabla_\nu \sigma^{\lambda \nu} 
	+ \widetilde{\kappa}_7 \nabla^\mu \theta \;,\label{eq:Kmu}\\
	\mathcal{K}^{\mu\nu} &= \widetilde{\eta}_1 \omega^{\lambda\langle \mu} \omega^{\nu\rangle}{}_\lambda 
	+ \widetilde{\eta}_2 \theta \sigma^{\mu\nu} + \widetilde{\eta}_3 \sigma^{\lambda \langle \mu} \sigma^{\nu\rangle}_\lambda + \widetilde{\eta}_4 \sigma^{\langle \mu}_\lambda \omega^{\nu\rangle \lambda}
	\nonumber \\ 
	&+ \widetilde{\eta}_5 I^{\langle\mu} I^{\nu \rangle} + \widetilde{\eta}_6 \dot{u}^{\langle\mu} \dot{u}^{\nu\rangle} 
	+ \widetilde{\eta}_7 I^{\langle\mu}\dot{u}^{\nu\rangle} + \widetilde{\eta}_8 \nabla^{\langle\mu} I^{\nu \rangle} \nonumber\\
	&+ \widetilde{\eta}_9 \nabla^{\langle \mu} \dot{u}^{\nu \rangle}\;,\label{eq:Kmunu}
\end{align}
where $I^\mu = \nabla^{\mu} \alpha$ was introduced.

Note that all coefficients appearing in Eqs.~\eqref{J}--\eqref{J_munu} and Eqs.~\eqref{eq:K}--\eqref{eq:Kmunu} are calculated using the $\mathcal{X}^{(\ell)}_{r0}$ and $\mathcal{Y}^{(\ell)}_{r0}$ notation in Appendix~\ref{app:2nd_order_coeff}.

%%%
\section{Exact collision matrices}
\label{sec:coll}

In this section, we provide exact expressions for the matrix elements of the linearized collision term assuming that the differential cross-section is constant. The transition rate from Eq.~\eqref{eq:W_general} now reads
\begin{equation}
 W_{\mathbf{kk}' \rightarrow \mathbf{pp}'} = s (2\pi)^5 \sigma_T 
 \delta(k^\mu + k'^\mu - p^\mu -p'^\mu)\;.\label{eq:W_HS}
 \end{equation}
We focus on the case of a massless, classical (Boltzmann) gas, such that 
\begin{equation}
 f_{0\k} = e^{\alpha -\beta E_\k} \;, \qquad 
\tilde{f}_{0\k} = 1 \;.
\end{equation}
Therefore the loss and gain terms introduced in Eqs.~\eqref{eq:loss_aux} and \eqref{eq:gain_aux} simplify to
\begin{align} 
	\mathcal{L}_{rn}^{(\ell)} &=
	\frac{\sigma_T}{(2\ell+1)} \int \d K \d K' f_{0\k} f_{0\k'} E_\k^{r} \frac{s}{2} \nonumber\\
	&\times k^{\langle \mu_1}\cdots k^{\mu_\ell\rangle}\left(E_\k^n k_{\langle\mu_1}\cdots k_{\mu_\ell\rangle} +  E_{\k'}^n k^{\prime}_{\langle\mu_1}\cdots k^{\prime}_{\mu_\ell\rangle} \right) \;, \label{eq:loss_aux_2} 
\end{align}
and
\begin{align}
    \mathcal{\mathcal{G}}_{rn}^{(\ell)} &=2\frac{\sigma_T  (2\pi)^5}{(2\ell+1)}
	\int \d P \d P' \d K \d K' f_{0\k} f_{0\k'} E_\k^{r} E_\p^n \frac{s}{2} \nonumber\\
	& \times \delta \left( k^{\mu} + k^{\prime \mu} - p^{\mu} - p^{\prime \mu} \right) k^{\langle \mu_1}\cdots k^{\mu_\ell\rangle} p_{\langle\mu_1}\cdots p_{\mu_\ell\rangle} \; .
	\label{eq:gain_aux_2}
\end{align}
We refer the reader to Appendices~\ref{app:proj}--\ref{app:gain_mat} for the details of 
the calculations. Specifically, Appendices \ref{app:proj} and \ref{app:PPprime} cover general techniques for solving the relevant collision integrals, while the results of Sect.~\ref{sec:coll_loss} are derived in Appendices \ref{app:loss} and \ref{app:loss_mat}. Appendices \ref{app:gain} and \ref{app:gain_mat} finally cover the results of Sect. \ref{sec:coll_gain}.

%%%
\subsection{The loss terms}
\label{sec:coll_loss}

The auxiliary loss terms defined in Eq.~\eqref{eq:loss_aux_2} can be obtained in closed form and read in the cases $\ell=0,1$, and $\ell=2$ as follows:
\begin{subequations}\label{eq:loss}
\begin{multline}
\mathcal{L}_{rn}^{(0)} =\frac{ P_0^2 \sigma_T}{4} \beta^{2-r-n}\Big[2\Gamma(r+n+3)\\
+\Gamma(r+3)\Gamma(n+3)\Big]\;,\label{eq:L_0}
\end{multline}
\begin{multline}
\mathcal{L}_{rn}^{(1)} =\frac{ P_0^2 \sigma_T}{36} \beta^{-r-n}\Big[-6\Gamma(r+n+5)\\
+\Gamma(r+4)\Gamma(n+4)\Big]\;,\label{eq:L_1}
\end{multline}
and
\begin{equation}
\mathcal{L}_{rn}^{(2)}=\frac{ P_0^2 \sigma_T}{15} \beta^{-2-r-n}\,\Gamma(r+n+7)\;.\label{eq:L_2}
\end{equation}
\end{subequations}
Now, inserting these expressions into Eq.~\eqref{eq:A_expansion_loss} and performing 
the summations, we obtain the loss part of the collision matrices,
\begin{subequations} \label{eq:A_loss}
\begin{align}
 \mathcal{A}^{(0),{\rm l}}_{rn} &=  \beta P_0 \sigma_T \left[ \delta_{nr} + 
 \frac{(r+1)!}{2} \delta_{n1} \beta^{1-r}\right] \;,\\
 \mathcal{A}^{(1),{\rm l}}_{rn} &=  \beta P_0 \sigma_T \left[ \delta_{nr} - 
 \frac{(r+2)!}{6} \delta_{n0} \beta^{-r}\right] \;,\\
 \mathcal{A}^{(2),{\rm l}}_{rn} &=  \beta P_0 \sigma_T \delta_{nr} \;.
\end{align}
\end{subequations}

%%%
\subsection{The gain terms}\label{sec:coll_gain}

Similarly, the auxiliary gain terms defined in Eq.~\eqref{eq:gain_aux_2} are given in closed form by
\begin{subequations}
\begin{multline}
 \mathcal{G}^{(0)}_{rn} = \frac{\sigma_TP_0^2\beta^{2-r-n}}{(n+1)(r+1)}\\\times 
 \left[\Gamma(4+n+r)-\Gamma(3+r)\Gamma(3+n)\right]\;,\label{gain0_main}
\end{multline}
\begin{multline}
    \mathcal{G}_{rn}^{(1)}= -\frac{ \sigma_T P_0^2 \beta^{-r-n}}{3(1+r)(2+r)(1+n)(2+n)} \\\times [(r+n+rn-3)\Gamma(6+n+r)\\
    +(3r+3n+rn+11)\Gamma(4+r)\Gamma(4+n)]\;,
    \label{eq:gain1_main}
\end{multline}
and
\begin{multline}
    \mathcal{G}_{rn}^{(2)}=\frac{2\sigma_T P_0^2 \beta^{-2-r-n}}{15(1+n)(2+n)(3+n)(1+r)(2+r)(3+r)}\\
     \times \Big\{ \big[64-6(r+n)+2(r^2+n^2)-3rn\\
    3(n^2 r+r^2 n)+r^2n^2 \big]\Gamma(8+r+n)\\
    -\big[22+4(r+n)+rn\big]\Gamma(6+r)\Gamma(6+n)\Big\}\label{eq:gain2_main}\;.
\end{multline}
\end{subequations}
Plugging these expressions into Eq.~\eqref{eq:A_expansion_gain}, the resulting gain contributions to the collision matrices read
\begin{subequations}
\begin{align}
\mathcal{A}_{0n}^{(0),{\rm g}} &=
 \frac{2 (-1)^n   \sigma_T P_0\beta^{1+n}}{(n+1)!} 
 \left[\delta_{n0}-S_n^{(0)}(N_0)\right]\;,\nonumber\\
 \mathcal{A}_{r>0,n\leq r}^{(0),{\rm g}} &= 
 \frac{2 \sigma_T P_0\beta^{1+n-r}(r+1)!}{ r (n+1)!}
 \left( 1 - \delta_{n0}\right)\;,\label{eq:A_g_0}
\end{align}
\begin{align}
 \mathcal{A}_{0n}^{(1),{\rm g}} &=
 \frac{16 (-1)^n \sigma_T P_0\beta^{1+n} }{(n+3)!}
 \Bigg[\frac{3}{4}\delta_{n0} -S_n^{(1)}(N_1)\Bigg]\;,\nonumber\\
 \mathcal{A}_{r>0,n \le r}^{(1),{\rm g}} &= 
 \frac{2 \sigma_T P_0 \beta^{1+n-r}(r+2)!}{ (n+3)!}
 \frac{n(r+4)-r}{r(r+1)}\;,\label{eq:A_g_1}
\end{align}
and
\begin{align}
\mathcal{A}^{(2),{\rm g}}_{0n} &= \frac{432 (-1)^n\sigma_T P_0\beta^{1+n}}{\lambda_{\rm mfp} (n+5)!} \left[\frac{5}{18}\delta_{n0}-S^{(2)}_n(N_2)\right]\;,\nonumber\\
\mathcal{A}^{(2),{\rm g}}_{r>0,n\le r} &= \frac{2\sigma_T P_0\beta^{1+n-r} (r+4)! (n+1)(9n+nr-4r)}{ (n+5)! r(r+1)(r+2)}\;,\label{eq:A_g_2}
\end{align}
while
\begin{equation}
	\mathcal{A}^{(\ell),{\rm g}}_{r>0,n > r} = 0\; .
\end{equation}
\end{subequations}
Here we defined the auxiliary sums
\begin{equation}
 S^{(\ell)}_n \left(N_\ell \right) \equiv \sum_{m = n}^{N_\ell} 
 \binom{m}{n} \frac{1}{(m+\ell)(m+\ell+1)}\;. \label{eq:sum_aux}
 \end{equation}

%%%
\subsection{The collision matrices}\label{sec:matrix_el:summary}

Collecting the results from the previous subsections, we can write down closed-form expressions for the elements $\mathcal{A}^{(\ell)}_{rn}$ of the collision matrices. 
In the case when $\ell=0$, we obtain
\begin{align}
\mathcal{A}^{(0)}_{00} &=\frac{1}{\lambda_{\text{mfp}}} \frac{N_0-1}{N_0+1}\;,\nonumber\\
 \mathcal{A}^{(0)}_{0,n>0} &= \frac{2(-\beta)^n}{\lambda_{\rm mfp} (n+1)!} 
 \left[S^{(0)}_n(N_0)-\frac{\delta_{n1}}{2}\right]\;,\nonumber\\
 \mathcal{A}^{(0)}_{r>0, n \le r} &= \frac{\beta^{n-r}(r+1)!}{\lambda_{\rm mfp}  (n+1)!} \left(\delta_{nr} + \frac{2}{r} \delta_{n0} + \delta_{n1} -\frac{2}{r}\right)\;,\nonumber\\
 \mathcal{A}_{r>0,n>r}^{(0)}&=0\;.\label{A_0}
\end{align}
Note that $\mathcal{A}_{1n}^{(0)}=\mathcal{A}_{2n}^{(0)}=0$, since the particle number and energy are conserved, while $\mathcal{A}^{(0)}_{r>0,0}=0$.

Similarly, when $\ell=1$ we have,
\begin{align}
 \mathcal{A}^{(1)}_{0n} &= \frac{16  (-\beta)^n}{\lambda_{\rm mfp}(n+3)!}
 \left[S^{(1)}_n(N_1) - \frac{\delta_{n0}}{2}\right],\nonumber\\
 \mathcal{A}_{r>0,n \le r}^{(1)} &= 
 \frac{ \beta^{n-r}(r + 2)!}{\lambda_{\rm mfp} (n+3)! r} 
 \left( 4n + nr -r \right) \nonumber \\
 & \times\left(\delta_{nr} + \delta_{n0} - 
 \frac{2}{r+1}\right),\nonumber\\
 \mathcal{A}_{r>0,n > r}^{(1)} &= 0,\label{A_1}
\end{align}
where $\mathcal{A}_{1n}^{(1)}=0$ reflects the momentum conservation in binary collisions.

Summarizing the results for $\ell=2$, we have
\begin{align}
 \mathcal{A}^{(2)}_{0n} &= \frac{432(-\beta)^n}{\lambda_{\rm mfp} (n+5)!} S^{(2)}_n(N_2)\;,\nonumber\\
 \mathcal{A}^{(2)}_{r>0,n\le r} &= \frac{ \beta^{n-r} (r+4)! (n+1)}{\lambda_{\rm mfp} (n+5)! r(r+1)} \left(9n+nr-4r \right)\nonumber\\
 & \times \left(\delta_{nr} - \frac{2}{r+2}\right)\;,\nonumber\\
 \mathcal{A}^{(2)}_{r>0,n > r} &= 0\;.\label{A_2}
\end{align}

All these collision matrices share a similar structure, in the sense that they are almost lower triangular matrices. 
In all cases, all entries appearing on the zeroth row are non-vanishing, most of them diverging when $N_\ell\to\infty$ with different degrees of severity. 
Furthermore, the matrices for tensor-rank $\ell\leq2$ have $2-\ell$ vanishing rows due to the conservation of  the particle number and of four-momentum in binary collisions. Note that the nonvanishing entries on the zeroth row imply that the moments corresponding to hydrodynamic variables, i.e., $\rho_0$, $\rho_0^\mu$, and $\rho_0^{\mu\nu}$, couple to all moments of the same tensor-rank, which was also a conclusion found in 
Ref.~\cite{Denicol:2022bsq} in the case of the $\lambda \varphi^4$-theory.

%%% Kappa Table %%%
\begin{table*}[htb!]
	\begin{tabular}{|c|c|c|c|c|c|c|}
		\hline
		Method &
		$\kappa$ & $\tau_V [\lambda_{\rm mfp}]$ & $\delta_{VV}[\tau_V]$ 
		&$\ell_{V\pi}[\tau_V] = \tau_{V\pi}[\tau_V]$
		& $\lambda_{VV}[\tau_V]$ 
		& $\lambda_{V\pi}[\tau_V]$  \\[2pt] \hline \hline
		$14$M &
		$1/12\sigma$
		& $9/4$
		& $1$
		& $\beta/20$
		& $3/5$ 
		& $\beta/20$ \\[2pt] \hline
		IReD &
		$0.15892/\sigma$
		& $2.0838$
		& $1$
		& $0.028371\beta$
		& $0.89862$ 
		& $0.069273\beta$ \\[2pt] \hline
		DNMR&
		$0.15892/\sigma$
		& $2.5721$
		& $1$
		& $0.11921\beta$
		& $0.92095$ 
		& $0.051709\beta$ \\[2pt] \hline
		cDNMR &
		$0.15892/\sigma$
		& $2.5721$
		& $1$
		& $0.098534\beta$
		& $0.92095$
		& $0.056878\beta$ \\[2pt] 
		\hline
	\end{tabular}
	\caption{The coefficient of diffusion and second-order transport coefficients in $\mathcal{J}^\mu$ for the particle-diffusion current $V^\mu$, evaluated in different approaches.}
	\label{tbl:diffusion}
\end{table*}

%%% Eta Table %%%
\begin{table*}[htb!]
	\begin{tabular}{|c|c|c|c|c|c|c|c|}
		\hline
		Method &
		$\eta$ & $\tau_\pi [\lambda_{\rm mfp}]$ & $\delta_{\pi\pi}[\tau_\pi]$ 
		&$\ell_{\pi V}[\tau_\pi]$
		& $\tau_{\pi V}[\tau_\pi]$ 
		& $\tau_{\pi \pi}[\tau_\pi]$
		& $\lambda_{\pi V}[\tau_\pi]$  \\[2pt] \hline \hline
		$14$M
		& $4/(3\sigma\beta)$
		& $5/3$
		& $4/3$
		& $0$
		& $0$
		& $10/7$ 
		& $0$ \\[2pt] \hline
		IReD 
		& $1.2676/(\sigma\beta)$
		& $1.6557$
		& $4/3$
		& $-0.56960/\beta$
		& $-2.2784/\beta$
		& $1.6945$ 
		& $0.20503/\beta$ \\[2pt] \hline
		DNMR \& cDNMR&
		$1.2676/(\sigma\beta)$
		& $2$
		& $4/3$
		& $-0.68317/\beta$
		& $-2.7327/\beta $
		& $1.6888$
		& $0.24188/\beta $ \\[2pt] \hline
	\end{tabular}
	\caption{Same as Table \ref{tbl:diffusion}, but for $\eta$ and the second-order transport coefficients in $\mathcal{J}^{\mu\nu}$ for the shear-stress tensor $\pi^{\mu\nu}$.}
	\label{tbl:shear}
\end{table*}

%%% KappaTilde Table %%%
\begin{table*}[htb!]
	\begin{tabular}{|c|c|c|c|c|}
		\hline
		Method &
		$\widetilde{\kappa}_1[\tau_V]$ 
		&$\widetilde{\kappa}_3[\tau_V]$
		&$\widetilde{\kappa}_5[\tau_V]$
		& $\widetilde{\kappa}_6[\tau_V]$ \\[2pt] \hline \hline
		DNMR
		&$0.050292$
		&$0.020115$
		&$-0.060345$
		&$-0.24395$ \\[2pt] \hline 
		cDNMR 
		&$0.050292$
		&$0.020115$
		&$-0.060345$
		&$-0.19152$ \\[2pt] 
		\hline
	\end{tabular}
	\caption{Second-order transport coefficients in $\mathcal{K}^\mu$ for the particle-diffusion current $V^\mu$ evaluated using the DNMR and corrected DNMR methods. The IReD and strict $14$M approaches lead to
		$\widetilde{\kappa}_1=\widetilde{\kappa}_3=\widetilde{\kappa}_5
		=\widetilde{\kappa}_6 = 0$.}
	\label{tbl:diffusion_Kn2}
\end{table*}

%%% EtaTilde Table %%%
\begin{table*}[hbt!]	
	\begin{tabular}{|c|c|c|c|c|c|c|c|c|c|}
		\hline
		Method &
		$\beta\widetilde{\eta}_1[\tau_\pi]$ 
		&$\beta\widetilde{\eta}_2[\tau_\pi]$
		&$\beta\widetilde{\eta}_3[\tau_\pi]$
		&$\beta\widetilde{\eta}_4[\tau_\pi]$
		&$\beta\widetilde{\eta}_5[\tau_\pi]$
		&$\beta\widetilde{\eta}_6[\tau_\pi]$
		&$\beta\widetilde{\eta}_7[\tau_\pi]$
		&$\beta\widetilde{\eta}_8[\tau_\pi]$
		&$\beta\widetilde{\eta}_9[\tau_\pi]$ \\[2pt] \hline \hline
		DNMR \& cDNMR
		&$-0.43647$
		&$0.14549$
		&$0.28867$
		&$-0.87294$
		&$-0.011466$
		&$-2.1824$
		&$-0.13454$
		&$0.033634$
		&$0.43647$ \\[2pt] \hline 
	\end{tabular}
	\caption{Same as Table \ref{tbl:diffusion_Kn2}, but for the second-order transport coefficients in $\mathcal{K}^{\mu\nu}$ for the shear-stress tensor $\pi^{\mu\nu}$.}
	\label{tbl:shear_Kn2}
\end{table*}

%%%
\section{Second-order transport coefficients}
\label{sec:2nd_order_coeff}

In this section, we compute all second-order transport coefficients from 
Eqs.~\eqref{J}--\eqref{eq:Kmunu} in the ultrarelativistic limit.  
The general expressions of these coefficients for arbitrary particle mass and various power-counting schemes are listed in Appendix~\ref{app:2nd_order_coeff}. 
All second-order transport coefficients are related to the inverse of the collision matrices $\tau^{(\ell)}_{rn}$, for which we obtained analytical expression only in the scalar case when $\ell = 0$. 
For $\ell = 1$ and $2$, we employed numerical computations to find the inverse of the collision matrices 
given in Eqs.~\eqref{A_1} and \eqref{A_2}. 

The numerical values were obtained through an extrapolation with respect to $1/N_2$ by computing the best fit parameters $a$, $b$, and $\nora{t}_\infty$ of the power law $\nora{t}(N_2) = \nora{t}_\infty + a N_2^{-b}$, where $\nora{t}$ denotes a generic transport coefficient with convergence value $\nora{t}_\infty$. The fits are done on data points up to 
$N_2 = 1000$ through \texttt{gnuplot} scripts that are included in the supplementary material to this paper.
All transport coefficients are listed to five significant digits, which is justified by the asymptotic standard deviation of the fit being of order $O(10^{-6})$ or lower.
We remark that the coefficients do not converge at the same speed. Specifically, we can estimate the values of $N_2$ for fixed relative differences between all transport coefficients and their respective convergence values as $N_2[O(10^{-4})]\simeq 16491$, $N_2[O(10^{-5})]\simeq 168329$, and $N_2[O(10^{-6})]\simeq 1718186$, respectively. These large numbers can be attributed mainly to the slow convergence of the coefficient $\zeta_4$ in the cDNMR approach. For contrast, IReD leads to $N_2[O(10^{-4})]\simeq 171$, $N_2[O(10^{-5})]\simeq 554$, and $N_2[O(10^{-6})]\simeq 1803$.

For the validation of our numerical computations against analytically solvable models, we verified that our computations reproduce the results of Ref.~\cite{Ambrus:2022vif}, where all transport coefficients were computed in the well-known relaxation-time approximation of Anderson and Witting~\cite{Anderson:1974a}.

In the following, all transport coefficients are computed involving the general power-counting scheme, in terms of $\mathcal{X}^{(\ell)}_{r0}$ and $\mathcal{Y}^{(\ell)}_{r0}$.
Henceforth as in Sect.~\ref{sec:DNMR:powerc}, we will report results for three power-counting schemes: DNMR, the corrected DNMR, and IReD. 
Differences between the DNMR and cDNMR methods appear only for the transport coefficients involving the functions $\mathcal{X}^{(\ell)}_{r0}$ with $r < 0$, or the functions $\mathcal{Y}^{(\ell)}_{r0}$. 
Conversely, the cDNMR and the IReD methods show discrepancies only for the coefficients involving $\mathcal{X}^{(\ell)}_{r0}$ and $\mathcal{Y}^{(\ell)}_{r0}$ with $r > 0$. 

Furthermore, in order to assess the magnitude of the higher-order corrections originating from the irreducible moments $\rho_r^{\mu_1 \cdots \mu_\ell}$ with $\ell \leq 2$ and $r\neq0$, we also list the values for the transport coefficients appearing in the $\mathcal{J}^{\mu_1\cdots\mu_\ell}$-terms for the lowest-possible truncation order of 14 dynamical moments (14M) contained in $N^{\mu}$ and $T^{\mu \nu}$, 
i.e., $N_0=2$, $N_1=1$, and $N_2=0$.

The computation of the transport coefficients is done via a \texttt{Mathematica} notebook, which can be found in the supplementary material to this article.

Since the bulk viscous pressure $\Pi$ vanishes in the ultrarelativistic limit, only the coefficients which are unrelated to it are computed in Sect.~\ref{sec:2nd_order_coeff:UR}. 
The remaining second-order coefficients involving the bulk viscosity are expanded up to leading order with respect to the particle mass $m_0$ in Sect.~\ref{sec:2nd_order_coeff:bulk}.

Finally, Sect.~\ref{sec:2nd_order_coeff:inv} ends with a discussion about the possible combinations 
of transport coefficients that remain invariant under the reshuffling between the ${\rm Kn}^2$ 
and ${\rm Re}^{-1} {\rm Kn}$ terms, as also considered in Ref.~\cite{Wagner:2022ayd}.

%%%
\subsection{Thermodynamic functions in the massless limit}
\label{sec:2nd_order_coeff:m0}

In this section we present the various thermodynamic functions necessary for the computation of 
the transport coefficients. Since $n_0 \sim \beta^{-3}$ and $P_0 \sim \beta^{-4}$, it follows that
\begin{equation}
 \mathcal{C}^{(\ell)}_{r0} \sim \beta^{-r} \;, \quad \Omega^{(\ell)}_{r0} \sim \beta^{-r}\;,\quad \mathcal{X}^{(\ell)}_{r0} \sim \beta^{-r} \;,
\end{equation}
while the mean free path $\lambda_{\rm mfp} = 1 / \sigma_T n_0 \sim \beta^3$, and thus, 
\begin{equation}
	 \tau^{(\ell)}_{nr} \sim \lambda_{\rm mfp} \beta^{n - r} \sim \beta^{3 + n - r} \;.
\end{equation}
The thermodynamic functions $\mathcal{H}$ and $\bar{\mathcal{H}}$ are given to leading order with respect to $m_0$ by
\begin{align}
	\mathcal{H}(\alpha,\beta) &\equiv \frac{n_0}{D_{20}} \left( h_0 J_{20} - J_{30}\right) 
	= m^2_0 \frac{\beta^2}{3}\;, \nonumber\\
	\bar{\mathcal{H}}(\alpha,\beta) &\equiv \frac{n_0}{D_{20}} \left( h_0 J_{10} - J_{20}\right) 
	= \frac{\beta}{3}\;. \label{eq:H_Hbar}
\end{align}
Furthermore, 
\begin{subequations}
\begin{align}
 \frac{G_{2r}}{D_{20}} &= \frac{\beta^{2-r}}{6} (1-r) (r+1)!\;, \\
 \frac{G_{3r}}{D_{20}} &= \frac{\beta^{1-r}}{2} (2-r) (r+1)!\;, \\ 
 \frac{\beta J_{r+2,1}}{e_0 + P_0} &= \frac{\beta^{1-r}}{24} (r+3)!\;.
\end{align}
\end{subequations}
These relations can be used to show that $\alpha^{(0)}_r$ vanishes in the massless limit. 
To leading order with respect to $m_0$, the $\alpha^{(\ell)}_r$ coefficients evaluate to
\begin{subequations}
\begin{align}
 \frac{\alpha^{(0)}_r}{m_0^2} &= \frac{\beta^{4-r} P}{36} r! (r-1) (r-2)\;, \\
 \alpha^{(1)}_r &= \frac{\beta^{1 - r} P}{24} (r+2)!(1 - r)\;, \\
 \alpha^{(2)}_r &= \frac{\beta^{-r} P}{30} (r+4)!\;.
\end{align}
\end{subequations}
Note that $\alpha^{(0)}_1 = \alpha^{(0)}_2 = \alpha^{(1)}_1 = 0$ for arbitrary mass.
We can thus derive the following relations:
\begin{gather}
 \frac{\beta}{m_0^4} \frac{\partial \zeta_r}{\partial \beta} =  (3 - r) \frac{\zeta_r}{m_0^4}\;, \nonumber\\
 \beta \frac{\partial \kappa_r}{\partial \beta} = -r \kappa_r\;, \quad 
 \beta \frac{\partial \eta_r}{\partial \beta} = -(r+1) \eta_r\;.
\end{gather}
This gives an identical behaviour for $\mathcal{Y}^{(\ell)}_{r0}$:
\begin{gather}
 \frac{\beta}{m_0^4} \frac{\partial \mathcal{Y}^{(0)}_{r0}}{\partial \beta} = (3-r) \frac{\mathcal{Y}^{(0)}_{r0}}{m_0^4}\;, \nonumber\\
 \beta \frac{\partial \mathcal{Y}^{(1)}_{r0}}{\partial \beta} = -r \mathcal{Y}^{(1)}_{r0}\;, \quad 
 \beta \frac{\partial \mathcal{Y}^{(2)}_{r0}}{\partial \beta} = -(r+1) \mathcal{Y}^{(2)}_{r0}\;.
\end{gather}

\subsection{Transport coefficients for the ultrarelativistic fluid} \label{sec:2nd_order_coeff:UR}

In this subsection, we summarize the second-order transport coefficients in the case of vanishing particle mass, by taking the appropriate limits in the formulas displayed in Appendix~\ref{app:2nd_order_coeff}. Since in this limit, the scalar sector involving the bulk viscous pressure does not play a role, we postpone the discussion of the transport coefficients governing the coupling to $\Pi$ to the next subsection.

We begin with the transport coefficients appearing in the equation for $V^\mu$, Eq.~\eqref{Vdot}. The coefficients for the $O({\rm Re}^{-1} {\rm Kn})$ terms appearing in $\mathcal{J}^\mu$, Eq.~\eqref{J_mu}, are
\begin{subequations}
	\label{eqs:coeff_n}
	\begin{align}
		\delta_{VV} &=  \sum_{r = 0,\neq 1}^{N_1} \! \tau_{0r}^{(1)} 
		\mathcal{X}_{r0}^{(1)} = \tau_V\;,\\
		\ell_{V\pi}&= \sum_{r=0,\neq 1}^{N_1} \! \tau_{0r}^{(1)} \! 
		\left[\frac{(r+3)!}{24}\beta^{1-r} - \mathcal{X}^{(2)}_{r-1,0}\right]\;, \\
		\tau_{V\pi}&= \ell_{V\pi}  \;,\\
		\lambda_{VV}&= \frac{1}{5} \! \sum_{r=0,\neq 1}^{N_1} \! \tau_{0r}^{(1)} \!
		(2r+3)\mathcal{X}^{(1)}_{r0} \;,\\
		\lambda_{V\pi}&= \frac{1}{4}\sum_{r=0,\neq 1}^{N_1} \! \tau_{0r}^{(1)} 
		(1-r) \mathcal{X}^{(2)}_{r-1,0} \;.
	\end{align}
\end{subequations}
These coefficients, together with the coefficient of diffusion $\kappa$ and relaxation time of diffusion $\tau_V$ introduced in Eqs.~\eqref{eq:toceffs_firstorder}, are computed in Table~\ref{tbl:diffusion}.

The $O({\rm Kn}^2)$ coefficients from $\mathcal{K}^\mu$ in Eq.~\eqref{eq:Kmu} are given by
\begin{subequations}
	\label{eqs:Kmu_coeff_massless}
	\begin{align}
		\widetilde{\kappa}_1 &= \!\sum_{r=0,\neq 1}^{N_1} \! \tau^{(1)}_{0r} \!\left[
		\frac{2(1 - r)}{5} \mathcal{Y}^{(1)}_{r0} + \frac{r}{2} \mathcal{Y}_{r-1,0}^{(2)} \right] \;,\\
		\widetilde{\kappa}_3 &= -\frac{2}{3}\sum_{r=0,\neq 1}^{N_1} \! \tau^{(1)}_{0r}
		\mathcal{Y}_{r0}^{(1)}  \;,\\
		\widetilde{\kappa}_5  &= 2\sum_{r=0,\neq 1}^{N_1} \! \tau_{0r}^{(1)} \mathcal{Y}_{r0}^{(1)} 
		= -3\widetilde{\kappa}_3\;, \\
		\widetilde{\kappa}_6  &= -2\sum_{r=0,\neq 1}^{N_1} \! \tau_{0r}^{(1)} 
		\mathcal{Y}_{r-1,0}^{(2)} \;.
	\end{align}
\end{subequations}
The numerical values of these  coefficients are given in Table~\ref{tbl:diffusion_Kn2} for the DNMR and cDNMR power-counting schemes. Note again that in the IReD power counting scheme, $\widetilde{\kappa}_i = 0$ by construction. As expected, DNMR and cDNMR disagree only for $\widetilde{\kappa}_6$, which involves the 
coefficient $\mathcal{Y}^{(2)}_{-1,0}$.
Note that here we excluded $\widetilde{\kappa}_2$, $\widetilde{\kappa}_4$ and $\widetilde{\kappa}_7$, since they vanish in the ultrarelativistic limit as $m_0^2 \rightarrow 0$. 
The leading-order contributions to the $\widetilde{\kappa}_4$ and $\widetilde{\kappa}_7$ are computed 
in the limit of small mass in Sect.~\ref{sec:2nd_order_coeff:bulk}. 

The $O({\rm Re}^{-1}{\rm Kn})$ coefficients in the relaxation equation for the shear-stress tensor~\eqref{pidot}, listed in Eq.~\eqref{J_munu}, are
\begin{subequations}
	\label{eqs:coeff_pi}
	\begin{align}
		\delta_{\pi\pi} &= \frac{4}{3} \sum_{r = 0}^{N_2} \! \tau_{0r}^{(2)} 
		\mathcal{X}_{r0}^{(2)} = \frac{4}{3} \tau_V \;,\\
		\ell_{\pi V} &= \frac{2}{5} \sum_{r = 0}^{N_2} \! \tau_{0r}^{(2)} \mathcal{X}^{(1)}_{r+1,0} \;,\\
		\tau_{\pi V} &= 4\ell_{\pi V}\;,\\
		\tau_{\pi\pi} &= \frac{2}{7}\sum_{r = 0}^{N_2} \! \tau_{0r}^{(2)} (2r+5)\mathcal{X}^{(2)}_{r0}\;,\\
		\lambda_{\pi V} &= -\frac{1}{10}\sum_{r = 0}^{N_2} \! \tau_{0r}^{(2)}  
		(r+1)  \mathcal{X}^{(1)}_{r+1,0}  \;.
	\end{align}
\end{subequations}
These are computed in Table~\ref{tbl:shear}.
The $O({\rm Kn}^2)$ coefficients appearing in $\mathcal{K}^{\mu\nu}$, introduced in Eq. \eqref{eq:Kmunu}, are 
\begin{subequations}
\label{eqs:Kmunu_coeff_massless}
	\begin{align}
		\widetilde{\eta}_1 &= 2\sum_{r=0}^{N_2} \! \tau^{(2)}_{0r} \mathcal{Y}^{(2)}_{r0} \; , \\
		\widetilde{\eta}_2 &= -\frac{1}{3} \widetilde{\eta}_1\;, \\
		\widetilde{\eta}_3 &= -\frac{2}{7}\sum_{r=0}^{N_2} \! \tau^{(2)}_{0r} (4r + 3)\mathcal{Y}^{(2)}_{r0} \; , \\
		\widetilde{\eta}_4 &= 2 \widetilde{\eta}_1 \;, \\
		\widetilde{\eta}_5 &= -\frac{1}{10} \sum_{r=0}^{N_2} \! \tau^{(2)}_{0r}
		(r+1)\mathcal{Y}_{r+1,0}^{(1)} \;, \\
		\widetilde{\eta}_6  &=5\widetilde{\eta}_1 \;, \\
		\widetilde{\eta}_7 &= -\frac{8}{5}\sum_{r=0}^{N_2} \! \tau_{0r}^{(2)} \! 
		\mathcal{Y}^{(1)}_{r+1,0} \;,\\
		\widetilde{\eta}_8 &= -\frac{1}{4}\widetilde{\eta}_7\;, \\
		\widetilde{\eta}_9 &= -\widetilde{\eta}_1 \;.
	\end{align}
\end{subequations}
Since none of the above coefficients involve $\mathcal{Y}^{(\ell)}_{r0}$ with negative $r$, both DNMR and cDNMR agree. The explicit values of these coefficients are summarized in Table~\ref{tbl:shear_Kn2}.

%%%
\subsection{Leading-order contributions to the transport coefficients coupling to the bulk viscous pressure} 
\label{sec:2nd_order_coeff:bulk}

Here we compute the leading order contributions of the remaining coefficients which couple to the bulk 
viscous pressure from Eqs.~\eqref{J}--\eqref{eq:Kmunu}. 
Note that we excluded $\widetilde{\kappa}_2$, since the evaluation of the leading-order correction to this coefficient requires the computation of $m_0^2$ corrections to the collision integral, which is beyond the scope of the present paper.

We begin with the transport coefficients appearing in the equation for the bulk viscous pressure $\Pi$, Eq.~\eqref{Pidot}. The $O({\rm Re}^{-1} {\rm Kn})$ coefficients appearing in Eq.~\eqref{J} are obtained by taking the massless limit of the expressions listed in Eqs.~\eqref{eqs:coeff_Pi_full}, and read
\begin{subequations}
	\label{eqs:coeff_Pi}
	\begin{align}
		\delta_{\Pi\Pi} &=\frac23 \tau_\Pi\;,\\
		\frac{\ell_{\Pi V}}{m_0^2} &=-\sum_{r=0,\neq 1,2}^{N_0}\frac{\tau_{0r}^{(0)}}{3}\left[\mathcal{X}^{(1)}_{r-1,0}+(r+1)!\frac{(r-2)}{2}\beta^{1-r}\right]\;,\\
		\frac{\tau_{\Pi V}}{m_0^2} &= -\frac{\ell_{\Pi V}}{m_0^2} \;,\\
		\frac{\lambda_{\Pi V}}{m_0^2} &= \frac{1}{12}\sum_{r=0,\neq 1,2}^{N_0}\tau_{0r}^{(0)} (r-1) \mathcal{X}_{r-1,0}^{(1)}\;,\\
		\frac{\lambda_{\Pi \pi}}{m_0^2} &= -\frac{1}{3}\sum_{r=0,\neq 1,2}^{N_0}\tau_{0r}^{(0)}(r-1)\left[\mathcal{X}_{r-2,0}^{(2)}-\frac{(r+1)!}{6}\beta^{2-r}\right]\;. 
	\end{align}
\end{subequations}
Note that here, the last four coefficients are divided by $m_0^2$ to extract their leading-order values.
The numerical values of these coefficients, together with the bulk viscosity $\zeta$ and bulk relaxation time $\tau_\Pi$, are listed in Table~\ref{tbl:bulk}.

The leading-order contribution of the terms of second order in the Knudsen number \eqref{eq:K} appearing in the equation of motion for the bulk viscous pressure are
\begin{subequations}
	\label{eqs:K_coeff}
	\begin{align}
		\frac{\widetilde{\zeta}_1}{m_0^4}&=\sum_{r=0,\neq 1,2}^{N_0} \tau^{(0)}_{0r}\frac{\mathcal{Y}^{(0)}_{r0}}{m_0^4}\;,\\
		\frac{\widetilde{\zeta}_2}{m_0^2}&=-\frac{2}{3}\sum_{r=0,\neq 1,2}^{N_0} \tau^{(0)}_{0r} (r-1) \mathcal{Y}_{r-2,0}^{(2)}\;,\\
		\frac{\widetilde{\zeta}_3}{m_0^4}&=\frac43 \frac{\widetilde{\zeta}_1}{m_0^4} \;,\\
		\frac{\widetilde{\zeta}_4}{m_0^2} &=-\frac{1}{12}\sum_{r=0,\neq 1,2}^{N_0} \tau^{(0)}_{0r} (r-1) \mathcal{Y}_{r-1,0}^{(1)}\;,\\
		\frac{\widetilde{\zeta}_5}{m_0^4}&= -5\frac{\widetilde{\zeta}_1}{m_0^4}\;,\\
		\frac{\widetilde{\zeta}_6}{m_0^2}&=-\frac{1}{3}\sum_{r=0,\neq 1,2}^{N_0} \tau^{(0)}_{0r}\mathcal{Y}^{(1)}_{r-1,0}\;,\\
		\frac{\widetilde{\zeta}_7}{m_0^2}&=-\frac{\widetilde{\zeta}_6}{m_0^2}\;,\\
		\frac{\widetilde{\zeta}_8}{m_0^4}&= \frac{\widetilde{\zeta}_1}{m_0^4}\;.
	\end{align}
\end{subequations}
These coefficients are collected in Table~\ref{tbl:bulk_Kn2}.

Next, we move on to Eq.~\eqref{Vdot} for the diffusion current. The coefficients appearing in Eq.~\eqref{J_mu} which are related to the bulk viscous pressure read:
\begin{subequations}
	\label{eqs:coeff_n_Pi}
	\begin{align}
		m_0^2\ell_{V\Pi}&=\sum_{r=0,\neq 1}^{N_1}\tau_{0r}^{(1)}\mathcal{X}_{r+1,0}^{(0)}\;,\\
		m_0^2\tau_{V\Pi}&=2m_0^2\ell_{V\Pi}\;,\\
		m_0^2\lambda_{V\Pi}&=\frac14\sum_{r=0,\neq 1}^{N_1}\tau_{0r}^{(1)}(1+r)\mathcal{X}^{(0)}_{r+1,0}\;.
	\end{align}
\end{subequations}
The leading-order contributions to the coefficients contained in the terms $\mathcal{K}^\mu$ which vanish in the ultrarelativistic limit are given by 
\begin{subequations}
	\label{eqs:Kmu_coeff_massless_2}
	\begin{align}
		\frac{\widetilde{\kappa}_4}{m^2_0}  &= \frac{\beta^2}{2}\widetilde{\kappa}_5-5\sum_{r=0,\neq 1}^{N_1} \! \tau_{0r}^{(1)} 
		\mathcal{Y}_{r+1,0}^{(0)}  \;, \\
		\frac{\widetilde{\kappa}_7}{m^2_0}  &= \sum_{r=0,\neq 1}^{N_1} \! \tau_{0r}^{(1)} 
		\mathcal{Y}_{r+1,0}^{(0)} -\frac{\beta^2}{6}\widetilde{\kappa}_5\; .
	\end{align}
\end{subequations}
Finally, in the case of the equation for the shear-stress tensor, the coefficient in Eq.~\eqref{J_munu} related to $\Pi$ is
\begin{equation}
	\label{eq:coeff_pi_Pi}
	m_0^2 \lambda_{\pi\Pi}=-\frac{2}{5}\sum_{r=0}^{N_2}\tau^{(2)}_{0r}(r+4)\mathcal{X}^{(0)}_{r+2,0}\;.
\end{equation}
There are no $O({\rm Kn}^2)$ coefficients to report in this case. Note that, in the $14$M approximation, where $N_2=0$, the coefficient $\lambda_{\pi\Pi}$ does not diverge when $m_0\to 0$. However, for all orders $N_2>0$, it diverges as $1/m_0^2$, which is why we list its value multiplied by the square of the mass.
The explicit values of the coefficients in Eqs.~\eqref{eqs:coeff_n_Pi}--\eqref{eq:coeff_pi_Pi} are listed in Table~\ref{tbl:LO}.

%%% Zeta Table %%%
\begin{table*}[htb!]
	\begin{tabular}{|c|c|c|c|c|c|c|c|}
		\hline 
		Method &
		$\zeta/m_0^4$ & $\tau_\Pi [\lambda_{\rm mfp}]$ & $\delta_{\Pi\Pi}[\tau_\Pi]$ 
		&$\ell_{\Pi V}[\tau_\Pi] /m_0^2$
		& $\tau_{\Pi V}[\tau_\Pi] /m_0^2$ 
		& $\lambda_{\Pi V} [\tau_\Pi] /m_0^2$ 
		& $\lambda_{\Pi\pi} [\tau_\Pi] /m_0^2$ \\[2pt] \hline \hline
		$14$M &
		$\beta^4/(18\sigma)$
		& $3$
		& $2/3$
		& $\beta/9$
		& $-\beta/9$
		& $-\beta/18$
		& $-7\beta^2/180$ \\[2pt] \hline
		IReD &
		$11\beta^4/(324\sigma)$
		& $(11+6\pi^2)/33$
		& $2/3$
		& $0.067077 \beta$
		& $-0.067077 \beta$
		& $-0.11638\beta$
		& $-0.051367 \beta^2$ \\[2pt] \hline
		DNMR&
		$11\beta^4/(324\sigma)$
		& $3$
		& $2/3$
		& $0.15415\beta$
		& $-0.15415\beta$
		& $-0.084570\beta$
		& $-0.067901 \beta^2$ \\[2pt] \hline
		cDNMR &
		$11\beta^4/(324\sigma)$
		& $3$
		& $2/3$
		& $0.12282\beta$
		& $-0.12282\beta$
		& $-0.092398\beta$
		& $-0.062583 \beta^2$ \\[2pt] 
		\hline
	\end{tabular}
	\caption{Same as Table \ref{tbl:diffusion}, but for $\zeta$ and the second-order transport coefficients in $\mathcal{J}$ for the bulk viscous pressure $\Pi$.}
	\label{tbl:bulk}
\end{table*}

%% ZetaTilde Table%%
\begin{table*}[htb!]
	\begin{tabular}{|c|c|c|c|c|c|c|c|c|}
		\hline
		Method 
        &$\widetilde{\zeta}_1 [\tau_\Pi] /m_0^4$ 
		&$\widetilde{\zeta}_2 [\tau_\Pi] /m_0^2$ 
		&$\widetilde{\zeta}_3 [\tau_\Pi] /m_0^4$ 
		&$\widetilde{\zeta}_4 [\tau_\Pi] /m_0^2$
		&$\widetilde{\zeta}_5 [\tau_\Pi] /m_0^4$ 
		&$\widetilde{\zeta}_6 [\tau_\Pi] /m_0^2$
		&$\widetilde{\zeta}_7 [\tau_\Pi] /m_0^2$ 
		&$\widetilde{\zeta}_8 [\tau_\Pi] /m_0^4$ \\[2pt] \hline \hline
		DNMR&
		$-0.0098705\beta^4$
		&$0.079777\beta$
		&$-0.013161\beta^4$
		&$-0.00032159\beta$
		&$0.049352\beta^4$
		&$-0.016937\beta$
		&$0.016937\beta$
		&$-0.0098705\beta^4$ \\[2pt] \hline
		cDNMR &
		$-0.0098705\beta^4$
		&$0.066291\beta$
		&$-0.013161\beta^4$
		&$-0.0015663\beta$
		&$0.049352\beta^4$
		&$-0.011958\beta$
		&$0.011958\beta$
		&$-0.0098705\beta^4$ 
        \\[2pt] 
		\hline
	\end{tabular}
	\caption{Same as Table \ref{tbl:diffusion_Kn2}, but for the second-order transport coefficients in $\mathcal{K}$ for the bulk viscous pressure $\Pi$.}
	\label{tbl:bulk_Kn2}
\end{table*}

%% LeadingOrder Table%%
\begin{table*}[htb!]
	\begin{tabular}{|c|c|c|c|c|c|c|}
		\hline
		Method
		&$m_0^2\ell_{V\Pi}[\tau_V]$ 
		&$m_0^2\tau_{V\Pi}[\tau_V]$ 
		&$m_0^2\lambda_{V\Pi}[\tau_V]$ 
		&$m_0^2\lambda_{\pi\Pi}[\tau_\pi]$
		&$\widetilde{\kappa}_4[\tau_V]/m_0^2$ 
		&$\widetilde{\kappa}_7[\tau_V]/m_0^2$ \\[2pt] \hline \hline
        14M &
		$0$
		&$0$
		&$0$
		&$0$
		&0
		&0 \\[2pt] \hline
		IReD &
		$-0.32062/\beta$
		&$-0.64124/\beta$
		&$-0.16367/\beta$
		&$-1.3938/\beta^2$
		&0
		&0 \\[2pt] \hline
		DNMR \& cDNMR &
		$-0.53325/\beta$
		&$-1.0665/\beta$
		&$-0.27211/\beta$
		&$-2.5303/\beta^2$
		&$-0.076600\beta^3$
		&$0.019343\beta^3$ \\[2pt] \hline
	\end{tabular}
	\caption{Coefficients in Eqs. \eqref{J_mu} and \eqref{J_munu} related to the bulk viscous pressure and leading-order contributions to the coefficients in \eqref{eq:Kmu} which diverge or vanish in the strict ultrarelativistic limit.}
	\label{tbl:LO}
\end{table*}

%%%
\subsection{Invariant combinations of transport coefficients}
\label{sec:2nd_order_coeff:inv}

It was already noticed in Ref.~\cite{Wagner:2022ayd} that there are various combinations of second-order transport coefficients that stay invariant regardless of the power-counting scheme. To keep the discussion in this section as general as possible, we consider the functions $\mathcal{X}^{(\ell)}_{r0}$ to be arbitrary and enforce Eq.~\eqref{eq:relation_XY} to determine 
\begin{equation}
 \mathcal{Y}^{(\ell)}_{r0} = \xi^{(\ell)}_0 (\mathcal{C}^{(\ell)}_{r0} - \mathcal{X}^{(\ell)}_{r0}) \;,
 \label{eq:Y_from_X}
\end{equation}
where $\mathcal{C}^{(\ell)}_{r>0,0} = \xi^{(\ell)}_r / \xi^{(\ell)}_0$ was introduced in Eq.~\eqref{eq:C_def} and since $\mathcal{F}_{-r,n}^{(\ell)} = \delta_{rn}$, therefore 
$\mathcal{C}^{(\ell)}_{r < 0,0} = \Gamma^{(\ell)}_{-r,0}$.

The combinations of transport coefficients that are invariant with respect to the power-counting method 
are those which have no explicit dependence on the essentially arbitrary functions $\mathcal{X}^{(\ell)}_{r0}$. 
In the more general case of massive particles, these combinations are listed in Table~II of Ref.~\cite{Wagner:2022ayd}. We now identify similar combinations in the case of massless particles. 
In order to do so, we compare the expressions for the leading-order contributions of the second-order transport coefficients belonging to terms of order $O(\mathrm{Kn}^2)$, 
i.e.,~Eqs. \eqref{eqs:Kmu_coeff_massless}, \eqref{eqs:Kmunu_coeff_massless}, \eqref{eqs:K_coeff}, and \eqref{eqs:Kmu_coeff_massless_2}, to the ones belonging to terms of order $O(\mathrm{Re}^{-1} \mathrm{Kn})$, i.e.,~Eqs. \eqref{eqs:coeff_n}, \eqref{eqs:coeff_pi}, \eqref{eqs:coeff_Pi}, \eqref{eqs:coeff_n_Pi}, and \eqref{eq:coeff_pi_Pi} and make use of Eq. \eqref{eq:Y_from_X} to eliminate $\mathcal{Y}^{(\ell)}_{r0}$ in favor of $\mathcal{X}^{(\ell)}_{r0}$. 

For the relaxation times, the invariant combinations are 
\begin{equation}
	\tau_\Pi +\frac{\widetilde{\zeta}_1}{\zeta}\;, \quad \tau_V+ \frac{\widetilde{\kappa}_5}{2\kappa}\;,\quad \tau_\pi +\frac{\widetilde{\eta}_1}{2\eta}\;,
\end{equation}
while for the second-order coefficients appearing in the equation of motion for $\Pi$, they read
\begin{equation}
	\ell_{\Pi V}- \frac{\widetilde{\zeta}_7}{\kappa}\;,\quad \tau_{\Pi V} -\frac{\widetilde{\zeta}_6}{\kappa}\;,\quad \lambda_{\Pi V} -\frac{\widetilde{\zeta}_4}{\kappa}\;,\quad\frac{\lambda_{\Pi\pi}}{m_0^2} +\frac{\widetilde{\zeta}_2}{2m_0^2\eta} \;.
\end{equation}
The invariant combinations for the coefficients in the equation for $V^\mu$ are given by
\begin{align}
	&\frac{\ell_{V\Pi}}{m_0^2} +\frac{\widetilde{\kappa}_7}{m_0^2\zeta}+\frac{\beta^2\widetilde{\kappa}_5}{6\zeta}\;,\quad 
    \ell_{V\pi} +\frac{\widetilde{\kappa}_6}{2\eta} \;,\nonumber\\
    &m_0^2 \tau_{V\Pi}-m_0^2\frac{\widetilde{\kappa}_4+3\widetilde{\kappa}_7}{\zeta}\;,\quad
    \tau_{V\pi} +\frac{\widetilde{\kappa}_6}{2\eta}\;,\nonumber\\
	& \lambda_{VV}+\frac{2\eta}{\kappa}\lambda_{V\pi} -\frac{4\widetilde{\kappa}_1-2\widetilde{\kappa}_5+\widetilde{\kappa}_6}{4\kappa}\;,
\end{align}
while those contained in the equation for $\pi^{\mu\nu}$ read
\begin{equation}
	\tau_{\pi\pi} +\frac{\widetilde{\eta}_1-\widetilde{\eta}_3}{2\eta}\;,\quad \tau_{\pi V} -\frac{\widetilde{\eta}_7}{\kappa}\;,\quad \ell_{\pi V} +\frac{\widetilde{\eta}_8}{\kappa}\;,\quad\lambda_{\pi V} +\frac{\widetilde{\eta}_5}{\kappa}\;.
\end{equation}
The above relations are in full agreement to the massless limit of the relations in Table II of Ref.~\cite{Wagner:2022ayd}, which  are valid for arbitrary mass and statistics. Note that, compared to that table, we do not list any relations for $\lambda_{V\Pi}$ and $\lambda_{\pi\Pi}$. Establishing such relations within the present framework requires the next-to leading order contributions in $m_0^2$ for the coefficients $\widetilde{\kappa}_3$, $\widetilde{\kappa}_5$, $\widetilde{\eta}_1$ and $\widetilde{\eta}_3$, which were not considered in this work.

%%%
\section{Exact results for the scalar sector}
\label{sec:bulk_exact}

In this section we discuss several analytical results derived from the collision matrix $\mathcal{A}^{(0)}_{rn}$ for the irreducible scalar moments. 
We derive exact results for the inverse matrix $\tau^{(0)}_{rn}$ in Sect.~\ref{sec:bulk_exact:tau}, 
while the first-order bulk viscosity coefficients $\zeta_r$ are computed in Sect.~\ref{sec:bulk_exact:zeta}. 
The relaxation times of bulk viscosity $\tau_{\Pi;r}$ are computed in Sect.~\ref{sec:bulk_exact:tauPi}. Finally, the scalar contribution to the deviation $\delta f_\bk$ from local equilibrium is discussed in Sect.~\ref{sec:bulk_exact:df0}.
Note that, since the bulk viscous pressure vanishes in the ultrarelativistic limit, 
one has to take appropriate care to derive the leading-order terms in an expansion in $m_0\beta$.
The respective calculations are detailed in Appendix \ref{app:bulk_calc}.

\subsection{The inverse collision matrix}
\label{sec:bulk_exact:tau}

The inverse collision matrix is given by
\begin{align}
    \tau^{(0)}_{00}&= \frac{1}{\mathcal{A}^{(0)}_{00}} = \lambda_{\text{mfp}} \frac{N_0+1}{N_0-1} \;,\nonumber\\
    \tau^{(0)}_{m>2, 2 < n \le m} &= \lambda_{\rm mfp} \beta^{n-m} \frac{(m+1)!(m-1)(m-2)}{(n+1)!(n-1)(n-2)} \nonumber\\
    &\times \left(\delta_{mn} + \frac{2}{m-2}\right)\;,\nonumber\\
    \tau_{0,n>0}^{(0)}&=-\frac{2\lambda_{\text{mfp}} (-\beta)^n}{(n-1)(n-2)(n+1)!}\nonumber\\
	&\times\binom{1+N_0}{n} \frac{(1+N_0-n)[N_0(n-2)+n]}{(N_0-1)N_0(N_0+1)}\;,\nonumber\\
    \tau^{(0)}_{m>0,0}&= \tau^{(0)}_{m>2, n > m} = 0\;,
    \label{eq:tau0_exact}
\end{align}
which then allows for the computation of the coefficient of bulk viscosity $\zeta_r$ and 
the relaxation times $\tau_{\Pi;r}$.

\subsection{Bulk viscosity coefficients} \label{sec:bulk_exact:zeta}

Considering Eqs. \eqref{NS_coeff}, we can write the coefficients of bulk viscosity as
\begin{equation}
	\frac{3}{m_0^4} \zeta_r= \frac{1}{m_0^2} \sum_{n=0,\neq 1,2}^{N_0} \tau^{(0)}_{rn} \alpha_n^{(0)}\;,\label{eq:zeta_def}
\end{equation}
where we divided by $m_0^4$ to obtain the leading-order contribution in the massless limit. 
Inserting the results for the inverse matrices $\tau^{(0)}_{rn}$, we obtain
\begin{align}
	\frac{\zeta_{r \ge 3}}{m_0^4}
	&= \frac{\lambda_{\rm mfp} P_0 \beta^{4-r}}{108 } (r-1) (r+1)!\left(2H_r - \frac{1}{1+r} - \frac{8}{3}\right)\;,\label{eq:zeta_r}
\end{align}
while $\frac{1}{m_0^4}\zeta_1 = \frac{1}{m_0^4} \zeta_2 = 0$.
In the above, $H_r \equiv  \sum_{n = 1}^r n^{-1}$ is the $r$-th harmonic number.
A calculation detailed in Appendix \ref{subsec:app_bulk_viscosity} yields the bulk viscosity as an exact expression in terms of $N_0$, namely
\begin{equation}
	\frac{1}{m_0^4}\zeta=P_0\beta^4 \lambda_{\text{mfp}}\frac{6+7N_0+11N_0^3}{324 N_0(N_0^2-1)}\;,\label{zeta_0_exact_tau}
\end{equation}
interpolating between the 14-moment approximation corresponding to $N_0 = 2$ and its convergence value when $N_0 \rightarrow \infty$:
\begin{equation}
	\left.\frac{\zeta}{m_0^4}\right|_{N_0 = 2} = \frac{P_0 \beta^4 \lambda_{\rm mfp}}{18 }\;, \;\;
	\lim_{N_0 \rightarrow \infty}\frac{\zeta}{m_0^4}= \frac{11 P_0 \beta^4 \lambda_{\rm mfp}}{324 }\;.
\end{equation}
The second equation above gives the exact value of the leading-order contribution to the bulk viscosity. Their ratio, $\zeta(N_0 \rightarrow \infty) / \zeta(N_0 = 2) = 11/18$ shows that including higher-order moments can lead to a decrease of the bulk viscosity of almost $40\%$.

%%%
\subsection{The relaxation time of the bulk viscous pressure}
\label{sec:bulk_exact:tauPi}

Depending on which power-counting method is considered, the relaxation times of the bulk viscous pressure take on different values 
\begin{align}
	\mathrm{DNMR \,  \& \, cDNMR:} \quad \tau_{\Pi} &\equiv \tau^{(0)}_{00}= \left[\chi^{(0)}_0\right]^{-1}\;,\\
 \mathrm{IReD:} \quad \tau_{\Pi;r} &\equiv \sum_{n=0,\neq 1,2}^{N_0} \tau^{(0)}_{rn} \frac{\zeta_n}{\zeta_r}\;,
\end{align}
where the eigenvalues $\chi_r^{(0)}$ are defined in Eq.~\eqref{eq:DNMR_diag}. Note that the IReD relaxation time of 
the bulk viscous pressure is denoted as $\tau_\Pi \equiv \tau_{\Pi;0}$.

Due to the fact that $\mathcal{A}^{(0)}_{r0}=0$ for $r>0$, the eigenvalues of the matrix $\mathcal{A}^{(0)}$ are given by its diagonal entries. 
The eigenvalues are solutions of the equation
\begin{equation}
	0=\det \left[\tau^{(0)} - \chi \mathbb{I} \right] = 
	\prod_{r = 0, \neq 1,2}^{N_0} \left[ \tau^{(0)}_{rr} - \chi \right]\;,
\end{equation}
where the diagonal entries $\tau^{(0)}_{rr}$ of the inverse collision matrix are given by
\begin{equation}
	\tau^{(0)}_{00} = \lambda_{\text{mfp}} \frac{N_0+1}{N_0-1}\;, \quad 
	\tau^{(0)}_{rr} = \lambda_{\text{mfp}}\frac{r}{r-2}\;,
\end{equation}
where we considered $r \ge 3$. In the case when $N_0 = 2$, there is a single eigenvalue equal to $\chi^{(0)}_0(N_0 = 2) \equiv 3 \lambda_{\rm mfp}$. For $N_0 > 2$, the largest eigenvalue corresponds to $\tau^{(0)}_{rr}$ with $r = 3$, being equal to $3\lambda_{\rm mfp}$, while 
$\tau^{(0)}_{00} > \lambda_{\rm mfp}$ becomes the lowest eigenvalue. 
Rearranging the above expressions in decreasing order gives the set of eigenvalues $[\chi^{(0)}_{r}]^{-1}$ as
\begin{equation}
	\left[\chi^{(0)}_0 \right]^{-1} = 3\lambda_{\rm mfp}\;, \quad 
	\left[\chi^{(0)}_{r \ge 3} \right]^{-1} = \lambda_{\rm mfp} \frac{r + 1}{r - 1}\;.
\end{equation}
Thus, the relaxation time of the bulk viscous pressure in the DNMR and cDNMR approaches becomes independent of $N_0 \ge 2$ and is given by
\begin{equation}
	\tau_{\Pi}=
	3\lambda_{\text{mfp}}\;.\label{eq:tauPiDNMR}
\end{equation}
Note that the inverse eigenvalues are bounded
\begin{equation}
	\lambda_{\text{mfp}}< \left[\chi_r^{(0)}\right]^{-1}\leq 3 \lambda_{\text{mfp}} \;,
\end{equation}
while from $\left[\chi_0^{(0)}\right]^{-1}=3\lambda_{\text{mfp}}$ and  $\left[\chi_3^{(0)}\right]^{-1}=2\lambda_{\text{mfp}} $
we see that there is a clear separation of scales.

A calculation provided in Appendix \ref{subsec:app_tau_IReD} yields the exact result for the relaxation time of the bulk viscous pressure in the IReD approach as a function of $N_0$,
\begin{multline}
	\tau_\Pi = \lambda_{\rm mfp} \Bigg\{\frac{11 + 6\pi^2}{33} - \frac{12}{11} \psi^{(1)}(N_0)+ \frac{2}{N_0 - 1} \\
	+ \frac{2}{6 + 7N_0 + 11N_0^3} \Big[ 
	(N_0 - 2)(3 + 5N_0) + \frac{6 \pi^2}{11} (1 + 3N_0) \\
	- \frac{32}{11} (1 + 3 N_0) \psi^{(1)}(N_0)\Big]
	\Bigg\}\;.\label{eq:tau_Pi_N0}
\end{multline}
When $N_0 \rightarrow \infty$, we arrive at
\begin{equation}
	\tau_\Pi = \lambda_{\rm mfp} \left(\frac{1}{3} + \frac{2\pi^2}{11}\right) \simeq 2.13 \lambda_{\rm mfp}\;.\label{eq:tau_Pi}
\end{equation}
The moments of higher orders on the other hand relax with
\begin{multline}
	\tau_{\Pi;r} = \frac{\lambda_{\rm mfp} }{2H_r - \frac{1}{r+1} - \frac{8}{3}} \Bigg[\frac{28r^2 + 33r + 11}{9(r^2 - 1)} \\
	- \frac{2(r+2)(5r-3)}{3r(r-1)} H_r + 2H_r^2 + 2H_{r,2}\Bigg]\;,\label{eq:tau_Pi_r_N0}
\end{multline}
where $H_{r,m} = \sum_{n = 1}^r n^{-m}$ is the generalized Harmonic number, with $H_r \equiv H_{r,1}$. At large $r$, the harmonic numbers $H_r$ and $H_{r,2}$ are given asymptotically as
\begin{equation}
 H_{r} = \ln r + \gamma + O(r^{-1}), \quad 
 H_{r,2} = \frac{\pi^2}{6} + O(r^{-1}),
\end{equation}
with $\gamma \simeq 0.577$ being the Euler-Mascheroni constant, such that that the highest-order moments relax without bounds for $r\to\infty$ as:
\begin{equation}
 \tau_{\Pi;r} = \frac{\lambda_{\rm mfp} (18 L^2 - 30 L + 28)}{6(3L - 4)} + O(r^{-1}).  
\end{equation}
where we introduced $L = \gamma + \ln r$. It can be seen that $\lim_{r \rightarrow \infty} \tau_{\Pi;r} \simeq \lambda_{\rm mfp} \ln r$, such that higher-order moments relax slower.

%%%
\subsection{Scalar correction to the equilibrium distribution function} \label{sec:bulk_exact:df0}

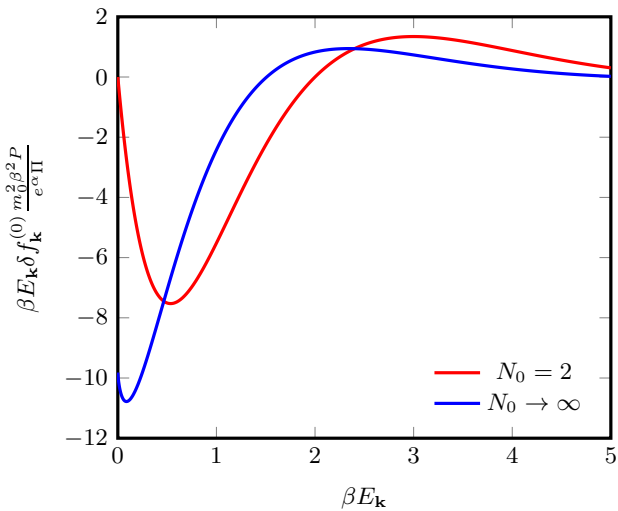
\begin{figure}
\begin{tikzpicture}[
/pgf/declare function={
        gam =  0.57721 56649 01532 86060 65120 90082 40243;
    }]
    \begin{axis}[enlargelimits=false,width=0.45\textwidth,height=0.4\textwidth,legend style={draw=none},legend pos= south east, very thick, yticklabel style = {/pgf/number format/fixed}, scaled y ticks=false, ymin=-12, ymax=2,xmin=0, xmax=5, ytick distance=2,
		xlabel=$\beta E_\k$,
		ylabel={$\beta E_\k \delta f_\k^{(0)} \frac{m_0^2\beta^2 P}{e^\alpha \Pi}$} ,
		]
		\addplot [red,smooth, domain=0:5, samples=1000] {-3*x*(6-x)*(2-x)*exp(-x)};
        \addplot [blue,smooth, domain=0:5, samples=1000] {-6*(
        18./11
        +(6./11)*x*(x-3)*ln(x)
        +(9./11)*x*(1-2*gam)
        +(2*x/11)*x*(3*gam-4)
        )*exp(-x)};
        \addlegendentry{$N_0=2$};
        \addlegendentry{$N_0\rightarrow\infty$};
		\end{axis}
    \end{tikzpicture}
    \caption{The scalar correction to local equilibrium, $\delta f_\bk^{(0)}$, as a function of $\beta E_\k$ in the lowest-order truncation $N_0=2$ [see Eq.~\eqref{eq:delta_f_0_14m}] and in the $N_0\to\infty$ limit [see Eq.~\eqref{eq:delta_f_0_final}], with red and blue lines respectively.}
    \label{fig:delta_f}
    \end{figure}

The results obtained in the present section allow us to estimate the scalar correction, $\ell = 0$, 
defined in Eq.~\eqref{eq:delta_f_expansion}, to local equilibrium,
\begin{equation}
 \delta f^{(0)}_\bk = f_{0\bk} \sum_{n =0}^{N_0} \rho_n \mathcal{H}^{(0)}_{n\bk} \, .
\end{equation}
Using the massless limit of $\mathcal{H}^{(0)}_{n\bk}$ from Eq.~\eqref{eq:UR_H}, 
the strict $14$-moment approximation $14$M, corresponding to $N_0 = 2$ leads to
\begin{equation}
	\delta f_\bk^{(0)}\rfloor_{N_0 = 2} \equiv -\frac{3\Pi}{m_0^2 \beta^2 P_0} (6 - \beta E_\bk)(2 -\beta E_\bk) f_{0\bk}.
 \label{eq:delta_f_0_14m}
\end{equation}
On the other hand, using Eq.~\eqref{IReD_matching} to express the non-dynamical moments according to the IReD approximation, we have
\begin{equation}
 \rho_n \simeq -\frac{3 \Pi}{m_0^2}\frac{\zeta_n}{\zeta_0}  \, ,
\end{equation}
such that now, $\delta f^{(0)}_\bk$ becomes 
\begin{equation}
 \delta f^{(0)}_\bk = -\frac{3\Pi}{m_0^2} \left[ \mathcal{H}^{(0)}_{\bk 0} + \sum_{n = 3}^{N_0} \frac{\zeta_n}{\zeta_0} \mathcal{H}^{(0)}_{\bk n}\right] f_{0\bk}\;.
 \label{eq:delta_f_0_start}
\end{equation}
After some algebra discussed in Appendix \ref{subsec:app_corr_dist}, we find the correction to be
\begin{multline}
 \delta f^{(0)}_\bk = -\frac{6 \Pi}{m_0^2 \beta^2 P_0} \Bigg[ \frac{18}{11 \beta E_\bk} + \frac{6}{11}(\beta E_\bk - 3) \ln (\beta E_\bk) \\
 + \frac{2\beta E_\bk}{11}(3\gamma - 4) + \frac{9}{11} (1 - 2\gamma) \Bigg] f_{0\bk}\;.
 \label{eq:delta_f_0_final}
\end{multline}
Fig.~\ref{fig:delta_f} shows a comparison between $\delta f^{(0)}_\bk$ computed in the 14-moment approximation, Eq.~\eqref{eq:delta_f_0_14m}, and the above expression. Contrary to the 14-moment approximation (indeed, to any finite-$N_0$ representation), the resummed result obtained in the $N_0 \rightarrow \infty$ limit exhibits terms that go like $E_\bk^{-1}$ and $\ln E_\bk$ in the infrared limit. One can conclude that the shape of $\delta f^{(0)}_\bk$ is very different in the resummed case compared to the $14$-moment truncation.

%%%
\section{Conclusion}\label{sec:conc}

In this work, we have analytically computed the linearized collision matrices for an ultrarelativistic gas of hard spheres, and determined the correlation structure of the moment equations. 
It was found that the collision matrices feature a nearly lower-triangular structure, coupling the moments of a given order $r>0$ to all lower-order ones. On the contrary, the irreducible moments of energy-rank zero, the primary dissipative quantities 
$\rho_{0} = -3 \Pi/m^2_0$,  $\rho_{0}^{\mu} = V^{\mu}$, and $\rho_{0}^{\mu \nu} = \pi^{\mu\nu}$,  couple to all higher-order ones included in the basis. 

Expressions for all first- and second-order transport coefficients that appear in different formulations of second-order fluid dynamics, i.e., DNMR, cDNMR, and IReD, have been obtained. 
The coefficients appearing in the terms of second-order in the Knudsen number are nonvanishing in both the DNMR and cDNMR approaches and we have computed them here for the first time. 
Even though they vanish in the strict 14-moment approximation, their convergence values are non-negligible. This is also evidenced by the fact that some of the second-order transport coefficients appearing in the terms of order $O(\mathrm{Kn}\,\mathrm{Re}^{-1})$ differ between the various power-counting methods in second-order fluid dynamics. 

{\color{blue} 
Furthermore, we obtained a closed-form expressions for the bulk viscosity and the relaxation time for the bulk viscous pressure. Compared to their values in the strict 14-moment approximation, there is a decrease of $39\%$ and $29\%$, respectively, as shown in Table~\ref{tbl:bulk}. 
Also, our results in Table~\ref{tbl:diffusion} show that the diffusion coefficient $\kappa$ increases its value from $1/12 \simeq 0.083$ in the 14-moment approximation to $0.16$, by almost $100\%$. Even though the shear viscosity $\eta$ exhibits only a $5\%$ decrease from $4/3$ to $1.27$ in Table~\ref{tbl:shear}, our results show that the inclusion of higher-order moments leads to sizeable changes for the second-order transport coefficients. 
The results reported in our study may serve as a reference for modeling high-energy ultrarelatvistic fluids with binary hard sphere interactions \cite{Karpenko:2013wva,Schenke:2010nt,Schenke:2010rr, Paquet:2015lta, Bouras:2009nn, Bouras:2009vs, Bouras:2010hm,Song:2007ux,Shen:2014vra}.
Furthermore, the computation of the collision integrals and corresponding transport coefficients presented here could easily be extended to the relativistic third-order theory of dissipative fluid dynamics introduced recently in 
Ref.~\cite{deBrito:2023tgb}.
}
%%%
\begin{acknowledgments}

The authors thank G.S.~Denicol, J.~Noronha, A.~Palermo, P.~Huovinen, D.H. Rischke and P. Aasha for fruitful 
discussions. 
D.W.\ acknowledges support by the Studienstiftung des deutschen Volkes 
(German Academic Scholarship Foundation), 
and support by the Research Cluster ELEMENTS (Project ID 500/10.006).
E.M.\ acknowledges support by the program Excellence Initiative--Research
University of the University of Wroc{\l}aw of the Ministry of Education 
and Science.
The authors gratefully acknowledge the support through a grant of the 
Ministry of Research, Innovation and Digitization, CNCS - UEFISCDI,
project number PN-III-P1-1.1-TE-2021-1707, within PNCDI III, as well as support by the Deutsche
Forschungsgemeinschaft (DFG, German Research Foundation) through the 
CRC-TR 211 ``Strong-interaction matter
under extreme conditions'' -- project number 315477589 -- TRR 211.
\end{acknowledgments}

\newpage 
\appendix
\section{Orthogonal polynomials for the ultrarelativistic ideal gas}
\label{sec:matrix_el:polys}

In this Appendix we follow Ref.~\cite{Ambrus:2022vif} and express the orthogonal polynomials 
$P^{(\ell)}_{\bk m}$ and $\mathcal{H}^{(\ell)}_{\bk m}$ as well as $\mathcal{F}^{(\ell)}_{rn}$
in the case of ultrarelativistic particles obeying classical Boltzmann statistics, i.e., $a=0$. 
The local-equilibrium distribution from Eq.~\eqref{f_0k} becomes,
\begin{equation}
	f_{0\k} \equiv e^{\alpha - \beta E_\bk} = g^{-1} \pi^2 P_0 \beta^4  e^{-\beta E_\bk},
	\label{f_0k_Juttner}
\end{equation}
where $E_\bk \equiv u_\mu k^\mu = \sqrt{-\Delta^{\alpha\beta} k_\alpha k_\beta} \equiv k^{0}$ is the particle energy in the comoving frame and 
\begin{equation}
	P_0 = g e^{\alpha}T^4/\pi^2  \label{P_0}
\end{equation} 
is the equilibrium pressure.

In the ultrarelativistic limit of a Boltzmann gas the thermodynamic integrals from Eqs.~\eqref{def:Inq} and \eqref{def:Jnq} reduce to 
\begin{equation}
	I_{nq} = J_{nq}  =  \frac{(n+1)!}{(2q+1)!!} \frac{P_0}{2\beta^{n-2}} \; . \label{I_nq_massless}
\end{equation}

Taking into account the orthogonality relation obeyed by the 
generalized Laguerre polynomials,
\begin{multline}
	\int_0^\infty \d x e^{-x} x^{2\ell + 1} 
	L_m^{(2\ell+1)}(x) L_n^{(2\ell+1)}(x) \\
	= \frac{(n+2\ell+1)!}{n!} \delta_{mn},
	\label{eq:Laguerre_ortho}
\end{multline}
the polynomial $P^{(\ell)}_{\bk m}(E_\bk)$ from Eq.~\eqref{eq:Hfunctions} is expressed in terms of 
the generalized Laguerre polynomials as
\begin{equation}
	P^{(\ell)}_{\bk m}(E_\bk) = \sqrt{\frac{m!(2\ell+1)!}{(m+2\ell+1)!}}
	L^{(2\ell+1)}_m(\beta E_\bk) .
	\label{eq:polys_Laguerre}
\end{equation}
Now using the explicit representation 
\begin{equation}
	L^{(2\ell + 1)}_m(x) = \sum_{n = 0}^m \frac{(m+2\ell+1)!}{(m-n)!(n+2\ell +1)!}
	\frac{(-1)^n x^n}{n!} , 
	\label{eq:Laguerre_coeffs}
\end{equation}
the expansion coefficients of 
$P^{(\ell)}_{\bk m} = \sum_{n =0}^m a_{mn}^{(\ell)} E_\bk^n$
are identified as
\begin{equation}
	a_{mn}^{(\ell)} = (-1)^n \beta^n \frac{\sqrt{m!(2\ell+1)!(m+2\ell+1)!}}
	{n!(m-n)!(n+2\ell+1)!}.
	\label{eq:Pm_coeffs}
\end{equation}
Furthermore setting $P_{\mathbf{k}0}^{(\ell)} \equiv a_{00}^{(\ell)} = 1$, the momentum-independent function $W^{(\ell)}$ from Eq.~\eqref{W_ell} evaluates to
\begin{equation}
	W^{(\ell)} = (-1)^\ell \frac{2\beta^{2\ell-2} (2\ell + 1)!!}{P_0 (2\ell + 1)!}\;.
	\label{eq:W_coeffs}
\end{equation}
Using these results the $\mathcal{H}^{(\ell)}_{\bk n}$  polynomial introduced 
in Eq.~\eqref{eq:Hfunctions} is expressed as
\begin{align}
	\mathcal{H}^{(\ell)}_{\bk n} &= 
	(-1)^{\ell+n} \frac{2 \beta^{2\ell + n-2}(2\ell+1)!!}
	{P_0\, (n+2\ell+1)! \ell! } \nonumber \\ 
	&\times \sum_{m = 0}^{N_\ell - n} \frac{(n+m)!}{n! m!} L^{(2\ell + 1)}_{m+n}(\beta E_\bk ) ,
	\label{eq:UR_H}
\end{align}
hence $\mathcal{F}_{rn}^{(\ell)}$ from Eq.~\eqref{F_rn} evaluates to \cite{Ambrus:2022vif}
\begin{align}
	\mathcal{F}_{rn}^{(\ell)} = \frac{(-1)^n}{(r+n)} \frac{\beta^{r+n} (2\ell + 1 - r)! (N_\ell + r)!}
	{n!(r-1)!(2\ell + 1 + n)! (N_\ell - n)!} .
	\label{eq:UR_F}
\end{align}

%%%
\section{Transport coefficients}
\label{app:2nd_order_coeff}

The coefficients appearing in Eq.~\eqref{J} are obtained by multiplying Eq.~\eqref{D_rho} by 
$-(m^2_0/3) \sum_{r=0,\neq 1,2}^{N_0} \tau^{(0)}_{0r}$. Then, replacing the irreducible moments using Eqs.~\eqref{rho_XY}-\eqref{rho_munu_XY} and collecting the corresponding terms, we obtain
\begin{subequations}
	\label{eqs:coeff_Pi_full}
	\begin{align}
		\delta_{\Pi\Pi} &= \!\sum_{r = 0,\neq 1,2}^{N_0} \! \tau_{0r}^{(0)} \!
		\left[ \frac{r+2}{3}\mathcal{X}^{(0)}_{r0} 
		+ \mathcal{H}\frac{\partial\mathcal{X}^{(0)}_{r0}}{\partial \alpha} + \bar{\mathcal{H}}\frac{\partial\mathcal{X}^{(0)}_{r0}}{\partial \beta}\right] \nonumber \\
		&- \frac{m^2_0}{3} \! \sum_{r = 0,\neq 1,2}^{N_0} \! \tau_{0r}^{(0)} 
		\left[(r-1) \mathcal{X}^{(0)}_{r-2,0} + \frac{G_{2r}}{D_{20}} \right]\;,\\
		\ell_{\Pi V} &= -\frac{m^2_0}{3} \!\sum_{r=0,\neq 1,2}^{N_0} \! \tau_{0r}^{(0)} \! \left[\mathcal{X}^{(1)}_{r-1,0} - \frac{G_{3r}}{D_{20}}\right]\;,\\ 
		\tau_{\Pi V} &= \frac{m^2_0}{3} \! \sum_{r=0,\neq 1,2}^{N_0} \! \tau_{0r}^{(0)} \!
		\left[r\mathcal{X}^{(1)}_{r-1,0} + \beta\frac{\partial\mathcal{X}^{(1)}_{r-1,0}}{\partial \beta} 
		- \frac{G_{3r}}{D_{20}}\right]  \;,\\ 
		\lambda_{\Pi V} &= -\frac{m^2_0}{3} \!\sum_{r=0,\neq 1,2}^{N_0} \! \tau_{0r}^{(0)} \!
		\left[\frac{\partial \mathcal{X}_{r-1,0}^{(1)}}{\partial \alpha} + \frac{1}{h_0}\frac{\partial \mathcal{X}_{r-1,0}^{(1)}}{\partial \beta} \right]\;,\\
		\lambda_{\Pi \pi} &= -\frac{m^2_0}{3} \!\sum_{r=0,\neq 1,2}^{N_0} \! \tau_{0r}^{(0)} \!
		\left[(r-1) \mathcal{X}_{r-2,0}^{(2)} + \frac{G_{2r}}{D_{20}}\right]\;. 
	\end{align}
\end{subequations}

The proper-time derivative and the gradient of $\mathcal{X}(\alpha,\beta)$ are expressed through
\begin{align} 
	\dot{\mathcal{X}} &\equiv \left[ \mathcal{H}(\alpha,\beta) \frac{\partial \mathcal{X}}{\partial \alpha} 
	+ \bar{\mathcal{H}}(\alpha,\beta) \frac{\partial \mathcal{X}}{\partial \beta} \right] \theta \; , \\
	\nabla^\mu\mathcal{X} &\equiv \left( \frac{\partial \mathcal{X}}{\partial \alpha} 
	+ \frac{1}{h_0} \frac{\partial \mathcal{X}}{\partial \beta} \right) \nabla^\mu \alpha 
	- \beta \frac{\partial \mathcal{X}}{\partial \beta} \dot{u}^{\mu} \; ,
\end{align}
where $\mathcal{H}(\alpha,\beta)$ and $\overline{\mathcal{H}}(\alpha,\beta)$ are given in Eq.~\eqref{eq:H_Hbar}.

Similarly to the scalar equation of motion, the coefficients appearing in Eq.~\eqref{J_mu} are found by multiplying Eq.~\eqref{D_rho_mu} by $\sum_{r=0,\neq 1}^{N_1} \tau^{(1)}_{0r}$, then using Eqs.~\eqref{rho_XY}-\eqref{rho_munu_XY} and finally collecting the corresponding terms
\begin{subequations}
	\label{eqs:coeff_n_full}
	\begin{align}
		\delta_{VV} &= \sum_{r = 0,\neq 1}^{N_1} \! \tau_{0r}^{(1)} \!
		\left[\frac{r+3}{3} \mathcal{X}_{r0}^{(1)} 
		+ \mathcal{H} \frac{\partial \mathcal{X}_{r0}^{(1)}}{\partial \alpha} 
		+ \bar{\mathcal{H}} \frac{\partial \mathcal{X}_{r0}^{(1)}}{\partial \beta} \right] \nonumber \\
		&-  \frac{m^2_0}{3} \! \sum_{r = 0,\neq 1}^{N_1} \! \tau_{0r}^{(1)} 
		(r-1)\mathcal{X}_{r-2,0}^{(1)} \; ,\\
		\ell_{V\Pi} &= \sum_{r=0,\neq 1}^{N_1}\! \tau_{0r}^{(1)} \!
		\left[\frac{\beta J_{r+2,1}}{e_0 + P_0} - \mathcal{X}_{r-1,0}^{(0)} + \frac{1}{m^2_0} \mathcal{X}_{r+1,0}^{(0)} \right]\;,\\
		\ell_{V\pi} &= \sum_{r=0,\neq 1}^{N_1} \! \tau_{0r}^{(1)} \! 
		\left[\frac{\beta J_{r+2,1}}{e_0 + P_0} - \mathcal{X}^{(2)}_{r-1,0}\right]\;,\label{eq:l_n_pi} \\
		\tau_{V\Pi} &= \sum_{r=0,\neq 1}^{N_1} \! \tau_{0r}^{(1)} \!
		\left[ \frac{\beta J_{r+2,1}}{e_0 + P_0} - r \mathcal{X}^{(0)}_{r-1,0} 
		- \beta \frac{\partial \mathcal{X}_{r-1,0}^{(0)}}{\partial \beta} \right] \nonumber \\
		&+ \frac{1}{m^2_0} \sum_{r=0,\neq 1}^{N_1} \! \tau_{0r}^{(1)} 
		\left[ (r+3)\mathcal{X}_{r+1,0}^{(0)} 
		+ \beta \frac{\partial \mathcal{X}_{r+1,0}^{(0)}}{\partial \beta} \right] \;, \\
		\tau_{V\pi} &= \sum_{r=0,\neq 1}^{N_1} \! \tau_{0r}^{(1)} \!
		\left[ \frac{\beta J_{r+2,1}}{e_0 + P_0} - r \mathcal{X}_{r-1,0}^{(2)} 
		- \beta \frac{\partial \mathcal{X}_{r-1,0}^{(2)}}{\partial \beta} \right]  \;,\\
		\lambda_{VV} &= \frac{1}{5} \! \sum_{r=0,\neq 1}^{N_1} \! \tau_{0r}^{(1)} 
		\left[(2r+3)\mathcal{X}^{(1)}_{r0} - m^2_0 (2r-2)\mathcal{X}^{(1)}_{r-2,0}\right]\;,\\
		\lambda_{V\Pi} &= \sum_{r=0,\neq 1}^{N_1} \! \tau_{0r}^{(1)} \!
		\left[ \frac{\partial \mathcal{X}^{(0)}_{r-1,0}}{\partial \alpha} 
		+ \frac{1}{h_0}\frac{\partial \mathcal{X}^{(0)}_{r-1,0}}{\partial \beta} \right] \nonumber \\
		&- \frac{1}{m^2_0}\sum_{r=0,\neq 1}^{N_1} \! \tau_{0r}^{(1)}  
		\left[ \frac{\partial \mathcal{X}^{(0)}_{r+1,0}}{\partial \alpha} 
		+ \frac{1}{h_0}\frac{\partial \mathcal{X}^{(0)}_{r+1,0}}{\partial \beta} \right]\;,\\
		\lambda_{V\pi} &= \sum_{r=0,\neq 1}^{N_1} \! \tau_{0r}^{(1)} \!
		\left[ \frac{\partial \mathcal{X}^{(2)}_{r-1,0}}{\partial \alpha} 
		+ \frac{1}{h_0}\frac{\partial \mathcal{X}^{(2)}_{r-1,0}}{\partial \beta} \right]\;.
	\end{align}
\end{subequations}

The coefficients of the shear-stress equation~\eqref{J_munu} follow after multiplying Eq.~\eqref{D_rho_munu} by $\tau^{(2)}_{0r}$ and summing from $r=0$ to $N_2$. Then, after some algebra we obtain the following results,
\begin{subequations}
	\label{eqs:coeff_pi_full}
	\begin{align}
		\delta_{\pi\pi} &= \sum_{r = 0}^{N_2} \! \tau_{0r}^{(2)} \!
		\left[ \frac{r+4}{3} \mathcal{X}_{r0}^{(2)} 
		+ \mathcal{H} \frac{\partial \mathcal{X}_{r0}^{(2)}}{\partial \alpha} 
		+ \bar{\mathcal{H}} \frac{\partial \mathcal{X}_{r0}^{(2)}}{\partial \beta} \right] \nonumber \\
		&- \frac{m^2_0}{3} \! \sum_{r = 0,\neq 1}^{N_1} \! \tau_{0r}^{(2)} 
		(r-1)\mathcal{X}_{r-2,0}^{(2)} \; ,\\
		\ell_{\pi V} &= \frac{2}{5} \sum_{r = 0}^{N_2} \! \tau_{0r}^{(2)} \!
		\left[ \mathcal{X}^{(1)}_{r+1,0} - m^2_0 \mathcal{X}^{(1)}_{r-1,0} \right]\;,\\
		\tau_{\pi V} &= \frac{2}{5} \sum_{r = 0}^{N_2} \! \tau_{0r}^{(2)} \! 
		\left[ (r+5)\mathcal{X}_{r+1,0}^{(1)} + \beta \frac{\partial \mathcal{X}_{r+1,0}^{(1)}}{\partial \beta}  \right] \nonumber \\ 
		&- \frac{2m^2_0}{5}\sum_{r = 0}^{N_2} \! \tau_{0r}^{(2)} \! 
		\left[ r \mathcal{X}_{r-1,0}^{(1)}+ \beta \frac{\partial \mathcal{X}_{r-1,0}^{(1)}}{\partial \beta}  \right]\;,\\
		\tau_{\pi\pi} &= \frac{2}{7}\sum_{r = 0}^{N_2} \! \tau_{0r}^{(2)} \! 
		\left[ (2r+5)\mathcal{X}^{(2)}_{r0} - m^2_0 (2r-2)\mathcal{X}^{(2)}_{r-2,0} \right]\;,\\
		\lambda_{\pi\Pi} &= \frac{2}{5} \sum_{r = 0}^{N_2} \! \tau_{0r}^{(2)} \!
		\left[(2r+3)\mathcal{X}^{(0)}_{r0} - m^2_0 (r-1) \mathcal{X}^{(0)}_{r-2,0} \right] \nonumber \\
		&- \frac{2}{5 m^2_0} \sum_{r = 0}^{N_2} \! \tau_{0r}^{(2)} (r+4) \mathcal{X}^{(0)}_{r+2,0} \;,\label{eq:lambdapiPi}\\
		\lambda_{\pi V} &= \frac{2}{5}\sum_{r = 0}^{N_2} \! \tau_{0r}^{(2)} \!
		\left[ \frac{\partial \mathcal{X}^{(1)}_{r+1,0}}{\partial \alpha} 
		+ \frac{1}{h_0}\frac{\partial \mathcal{X}^{(1)}_{r+1,0}}{\partial \beta} \right] \nonumber \\
		&- \frac{2m^2_0}{5}\sum_{r = 0}^{N_2} \! \tau_{0r}^{(2)} \!
		\left[ \frac{\partial \mathcal{X}^{(1)}_{r-1,0}}{\partial \alpha} 
		+ \frac{1}{h_0}\frac{\partial \mathcal{X}^{(1)}_{r-1,0}}{\partial \beta} \right] \; .
	\end{align}
\end{subequations}
Note that using Eqs.~\eqref{DNMR_X} in these transport coefficients leads to the results of Ref.~\cite{Denicol:2012cn},\footnote{Please note that there is a sign error in Ref. \cite{Denicol:2012cn} related to the term on the second line in Eq.~\eqref{eq:lambdapiPi}.}
while using Eqs.~\eqref{IRed_X} correspond to the results of Ref.~\cite{Wagner:2022ayd}.

The transport coefficients from Eq.~\eqref{eq:K} are proportional to $\mathcal{Y}^{(\ell)}$ and evaluate to
\begin{subequations}
	\label{eqs:K_coeff_full}
	\begin{align}
		\widetilde{\zeta}_1 &= \sum_{r=0,\neq 1,2}^{N_0} \! \tau^{(0)}_{0r} 
		\mathcal{Y}^{(0)}_{r0}\;,\\
		\widetilde{\zeta}_2 &= -\widetilde{\zeta}_1 -\frac{2m^2_0}{3}\sum_{r=0,\neq 1,2}^{N_0} \! \tau^{(0)}_{0r} (r-1) \mathcal{Y}_{r-2,0}^{(2)}\;,\\
		\widetilde{\zeta}_3 &= \sum_{r=0,\neq 1,2}^{N_0} \! \tau^{(0)}_{0r} \! 
		\left[ \frac{(r+1)}{3} \mathcal{Y}_{r0}^{(0)} + 
		\mathcal{H} \frac{\partial \mathcal{Y}_{r0}^{(0)}}{\partial \alpha} 
		+ \bar{\mathcal{H}} \frac{\partial \mathcal{Y}_{r0}^{(0)}}{\partial \beta} \right]  \nonumber \\
		&- \frac{m^2_0}{3} \! \sum_{r = 0,\neq 1,2}^{N_0} \! \tau_{0r}^{(0)} 
		(r-1) \mathcal{Y}^{(0)}_{r-2,0} \;,\\
		\widetilde{\zeta}_4  &= \frac{m^2_0}{3} \! \sum_{r=0,\neq 1,2}^{N_0} \! \tau_{0r}^{(0)} \!
		\left[\frac{\partial \mathcal{Y}_{r-1,0}^{(1)}}{\partial \alpha} + \frac{1}{h_0}\frac{\partial \mathcal{Y}_{r-1,0}^{(1)}}{\partial \beta} \right]\;,\\
		\widetilde{\zeta}_5  &= \sum_{r=0,\neq 1,2}^{N_0} \! \tau_{0r}^{(0)} 
		\mathcal{Y}^{(0)}_{r0} \left[2 + \frac{\beta J_{30}}{(e_0+ P_0)} \right] \;, \\
		\widetilde{\zeta}_6 &= -\sum_{r=0,\neq 1,2}^{N_0} \! \tau_{0r}^{(0)} 
		\mathcal{Y}^{(0)}_{r0} \frac{\mathcal{H} D_{20}}{(e_0+ P_0)^2} \nonumber \\ 
		&-\frac{m^2_0}{3} \!  \sum_{r=0,\neq 1,2}^{N_0} \! \tau_{0r}^{(0)} \!
		\left[r \mathcal{Y}^{(1)}_{r-1,0} 
		+ \beta \frac{\partial \mathcal{Y}^{(1)}_{r-1,0}}{\partial \beta} \right]  \;,\\
		\widetilde{\zeta}_7 &= \frac{m^2_0}{3} \! \sum_{r = 0,\neq 1,2}^{N_0} \! \tau_{0r}^{(0)} 
		\mathcal{Y}^{(1)}_{r-1,0} \;,\\
		\widetilde{\zeta}_8 &= 	\widetilde{\zeta}_1\;.
	\end{align}
\end{subequations}

The coefficients from Eq.~\eqref{eq:Kmu} are given by
\begin{subequations}
	\label{eqs:Kmu_coeff_full}
	\begin{align}
		\widetilde{\kappa}_1 &= -2\!\sum_{r=0,\neq 1}^{N_1} \! \tau^{(1)}_{0r} \!\left[
		\frac{(r - 1)}{5} \mathcal{Y}^{(1)}_{r0} +
		\frac{\partial \mathcal{Y}_{r-1,0}^{(2)}}{\partial \alpha} + \frac{1}{h_0}\frac{\partial \mathcal{Y}_{r-1,0}^{(2)}}{\partial \beta} \right] \nonumber \\
		&+ \frac{2m^2_0}{5}\sum_{r=0,\neq 1}^{N_1} \! \tau^{(1)}_{0r} (r - 1) 
		\mathcal{Y}^{(1)}_{r-2,0} \;,\\
		\widetilde{\kappa}_2 &= 2\sum_{r=0,\neq 1}^{N_1} \! \tau^{(1)}_{0r} \!
		\left[ r \mathcal{Y}_{r-1,0}^{(2)} + \beta \frac{\partial \mathcal{Y}_{r-1,0}^{(2)}}{\partial \beta} \right]\;,\\
		\widetilde{\kappa}_3 &= -\sum_{r=0,\neq 1}^{N_1} \! \tau^{(1)}_{0r} \! 
		\left[ \frac{(r+2)}{3} \mathcal{Y}_{r0}^{(1)} + 
		\mathcal{H} \frac{\partial \mathcal{Y}_{r0}^{(1)}}{\partial \alpha} 
		+ \bar{\mathcal{H}} \frac{\partial \mathcal{Y}_{r0}^{(1)}}{\partial \beta} \right]  \nonumber \\
		&- \sum_{r=0,\neq 1}^{N_1} \! \tau_{0r}^{(1)} \mathcal{Y}_{r0}^{(1)}
		\left[\frac{\partial \mathcal{H} }{\partial \alpha} + \frac{1}{h_0}\frac{\partial \mathcal{H}}{\partial \beta} \right] \nonumber \\
		&- \sum_{r=0,\neq 1}^{N_1} \! \tau_{0r}^{(1)} \!
		\left[\frac{\partial \mathcal{Y}_{r-1,0}^{(0)}}{\partial \alpha} + \frac{1}{h_0}\frac{\partial \mathcal{Y}_{r-1,0}^{(0)}}{\partial \beta} \right] \nonumber \\
		&+ \frac{1}{m^2_0}\sum_{r=0,\neq 1}^{N_1} \! \tau_{0r}^{(1)} \!
		\left[\frac{\partial \mathcal{Y}_{r+1,0}^{(0)}}{\partial \alpha} + \frac{1}{h_0}\frac{\partial \mathcal{Y}_{r+1,0}^{(0)}}{\partial \beta} \right] \nonumber \\
		&+ \frac{m^2_0}{3} \! \sum_{r = 0,\neq 1}^{N_1} \! \tau_{0r}^{(1)} 
		(r-1) \mathcal{Y}^{(1)}_{r-2,0} \;,\\
		\widetilde{\kappa}_4  &= \sum_{r=0,\neq 1}^{N_1} \! \tau_{0r}^{(1)} \! 
		\left[r \mathcal{Y}_{r-1,0}^{(0)} + \mathcal{Y}_{r0}^{(1)} \frac{\partial (\beta \mathcal{H})}{\partial \beta} 
		+ \beta \frac{\partial \mathcal{Y}_{r-1,0}^{(0)}}{\partial \beta} \right] \nonumber \\
		&- \frac{1}{m^2_0} \! \sum_{r = 0,\neq 1}^{N_1} \! \tau_{0r}^{(1)} \!
		\left[ (r+3) \mathcal{Y}^{(0)}_{r+1,0} 
		+ \beta \frac{\partial \mathcal{Y}_{r+1,0}^{(0)}}{\partial \beta}\right] \;, \\
		\widetilde{\kappa}_5  &= 2\sum_{r=0,\neq 1}^{N_1} \! \tau_{0r}^{(1)} \mathcal{Y}_{r0}^{(1)} \; , \\
		\widetilde{\kappa}_6  &= -2\sum_{r=0,\neq 1}^{N_1} \! \tau_{0r}^{(1)} 
		\mathcal{Y}_{r-1,0}^{(2)} \; , \\
		\widetilde{\kappa}_7  &= -\sum_{r=0,\neq 1}^{N_1} \! \tau_{0r}^{(1)} \!
		\left[\mathcal{Y}_{r-1,0}^{(0)} + \mathcal{H}\mathcal{Y}_{r,0}^{(1)} \right] \nonumber \\
		&+ \frac{1}{m^2_0} \! \sum_{r = 0,\neq 1}^{N_1} \! \tau_{0r}^{(1)} \mathcal{Y}^{(0)}_{r+1,0}  \; .
	\end{align}
\end{subequations}

Finally the coefficients from Eq.~\eqref{eq:Kmunu} are
\begin{subequations}
	\label{eqs:Kmunu_coeff_full}
	\begin{align}
		\widetilde{\eta}_1 &= 2\sum_{r=0}^{N_2} \! \tau^{(2)}_{0r} \mathcal{Y}^{(2)}_{r0} \; , \\
		\widetilde{\eta}_2 &= -\frac{2}{3}\sum_{r=0}^{N_2} \! \tau^{(2)}_{0r} \!
		\left[ (r + 1) \mathcal{Y}^{(2)}_{r0} - m^2_0(r-1) \mathcal{Y}^{(2)}_{r-2,0}  \right]
		\nonumber \\
		&-2\sum_{r=0}^{N_2} \! \tau^{(2)}_{0r} \!\left[ \mathcal{H}\frac{\partial\mathcal{Y}^{(2)}_{r0}}{\partial \alpha} + \bar{\mathcal{H}}\frac{\partial\mathcal{Y}^{(2)}_{r0}}{\partial \beta}\right] \nonumber \\
		&- \frac{2}{5}\sum_{r=0}^{N_2} \! \tau^{(2)}_{0r} \! 
		\left[ (2r+3)\mathcal{Y}^{(0)}_{r0} - m^2_0(r - 1)\mathcal{Y}^{(0)}_{r-2,0}\right] \nonumber \\ 
		&+ \frac{2}{5m^2_0}\sum_{r=0}^{N_2} \! \tau^{(2)}_{0r} \! (r+4) \mathcal{Y}^{(0)}_{r+2,0}\; , \\
		\widetilde{\eta}_3 &= -\frac{2}{7}\sum_{r=0}^{N_2} \! \tau^{(2)}_{0r} \!
		\left[(4r + 3)\mathcal{Y}^{(2)}_{r0} 
		- m^2_0 (4r - 4)\mathcal{Y}^{(2)}_{r-2,0} \right]\; , \\
		\widetilde{\eta}_4 &= 2 \widetilde{\eta}_1 \; , \\
		\widetilde{\eta}_5 &= \frac{2}{5}\sum_{r=0}^{N_2} \! \tau^{(2)}_{0r} \!
		\left[\frac{\partial \mathcal{Y}_{r+1,0}^{(1)}}{\partial \alpha} + \frac{1}{h_0}\frac{\partial \mathcal{Y}_{r+1,0}^{(1)}}{\partial \beta} \right] \nonumber \\
		&- \frac{2m^2_0}{5}\sum_{r=0}^{N_2} \! \tau^{(2)}_{0r} \!
		\left[\frac{\partial \mathcal{Y}_{r-1,0}^{(1)}}{\partial \alpha} + \frac{1}{h_0}\frac{\partial \mathcal{Y}_{r-1,0}^{(1)}}{\partial \beta} \right] \; , \\
		\widetilde{\eta}_6  &= 2\sum_{r=0}^{N_2} \! \tau_{0r}^{(2)} 
		\mathcal{Y}^{(2)}_{r0} \left[2 + \frac{\beta J_{30}}{(e_0+ P_0)} \right] \;, \\
		\widetilde{\eta}_7 &= \sum_{r=0}^{N_2} \! \tau_{0r}^{(2)} \! \left[
		\mathcal{Y}^{(2)}_{r0} \frac{2\mathcal{H} D_{20}}{(e_0+ P_0)^2} 
		- \frac{2(r+5)}{5} \mathcal{Y}^{(1)}_{r+1,0} \right]\nonumber \\ 
		&+ \frac{2m^2_0}{5} \!  \sum_{r=0}^{N_2} \! \tau_{0r}^{(2)} \!
		\left[r \mathcal{Y}^{(1)}_{r-1,0} 
		+ \beta \frac{\partial \mathcal{Y}^{(1)}_{r-1,0}}{\partial \beta} 
		- \frac{\beta}{m^2_0}\frac{\partial \mathcal{Y}^{(1)}_{r+1,0}}{\partial \beta} \right] \;,\\
		\widetilde{\eta}_8 &= \frac{2}{5}\sum_{r=0}^{N_2} \! \tau_{0r}^{(2)} \! \left[
		\mathcal{Y}^{(1)}_{r+1,0} - m^2_0 \mathcal{Y}^{(1)}_{r-1,0} \right] \; , \\
		\widetilde{\eta}_9 &= -\widetilde{\eta}_1 \; .
	\end{align}
\end{subequations}
Note that using Eqs.~\eqref{DNMR_Y} in these transport coefficients leads to the results listed in Appendix I of Ref.~\cite{Molnar.2014}. However, in the formulation used in that reference, the contributions that stem from the coefficients $\mathcal{Y}_{r0}^{(\ell)}$ with $r<0$ were not considered, cf. the discussion after Eq. (24) in Ref.~\cite{Wagner:2022ayd}.

%%%
\section{Reference frames and projection operators}
\label{app:proj}

In order to calculate the collision matrix, it is beneficial to define the total momentum 
involved in binary collisions
\begin{equation}
	P^\mu_T \equiv  k^{\mu} + k'^{\mu} = p^{\mu} + p'^{\mu} \; .
	\label{P_tot}
\end{equation}
Its squared norm corresponds to the Mandelstam variable $s \equiv P^\mu_T P_{T,\mu}$. 
The projection operator orthogonal to the total momentum, 
i.e., $\Delta^{\mu \nu}_{T} P_{T,\nu} = 0$, is
\begin{equation}
	\Delta^{\mu \nu}_{T} \equiv  g^{\mu \nu} - \frac{P^{\mu}_T P^{\nu}_T}{s}\;.
	\label{Delta_munu_P}
\end{equation}
Using these definitions the particle momentum can be decomposed with respect to the total 
momentum and the corresponding projection operator as 
\begin{equation}
	p^\mu = P^\mu_T \frac{(P^{\nu}_{T} p_\nu )}{s} + \Delta_T^{\mu \nu} p_\nu \;.
	\label{p_mu_Delta_decomp}
\end{equation}
Furthermore it is useful to define the center-of-momentum (CM) frame where the total momentum is 
$P^\mu_{T}\overset{\mathrm{CM}}{=} \left(\sqrt{s}, \mathbf{0}\right)$, such that
\begin{align}
	P^0_{T} &\overset{\mathrm{CM}}{=}  k^{0} + k'^{0} = p^{0} + p'^{0}= \sqrt{s}\; , \\
	\mathbf{P}_{T} &\overset{\mathrm{CM}}{=} \mathbf{k} + \mathbf{k}' = \mathbf{p} + \mathbf{p}' = \mathbf{0} \;.
\end{align}
However, in the CM-frame the fluid four-flow vector is 
$u^\mu \overset{{\mathrm{CM}}}{=} (u^0, \mathbf{u})$, hence it follows that 
\begin{equation}
P_{T}^\mu u_\mu  \overset{{\mathrm{CM}}}{=} \sqrt{s} u^0 \;, \label{su0_CM}
\end{equation} 
while the normalization condition $u^\mu u_{\mu}=1$ yields
\begin{equation}
\sqrt{\left(P_{T}^\mu u_\mu \right)^2 - s} \overset{{\mathrm{CM}}}{=} \sqrt{s} u  \;, \label{su_CM}
\end{equation}
where we denoted $u\equiv|\mathbf{u}|$. 

In the local rest (LR) frame, where $u^\mu \overset{{\mathrm{LR}}}{=} (1,\mathbf{0})$, we have 
the following representation of the invariant scalars
\begin{equation} 
P_{T}^\mu u_\mu\overset{{\mathrm{LR}}}{=} k^{0} + k^{\prime 0} = p^{0} + p^{\prime 0}  \; ,
\label{su0_LR}
\end{equation}
and
\begin{equation}
\sqrt{\left(P_{T}^\mu u_\mu \right)^2 - s}\overset{{\mathrm{LR}}}{=}|\mathbf{k} + \mathbf{k}^\prime| = |\mathbf{p} + \mathbf{p}^\prime|  \; . 
\label{su_LR}
\end{equation}
In the ultrarelativistic limit, $k^{\mu} k_{\mu} = m^2_0 \rightarrow 0$, and hence 
$k^0 = |\mathbf{k}|\equiv k$, while
\begin{equation}
	s \equiv 2k^\mu k^{\prime}_\mu \overset{\mathrm{LR}}{=} 2 k k^{\prime} \left(1 - \cos \theta_{kk^\prime}\right) \; , 
	\label{s_LR}
\end{equation}
where $\theta_{kk^\prime}$ is the center-of-mass angle between the colliding particles with momenta $\mathbf{k}$ and $\mathbf{k}^\prime$ in the LR-frame.

Similarly to the projection operator in Eq. (\ref{Delta_munu_P}), we introduce another four-vector, $z_T^\mu$, in the CM-frame that is orthogonal to $P^\mu_T$, i.e., $z_T^\mu P_{T,\mu} =0$,
\begin{equation}
	z_T^\mu \equiv  u^\mu - P_T^\mu \frac{(P^{\nu}_T u_\nu)}{s} \overset{\mathrm{CM}}{=}  \left(0,\mathbf{u}\right) \; ,
	\label{z_mu}
\end{equation}
normalized as $z_T^\mu z_{T,\mu} \equiv 1 - (P_T^\mu u_\mu)^2/s \overset{\mathrm{CM}}{=} -\mathbf{u}^2$. 
Hence, using Eq.~\eqref{z_mu} we also obtain that
\begin{equation}
	E_\p \equiv u^\mu p_{\mu} = z_{T}^{\mu} p_{\mu} + P^{\mu}_{T} u_\mu \frac{(P^\nu_T p_\nu )}{s} \; .
	\label{E_p_z}
\end{equation}
The underlying space-like unit vector, $l^{\mu}_T \overset{\mathrm{CM}}{=}  (0,\mathbf{u}/u)$ constructed from $z^{\mu}_{T}$, is also orthogonal to the total momentum, $P^\mu_T l_{T,\mu }  = 0$, and it is defined in a covariant fashion as
\begin{equation}
	l^{\mu}_T \equiv \frac{z^{\mu}_T}{u} = \frac{u^{\mu}}{u} - P_T^\mu \frac{(P^{\nu}_T u_\nu)}{su} \;.
	\label{l_T_mu}
\end{equation}
With the help of this new space-like four-vector a new symmetric and traceless projection operator, 
similarly as in anisotropic fluid dynamics, see for example Ref.~\cite{Molnar:2016vvu}, can be constructed. Here besides the usual space-like projection operator a new projection onto the two-dimensional subspace that is orthogonal to both $P^\mu_T$ and $l^\mu_T$ is defined as,
\begin{equation}
	\Xi^{\mu \nu}_T \equiv g^{\mu \nu} - \frac{P^{\mu}_T P^{\nu}_T}{s} + l^{\mu}_T l^{\nu}_T 
	= \Delta^{\mu \nu}_T + l^{\mu}_T l^{\nu}_T  \;,
\end{equation}
where $\Xi^{\mu \nu}_T P_{T,\nu} = \Xi^{\mu \nu}_T l_{T,\nu} = 0$, 
while $\Xi^{\mu \nu}_T g_{\mu \nu} = 2$.
Using these projectors, the particle momentum can be decomposed with respect to $P^\mu_T$, $l^\mu_T$ 
and $\Xi^{\mu \nu}_T$ as
\begin{equation}
	p^\mu = P^\mu_T \frac{(P^{\nu}_{T} p_\nu )}{s} - l^\mu_T (l^{\nu}_{T} p_\nu )  
	+ \Xi^{\mu \nu}_T p_\nu  \; .
	\label{p_mu_Xi_decomp}
\end{equation}

%%%
\section{The \boldmath\texorpdfstring{$P$}{P} and \texorpdfstring{$P'$}{P'} integrals}
\label{app:PPprime}

In order to evaluate the collision matrix, i.e., Eqs.~\eqref{eq:loss_aux} and~\eqref{eq:gain_aux}, we have to compute the following type of momentum integrals,
\begin{equation}
    \mathcal{P}^{\mu_{1}\cdots \mu_{n}}_{i} \equiv \frac{1}{2}\int \d P \d P' W_{\mathbf{kk}\prime \rightarrow \mathbf{pp}\prime }  E_\p^i  p^{\mu_{1}}\cdots p^{\mu_{n}} \;. \label{eq:P_def}
\end{equation}
It is beneficial to first introduce the following auxiliary integral from Refs.~\cite{Denicol:2012cn,Molnar.2014},
\begin{align}
	\Theta^{\mu_{1}\cdots \mu_{n}} &\equiv \frac{1}{2}\int \d P \d P' 
	W_{\mathbf{kk}\prime \rightarrow \mathbf{pp}\prime } 
	p^{\mu_{1}}\cdots p^{\mu_{n}}  \notag \\
	&=\sum_{q=0}^{\left[ n/2\right] }\left( -1\right) ^{q} b_{nq}
	\mathcal{B}_{nq} \notag \\
	&\times \Delta_{T}^{\left( \mu_{1}\mu_{2}\right. }\cdots 
	\Delta_{T}^{\mu _{2q-1}\mu_{2q}}\,P_{T}^{\mu_{2q+1}}\cdots 
	P_{T}^{\left. \mu_{n}\right) } \; ,  \label{Theta_n}
\end{align}
where we used Eq.~\eqref{p_mu_Delta_decomp} repeatedly to replace $p^{\mu_{1}}\cdots p^{\mu_{n}}$.
Here, $n$, $q$ are natural numbers while the sum runs up to $[n/2]$ 
denoting the largest integer which is less than or equal to $n/2$. 
The symmetrized tensors 
$\Delta^{\left( \right. }_T\cdots  P_T^{\left. \right) }$ are counted by 
$b_{nq} \equiv \frac{n!}{2^{q}q!\left( n-2q\right) !}$, while the $\mathcal{B}_{nq}$ coefficients are
\begin{align}
	\mathcal{B}_{nq} &\equiv \frac{(-1)^{q}}{\left( 2q+1\right) !!}
	\frac{1}{2}\int \d P \d P' W_{\mathbf{kk}\prime \rightarrow \mathbf{pp}\prime }  \notag \\
	&\times \left( \frac{P_{T}^{\mu} p_{\mu}}{\sqrt{s}} \right)^{n-2q}
	\left( \Delta_{P}^{\alpha \beta } p_{\alpha }p_{\beta }\right)^{q} \; .
	\label{B_nq}
\end{align}
In evaluating $\mathcal{B}_{nq}$, we changed $p^\mu$ and $p'^\mu$ to the CM frame defined by $P^\mu_T = k^\mu + k'^\mu$, such that 
$P^{\mu}_T p_\mu = s/2$ and $\Delta_{P}^{\alpha \beta }p_{\alpha }p_{\beta }=-s/4$.
The integrals in Eq.~\eqref{eq:P_def} are then obtained via
\begin{equation}
\mathcal{P}_i^{\mu_1\cdots\mu_\ell}=u_{\nu_1}\cdots u_{\nu_i} \Theta^{\nu_1\cdots\nu_i \mu_1\cdots\mu_\ell} \;. \label{eq:P_to_Theta}
\end{equation}

Even though the integral $\mathcal{P}^{\mu_1\cdots\mu_n}_i$ could in principle be evaluated via 
Eq.~\eqref{eq:P_to_Theta}, doing so is rather complicated.
Instead, it is more sensible to use the decomposition from Eq.~\eqref{E_p_z} to write
\begin{align}
	\mathcal{P}^{\mu_{1}\cdots \mu_{n}}_{i} &= \frac{1}{2}\int \d P \d P' W_{\mathbf{kk}\prime \rightarrow \mathbf{pp}\prime }   p^{\mu_{1}}\cdots p^{\mu_{n}}  \notag \\
	&\times \left(P^{\nu}_{T} u_\nu  \frac{(P^\mu_T p_\mu )}{s} + z^\mu_T p_{\mu}\right)^i \; ,
\end{align}
from where it is clear that the tensor structure of $\mathcal{P}^{\mu_1\cdots\mu_n}_i$ can only consist of the tensors $l_T^\mu$, $P_T^\mu$ and $\Xi_T^{\mu\nu}$.

Now, in the CM-frame we express $z^\mu_T p_{\mu} = u (l^\mu_T p_{\mu}) 
= (\sqrt{s} u /2)\cos\theta_{pu}$, where $\theta_{pu}$ is the angle between $\mathbf{p}$ and $\mathbf{u}$. Furthermore, we have 
$P^{\mu}_T u_\mu = E_\p + E_{\p'} = \sqrt{s}u^0$, and using the binomial formula 
we obtain
\begin{align}
E_\p^i = \sum_{j=0}^{i} \binom{i}{j} \frac{ (u^0)^{i-j}}{u^{-j}} 
\left( \frac{P^\mu_T p_\mu }{\sqrt{s}} \right)^{i-j}  \left(l^\mu_T p_{\mu} \right)^{j} \; .
\end{align}
Subsequently, we expand the integral $\mathcal{P}^{\mu_1\cdots\mu_n}_i$ in terms of the tensors $l_T^\mu$, $P_T^\mu$ and $\Xi_T^{\mu\nu}$,
\begin{gather}
	\mathcal{P}^{\mu_{1}\cdots \mu_{n}}_{i} \equiv 
	\sum_{q=0}^{\left[ n/2\right]} \sum_{m=0}^{n-2q} \left( -1\right)^{q} b_{nmq}
	\mathcal{D}^{(i)}_{nmq} \notag \\
	\times \Xi_{T}^{\left( \mu_{1}\mu_{2}\right. }\cdots \Xi_{T}^{\mu _{2q-1}\mu_{2q}}\, 
	l_{T}^{\mu_{2q+1}}\cdots l_{T}^{ \mu_{2q+m} } P_{T}^{\mu_{2q+m+1}}\cdots 
	P_{T}^{\left. \mu_{n}\right) } \;,  \label{eq:P_eval}
\end{gather}
where $b_{nmq}\equiv n! / [2^q q! m ! (n-2q-m)!]$ counts the number of tensor symmetrizations and the coefficients $\mathcal{D}^{(i)}_{nmq}$ are defined as
\begin{gather}
	\mathcal{D}^{(i)}_{nmq} \equiv \frac{(-1)^{q+m}}{\left( 2q\right) !!}
	\frac{1}{2}\int \d P \d P' W_{\mathbf{kk}\prime \rightarrow \mathbf{pp}\prime }  
	\left( \Xi_{T}^{\alpha \beta } p_{\alpha }p_{\beta }\right)^{q} \notag \\
	\times \sum_{j=0}^{i} \binom{i}{j} \frac{ (u^0)^{i-j}}{u^{-j}}  \left( \frac{P_{T}^{\mu} p_{\mu}}{\sqrt{s}}\right)^{n+i-m-j-2q} 
	\left( l_{T}^{\mu} p_{\mu}\right)^{m+j} \;,
	\label{D_nmq}
\end{gather}
where the double factorial for even numbers is $(2q)!!\equiv 2^q q!$.
To evaluate these coefficients, we note that 
$l_{T}^{\mu} p_{\mu} = (E_\p - E_{\p'})/(2u)$, hence in the CM-frame, 
$s=  2p^{\mu} p'_{\mu} = 4p^2$ and 
$l_{T}^{\mu} p_{\mu} = \mathbf{p} \cdot \mathbf{u}/u = p \cos\theta_{pu}$, while
$\Xi^{\alpha \beta}_T p_\alpha p_\beta = -s/4 + (l_{T}^{\mu} p_{\mu})^2 
= -p^2 \sin^2 \theta_{pu}$. 
In the ultrarelativistic limit for a constant cross-section we then have
\begin{gather}
	\mathcal{D}^{(i)}_{nmq} = \frac{(-1)^m}{4\left( 2q \right) !!}
	\frac{\sigma_T }{2^{n+1+i}}  s^{q+1 + \frac{m}{2}}  \nonumber \\ 
	\times \sum_{j=0}^{i} \binom{i}{j} \left[ 1 + (-1)^{m+j} \right]
	\left(\sqrt{s}u^0 \right)^{i-j} \left(\sqrt{s}u \right)^{j} \nonumber \\
	 \times B\left(\frac{m}{2} + \frac{j}{2}+\frac{1}{2}, q + 1\right) \;,
	\label{D_nmq_final}
\end{gather}
where $B(i,j)\equiv \Gamma(i)\Gamma(j)/\Gamma(i+j)$ denotes the Euler Beta function. Note that $\mathcal{D}^{(0)}_{n0q}=\mathcal{B}_{nq}$, as expected.
We now evaluate Eq. \eqref{eq:P_eval} for the cases $n=0$, $n=1$, and $n=2$, corresponding to the scalar, vector, and tensor cases, respectively. 

In the scalar case, $n=0$, such that $q=m=0$, and thus
\begin{align}
	\mathcal{P}_{i} &\equiv \frac{1}{2} \int \d P \d P' W_{\mathbf{kk}' \rightarrow \mathbf{pp}' }  
	E_\p^i =\mathcal{D}_{000}^{(i)} \nonumber\\
	&= \frac{\sigma_T}{(i+1)2^{i+2} } \frac{(\sqrt{s})^{i+2}}{u} 
 \left[\left(u^0 + u \right)^{i+1} - \left(u^0 - u\right)^{i+1} \right]\;.
\label{P_n}
\end{align}
In particular, for $i=0$ we obtain the following identity,
\begin{align}
\int \d P \d P' \delta \left( k^{\mu} + k^{\prime \mu} - p^{\mu} - p^{\prime \mu} \right) 
= \frac{1}{(2\pi)^{5}} \;.
\end{align}

Note that the latter scalar integrals can be evaluated in other ways as in Refs.~\cite{Bazow.2016,Bazow:2016oky},
\begin{align}
	\mathcal{P}_{i} &\equiv \sigma_T \frac{(2\pi)^5}{(2\pi)^6} \frac{1}{2}\int_{-\infty}^{\infty} 
	\frac{\d^3 \mathbf{p}}{p^0} \int_{-\infty}^{\infty}  \frac{\d^3 \mathbf{p}'}{p'^0} 
	E_\p^i \nonumber \\
	&\times	s \, \delta \left( \sqrt{s} - (p^0 + p'^0) \right) 
	\delta \left(\mathbf{p} + \mathbf{p'} \right) \nonumber \\
	&= 2\sigma_T  \int_{0}^{\infty} \d p p^{i+2}
	\delta \left( \sqrt{s} - 2p  \right) \int_{-1}^{1} \d x \left(u^{0} - ux\right)^i \; ,
	\label{def_P_n}	
\end{align}
where $s= (2p^0)^2 = (2p)^2$ and $x = \cos \theta_{pu}$. 

In the vector case, $n=1$ and hence $q=0$ and $m={0,1}$, 
\begin{align}
	\mathcal{P}^{\mu}_{i} &\equiv \frac{1}{2}\int \d P \d P' W_{\mathbf{kk}\prime \rightarrow \mathbf{pp}\prime }  E_\p^i  p^{\mu} \nonumber \\
	&= \mathcal{D}^{(i)}_{100} P^{\mu}_T + \mathcal{D}^{(i)}_{110} l^{\mu}_T \; ,
	\label{P_i_mu}
\end{align}
where the corresponding coefficients are
\begin{align}
	\mathcal{D}^{(i)}_{100} &= 
	\frac{\sigma_T}{(i+1)2^{i+3} } \frac{(\sqrt{s})^{i+2}}{u} \nonumber \\
	&\times \left[\left(u^0 + u \right)^{i+1} 
	- \left(u^0 - u\right)^{i+1} \right] \; ,
	\label{D_100_final}
\end{align}
and 
\begin{align}
	\mathcal{D}^{(i)}_{110} &= 
	\frac{\sigma_T}{(i+2)2^{i+3} } \frac{ (\sqrt{s})^{i+3}}{u} \nonumber \\
	&\times \left\{ \frac{u^0}{u (i+1)}\left[\left(u^0 + u \right)^{i+1} - \left(u^0 - u\right)^{i+1} \right] \right. \nonumber \\
	&- \left. \left[\left(u^0 + u \right)^{i+1} + \left(u^0 - u\right)^{i+1} \right] \right\} \;. 
	\label{D_110_final}
\end{align}
For the tensor case, $n=2$, and we have $q={0,1}$ and $m={0,1}$, leading to the following decomposition,
\begin{align}
	\mathcal{P}^{\mu \nu}_{i} &\equiv \frac{1}{2}\int \d P \d P' W_{\mathbf{kk}\prime \rightarrow \mathbf{pp}\prime }  E_\p^i  p^{\mu} p^{\nu} \notag \\ 
	&= \mathcal{D}^{(i)}_{200} P^{\mu}_T P^{\nu}_T 
	+ 2 \mathcal{D}^{(i)}_{210} P^{(\mu}_T l^{\nu)}_T +\mathcal{D}^{(i)}_{220} l^{\mu}_T l^{\nu}_T - \mathcal{D}^{(i)}_{201} \Xi^{\mu \nu}_{T} \; ,
	\label{P_i_munu}
\end{align}
where the coefficients are
\begin{equation}
	\mathcal{D}^{(i)}_{200} = \frac{1}{2}\mathcal{D}^{(i)}_{100} , \quad
	\mathcal{D}^{(i)}_{210} = \frac{1}{2}\mathcal{D}^{(i)}_{110} \; ,
	\label{D_200_D210_final}
\end{equation}
as well as
\begin{align}
	\mathcal{D}^{(i)}_{220} &= 
	\frac{\sigma_T}{(i+3)2^{i+4} } \frac{ (\sqrt{s})^{i+4}}{u} \left\{ \left(1 + \frac{2(u^0)^2}{u^2 (i+1)(i+2)} \right) \right. \nonumber \\
	&\left.  \times \left[\left(u^0 + u \right)^{i+1} - \left(u^0 - u\right)^{i+1} \right] \right.
	\nonumber \\
	&\left. - \frac{2u^0}{u(i+2)}\left[\left(u^0 + u \right)^{i+1} + \left(u^0 - u\right)^{i+1} \right] \right\} \; ,
	\label{D_220_final}
\end{align}
and
\begin{align}
	\mathcal{D}^{(i)}_{201} &= 
	-\frac{\sigma_T}{(i+1)(i+3)2^{i+4} } \frac{ (\sqrt{s})^{i+4}}{u^2} \nonumber \\
	&\times \bigg\{ \frac{u^0}{u (i+2)}\left[\left(u^0 + u \right)^{i+2} - \left(u^0 - u\right)^{i+2} \right] 	\nonumber \\
	&- \left[\left(u^0 + u \right)^{i+2} + \left(u^0 - u\right)^{i+2} \right] \bigg\} \; . 
	\label{D_201_final}
\end{align}

%%%
\section{Computation of the loss terms}\label{app:loss}

In this section we compute the loss terms $\mathcal{L}_{rn}^{(\ell)}$ defined in 
Eq.~\eqref{eq:loss_aux} for $\ell=0,1,2$. 
These integrals are Lorentz scalars and thus can be evaluated in any frame. Here 
we choose the LR-frame of the fluid, where $E_\k \equiv k^\mu u_\mu  
\overset{\mathrm{LR}}{=} k^0 = \sqrt{k^2+m^2_0}$. In the following, we will omit the notation ``LR'' for brevity.

In spherical coordinates, $\d K=\frac{1}{(2\pi)^3} \frac{k^2}{k^0} \sin\theta \d k \d\theta \d\varphi$, 
where $k\in \left[ 0,\infty \right) $, $\theta \in \left[ 0,\pi \right] $, 
and $\varphi \in \left[ 0,2\pi \right) $.
Furthermore, by choosing the orientation of $\k^\prime$ parallel to the $z$-axis, 
the angle between the colliding particles $\theta_{kk^\prime}$ 
is equivalent to the elevation angle $\theta = \arccos(k^z/k)$.
Substituting now $k^0 = k$, $k'^{0} = k'$, and 
$s/2 = kk' \left(1 - x\right)$, with $x \equiv \cos \theta_{kk'}$,
the loss term for $\ell=0$ yields
\begin{align}
\mathcal{L}_{rn}^{(0)} &\equiv \sigma_T \int \d K \d K'f_{0\k} f_{0\k'} \frac{s}{2} 
E_\k^r \left(E_\k^n+E_{\k'}^n\right) \nonumber\\
&=\sigma_T \frac{g^2 e^{2\alpha}}{8\pi^4} \int_0^\infty \d k k^{r+2} 
\int_{-1}^{1} \d x \left(1 - x\right)\nonumber \\
&\times \int_0^\infty \d k' k'^2 e^{-\beta (k+k')} \left(k^n+k'^n \right) \nonumber \\
&=  \frac{\sigma_T P_0^2 \beta^{2-r-n}}{4} \nonumber \\ 
&\times \big[2\Gamma(r+n+3) +\Gamma(r+3)\Gamma(n+3)\big] \;, 
\end{align}
where we used Eqs.~(\ref{f_0k_Juttner}--\ref{P_0}), as well as the definition of the 
Gamma function to compute the integrals  
\begin{equation}
\int_0^\infty \! \d y \int_0^\infty \! \d y' e^{-y-y'}  y^{r+a} y'^b = \Gamma(r+a+1) \Gamma(b + 1) \;.
\label{Gamma_int}
\end{equation}

For computing the result for $\ell=1$ that can be found in the same way, we note that in the LR frame we have
\begin{align} \label{kk}
k^{\langle\mu\rangle}k_\mu &\equiv k^{\langle\mu\rangle}k_{\langle\mu\rangle} = -k^2 \; , \\ k^{\langle\mu\rangle}k'_\mu &\equiv k^{\langle\mu\rangle} k'_{\langle\mu\rangle} = -kk' x \; . \label{kk'}
\end{align}
Using these results, Eq.~\eqref{eq:loss_aux} for $\ell=1$ yields
\begin{align}
\mathcal{L}_{rn}^{(1)} &\equiv \frac{\sigma_T}{3} \int \d K \d K'f_{0\k} f_{0\k'} 
\frac{s}{2} E_\k^r k^{\langle\mu\rangle} \left(E_\k^nk_\mu+E_{\k'}^n k'_\mu\right) \nonumber\\
&=-\frac{\sigma_T}{3} \frac{g^2 e^{2\alpha}}{8\pi^4} \int_0^\infty \d k k^{r+3} \int_{-1}^1 \d x (1-x)\nonumber\\
&\times \int_0^\infty \d k' k'^2 e^{-\beta (k+k')} \left(k^{n+1}+k'^{n+1}x\right) \nonumber\\
&= -\frac{\sigma_T P_0^2 \beta^{-r-n} }{36} \nonumber\\ 
&\times \big[6\Gamma(r+n+5) - \Gamma(r+4)\Gamma(n+4)\big] \; . 
\end{align}

In the case when $\ell=2$, we make use of the following identities
\begin{align}\label{kkkk}
k^{\langle\mu}k^{\nu\rangle} k_{\mu}k_\nu &\equiv k^{\langle\mu} k^{\nu\rangle} k_{\langle\mu}k_{\nu\rangle} = \frac{2}{3}k^4  \; , \\
k^{\langle\mu}k^{\nu\rangle}k'_{\mu}k'_\nu &\equiv k^{\langle\mu}k^{\nu\rangle} k'_{\langle\mu}k'_{\nu\rangle} = k^2 k'^2 \left(x^2 - \frac{1}{3} \right) \; ,
\label{kkk'k'}
\end{align}
where $k^{\langle\mu}k^{\nu\rangle} \equiv \Delta^{\mu \nu}_{\alpha \beta} k^{\alpha} k^{\beta} 
= k^{\langle \mu \rangle} k^{\langle \nu \rangle} 
- k^{\langle \alpha \rangle} k_{\langle \alpha \rangle} \Delta^{\mu \nu} /3$ 
and $\Delta^{\mu \nu}_{\alpha \beta} = \frac{1}{2}\left(\Delta^{\mu}_{\alpha} \Delta^{\nu}_{\beta}
+ \Delta^{\mu}_{\beta} \Delta^{\nu}_{\alpha} \right) 
- \frac{1}{3} \Delta^{\mu \nu} \Delta_{\alpha \beta}$. 
The corresponding loss term now reads,
\begin{align}
\mathcal{L}_{rn}^{(2)} &\equiv\frac{\sigma_T}{5} \int \d K \d K'f_{0\k} f_{0\k'} \frac{s}{2} 
E_\k^r k^{\langle\mu} k^{\nu\rangle}\nonumber\\
&\times \left(E_\k^nk_{\langle\mu}k_{\nu\rangle}+E_{\k'}^n k'_{\langle\mu}k'_{\nu\rangle}\right)\nonumber\\
&= \frac{\sigma_T}{5} \frac{g^2 e^{2\alpha}}{24\pi^4} \int_0^\infty \d k k^{r+4} \int_{-1}^1 \d x (1-x)\nonumber\\
&\times \int_0^\infty \d k' k'^2 e^{-\beta (k+k')} \left[2k^{n+2}+k'^{n+2}(3x^2-1)\right] \nonumber\\
&= \frac{\sigma_T P_0^2 \beta^{-2-r-n}}{15}\Gamma(r+n+7) \;. 
\end{align}

These results for $\ell=0,1,2$, corresponding to Eqs. \eqref{eq:loss}, can be put in a unitary form using the following expression
\begin{align}
\mathcal{L}_{rn}^{(\ell)} &= \frac{\sigma_T P_0^2 \beta^{2-2\ell-r-n}}{2} \left[ \frac{(-1)^\ell \ell!}{(2\ell + 1)!!} \Gamma(r + n + 2\ell + 3) \right. \nonumber\\
&\left. + \mathcal{B}^{(\ell)} \Gamma(r + \ell + 3)\Gamma(n + \ell + 3) \right] \;.
\label{eq:loss_aux_unit}
\end{align}
where we introduced the coefficient $\mathcal{B}^{(\ell)}=\left\{1/2,1/18,0 \right\}$ for 
$\ell = \left\{0,1,2\right\}$, respectively.

%%%
\begin{widetext}

\section{Computation of the gain terms}\label{app:gain}

In this section we compute the gain terms $\mathcal{G}_{rn}^{(\ell)}$ defined in Eq.~\eqref{eq:gain_aux} for $\ell=0$, $\ell=1$ and $\ell=2$.

%%%
\subsection{Gain terms for \boldmath\texorpdfstring{$\ell = 0$}{l=0}}\label{sec:coll:gain_0}

Considering Eq.~\eqref{eq:gain_aux} in the case when $\ell=0$ we obtain
\begin{align}
	\mathcal{\mathcal{G}}_{rn}^{(0)} &\equiv
	2\sigma_T (2\pi)^5 \! \int \d P \d P' \d K \d K' f_{0\k} f_{0\k'} E_\k^{r} E_\p^n 
	\frac{s}{2} 
	\delta \left( k^{\mu} + k^{\prime \mu} - p^{\mu} - p^{\prime \mu} \right) \nonumber \\
	&= 2 \! \int \d K \d K' f_{0\k} f_{0\k'} E_\k^{r} \, \mathcal{P}_{n} \; ,  \label{eq:gain_0}
\end{align}
where the $P$ and $P'$ integrals in the center-of-momentum frame are given in Eq.~\eqref{P_n},
and hence
\begin{align}
\mathcal{G}_{rn}^{(0)} &=\frac{\sigma_T }{(n+1)2^{n+1}} \int \d K \d K' f_{0\k}f_{0\k'} 
E_\k^r \, \frac{(\sqrt{s})^{n+2}}{u} 
 \left[\left(u^0 + u \right)^{n+1} 
- \left(u^0 - u\right)^{n+1} \right] \; . \label{eq:gain_0_step1}
\end{align}

The next step consists in evaluating the remaining $K$ and $K'$ integrals in the 
LR-frame of the fluid. Here we recall Eqs.~\eqref{f_0k_Juttner}, \eqref{su0_LR} and \eqref{su_LR} and note that  in the LR-frame $E_\k\equiv k^0=k$ and $s=2kk'(1-x)$ are the massless limits. 
Using these relations we get
\begin{align}\label{eq:gain_0_pintegrated}
    \mathcal{G}_{rn}^{(0)} &= \frac{\sigma_TP_0^2\beta^{8} }{(n+1)2^{n+4}} 
     \int_0^\infty \d k \int_0^\infty \d k' e^{-\beta(k+k')} \nonumber\\ 
    &\times \int_{-1}^1 \d x \frac{2k^{r+2}k'^2(1-x)}{|\k+\k'|} 
    \left[ \left(k+k'+|\k+\k'| \right)^{n+1} - \left(k+k'-|\k+\k'|\right)^{n+1} \right] \; , 
\end{align}
where $|\k+\k'| = \sqrt{k^2+k'^2 + 2kk'x}$. 
Next we change the integration variables to 
\begin{equation}
y = \beta k ,  \qquad  
y' = \beta k' , 
\end{equation}
and we introduce a new angular integration variable,
\begin{equation}
z \equiv \frac{|\k+\k'|}{k+k'} = \frac{\sqrt{y^2+y'^2+2yy'x}}{y+y'}
=\frac{u}{u^0} \; .
\end{equation}
Noting the following useful relation, $s \beta^2  \equiv 2yy'(1-x)=(y+y')^2(1-z^2)$, 
it follows that 
\begin{equation}
u^{0} = \frac{1}{\sqrt{1 - z^2}} \;, \qquad u = \frac{z}{\sqrt{1 - z^2}} \;, \label{eq:u_z_relation}
\end{equation}
and hence
\begin{equation}
\d x \equiv z \frac{(y + y')^2}{yy'} \d z = 2 z \frac{(1-x)}{(1-z^2)} \d z .
\end{equation}
\textcolor{blue}{We remind the reader that Eq. \eqref{eq:u_z_relation} is a consequence of the fact that the four-velocity, since it is evaluated in the CM frame, is dependent on the momentum.}
With these substitutions, the integral becomes
\begin{align}
    \mathcal{G}_{rn}^{(0)} &=\frac{\sigma_T P_0^2\beta^{2-r-n}}{(n+1)2^{n+4}}
    \int_0^\infty \d y \! \int_0^\infty \d y' e^{-y-y'} y^r (y+y')^{n+4}  
    \int_{\frac{|y-y'|}{y+y'}}^1 \d z \left(1 - z^2\right)
    \left[ (1+z)^{n+1}-(1-z)^{n+1} \right]\nonumber\\
    &= \frac{ \sigma_T P_0^2\beta^{2-r-n}}{(n+1)} \int_0^\infty \d y \int_0^\infty \d y' e^{-y-y'} y^{r}\left[\frac{(y+y')^{n+4}}{(n+3)(n+4)}-(y+y')\frac{y^{n+3}+y'^{n+3}}{(n+3)}
    +\frac{y^{n+4}+y'^{n+4}}{(n+4)}\right]\label{eq:gain_0_term_evalz}\;.
\end{align}

The first term of the integral is computed by changing the integration variable $y'$ to 
$x = y + y'$, such that the range for $y$ becomes $[0,x]$:
\begin{equation}
    \int_0^\infty \d y \int_0^\infty \d y' e^{-y-y'} (y+y')^{n+4} y^r  = \int_0^\infty \d x  e^{-x} x^{n+4} \int_0^x \d y  y^r 
    = \frac{\Gamma(n + r+ 6)}{r + 1}\;.
\end{equation}
The remaining terms under the integrals are computed straightforwardly with respect to 
$y$ and $y'$ in terms of the Gamma function, and the final result is Eq. \eqref{gain0_main},
\begin{align}
\mathcal{G}^{(0)}_{rn} &= \frac{\sigma_T P_0^2\beta^{2-r-n}}{(1+n)(1+r)} 
\left[\Gamma(4+n+r)-\Gamma(3+r)\Gamma(3+n)\right]  \;. \label{gain0}
\end{align}
Note that this expression has a finite limit when $r\to -1$,
\begin{equation}
 \mathcal{G}^{(0)}_{-1,n} = \frac{\sigma_T P_0^2 \beta^{3-n}}{(n+1)} \Gamma(n+3) 
 \left[\psi(3+n) - \psi(2) \right] \;.
 \label{gain_0_r-1}
\end{equation}
where $\psi^{(0)}(z) \equiv \psi(z) = \d \ln\Gamma(z)/ \d z$ denotes the digamma function.

%%%	
\subsection{Gain terms for \boldmath\texorpdfstring{$\ell = 1$}{l=1}}\label{sec:coll:gain_1}

The gain term, Eq.~\eqref{eq:gain_aux}, for $\ell=1$ reads
\begin{align}
	\mathcal{\mathcal{G}}_{rn}^{(1)} &\equiv
	\frac{2\sigma_T }{3}(2\pi)^5 \! \int \d P \d P' \d K \d K' f_{0\k} f_{0\k'} 
	E_\k^{r} E_\p^n  p^{\mu} k_{\langle \mu \rangle} \frac{s}{2} 
	\delta \left( k^{\mu} + k^{\prime \mu} - p^{\mu} - p^{\prime \mu} \right) \nonumber \\ 
	&= \frac{2}{3} \! \int \d K \d K' f_{0\k} f_{0\k'} E_\k^{r} \, 
	\mathcal{P}^{\mu}_n k_{\langle \mu \rangle} 
	\label{eq:gain_1} \;.
\end{align}
Recalling the result for the $P$ and $P'$ integrals from Eq.~\eqref{P_i_mu} together 
with Eq.~\eqref{l_T_mu}, we obtain 
\begin{align}
	\mathcal{\mathcal{G}}_{rn}^{(1)} 
	&= \frac{2}{3} \! \int \d K \d K' f_{0\k} f_{0\k'} E_\k^{r} 
	P^{\mu}_T k_{\langle \mu \rangle} 
	\left[\mathcal{D}^{(n)}_{100}- \mathcal{D}^{(n)}_{110} \frac{(P^{\nu}_T u_\nu)}{s u} \right]  \nonumber \\
	&=-\frac{\sigma_T}{3(n+1) 2^{n+2}} \! \int \d K \d K' f_{0\k} f_{0\k'} 
	E_\k^{r} P^{\mu}_T k_{\langle \mu \rangle} \frac{1}{(\sqrt{s}u)^2} (\sqrt{s})^{n+4} \nonumber \\
	&\times \left\{ \frac{u^0}{(n+2)u} \left[\left(u^0 + u \right)^{n+2} 
	- \left(u^0 - u\right)^{n+2} \right] 
 	- \left[\left(u^0 + u \right)^{n+2} + \left(u^0 - u\right)^{n+2} \right] \right\}\;.
	\label{eq:gain_1_step1}
\end{align}
In order to perform the $KK'$-integration, we apply Eqs.~\eqref{su0_LR}, \eqref{su_LR}, and \eqref{s_LR} and express  $k^{\langle\mu\rangle}P_{T\mu} =-k(k+k'x)$ by using Eqs.~\eqref{kk} and \eqref{kk'}, 
where $x = \cos\theta_{kk'}$  denotes the cosine of the angle between $\k$ and $\k'$. 
Thus, after these replacements, we get
\begin{align}\label{eq:gain_1_pintegrated}
	\mathcal{G}_{rn}^{(1)} &= -\frac{\sigma_TP_0^2\beta^{8} }{3(n+1) 2^{n+5}} 
	 \int_0^\infty \d k \int_0^\infty \d k' e^{-\beta(k+k')}
	 \int_{-1}^1 \d x \frac{2k^{r+2}k'^2 (k^2 + k k'x) (1-x)}{|\k+\k'|^2} \nonumber\\ 
	&\times \left\{  \frac{(k+k')}{|\k+\k'| (n+2)} \left[ \left(k+k'+ |\k+\k'| \right)^{n+2} - \left(k+k'-|\k+\k'|\right)^{n+2} \right] \right. \nonumber\\ 
	&\left. - \left[ \left(k+k'+ |\k+\k'| \right)^{n+2} + 
	\left(k+k'-|\k+\k'|\right)^{n+2} \right] \right\} \; .
\end{align}
Changing the variables to $y$, $y'$ and $z$ as before and expressing 
$y^2 + yy'x = (y+y')[y-y' + z^2(y+y')]/2$, we obtain
\begin{align}
	\mathcal{G}_{rn}^{(1)} &=-\frac{\sigma_T P_0^2\beta^{-r-n}}{3(n+1)2^{n+6}}
	\int_0^\infty \d y \! \int_0^\infty \d y' e^{-y-y'} y^r (y+y')^{n+5} 
	\int_{\frac{|y-y'|}{y+y'}}^1  \d z \left(1 - z^2\right) 
	\left[ (y - y') + (y+y') z^2\right] \nonumber \\ 
	&\times \frac{1}{z}\left\{ \frac{1}{z(n+2)} \left[ (1+z)^{n+2}-(1-z)^{n+2} \right] 
	- \left[ (1+z)^{n+2} + (1-z)^{n+2} \right] \right\} \; .
	\label{eq:gain_term_evalz}
\end{align}
Employing similar steps as in the $\ell=0$ case, we arrive at Eq. \eqref{eq:gain1_main},
\begin{align}
	\mathcal{G}_{rn}^{(1)} &= -\frac{ \sigma_T P_0^2 \beta^{-r-n}}{3(1+n)(2+n)(1+r)(2+r)} 
	\big[\Gamma(6+n+r)(r+n+rn-3) + \Gamma(4+r)\Gamma(4+n) (3r+3n+rn+11) \big] \;.
	\label{eq:gain1_final}
\end{align}
As in the $\ell=0$ case, this expression has a finite limit for $r\to -1$, 
\begin{equation}
	\mathcal{G}_{-1,n}^{(1)} = -\frac{ \sigma_T P_0^2 \beta^{1-n} }{3(1+n)(2+n)} 
	\Big\{(n+5)! -2(n+3)! -4(n+4)!\big[\psi(n+5)-\psi(2)\big] \Big\}\;.
	\label{gain_1_r-1}
\end{equation}

%%%
\subsection{Gain terms for \boldmath\texorpdfstring{$\ell=2$}{l=2}}\label{sec:coll:gain_2}

The gain term related to the tensor moments $\ell = 2$ from Eq.~\eqref{eq:gain_aux} reads 
\begin{align}
	\mathcal{\mathcal{G}}_{rn}^{(2)} &\equiv
	\frac{2\sigma_T }{5}(2\pi)^5 \! \int \d P \d P' \d K \d K' f_{0\k} f_{0\k'} 
	E_\k^{r} E_\p^n
	 p^{\mu} p^{\nu} k_{\left\langle \mu \right.} k_{\left. \nu \right\rangle} \frac{s}{2} 
	\delta \left( k^{\mu} + k^{\prime \mu} - p^{\mu} - p^{\prime \mu} \right) \nonumber \\ 
	&= \frac{2}{5} \! \int \d K \d K' f_{0\k} f_{0\k'} E_\k^{r} \, 
	\mathcal{P}^{\mu \nu}_n k_{\left\langle \mu \right.} k_{\left. \nu \right\rangle} \; .
	\label{eq:gain_2} 
\end{align}
Using the result for the $P$ and $P'$ integrals from Eq.~\eqref{P_i_munu} we obtain
\begin{align}
	\mathcal{\mathcal{G}}_{rn}^{(2)} 
	&\equiv \frac{2}{5} \! \int \d K \d K' f_{0\k} f_{0\k'} E_\k^{r} 
	P^{\mu}_T P^{\nu}_T k_{\left\langle \mu \right.} k_{\left. \nu \right\rangle} \nonumber \\
	&\times  \left[\mathcal{D}^{(n)}_{200} - 2\mathcal{D}^{(n)}_{211} \frac{(P^{\nu}_T u_\nu)}{s u} 
	+ \mathcal{D}^{(n)}_{220} \frac{(P^{\nu}_T u_\nu)^2}{s^2 u^2} 
	+ \mathcal{D}^{(n)}_{201} \left( \frac{1}{s} - \frac{(P^{\nu}_T u_\nu)^2}{s^2 u^2} \right) \right] \nonumber \\
	&=\frac{\sigma_T}{5(n+1)(n+3) 2^{n+3}} \! \int \d K \d K' f_{0\k} f_{0\k'} 
	E_\k^{r}  P^{\mu}_T P^{\nu}_T k_{\left\langle \mu \right.} k_{\left. \nu \right\rangle} \frac{1}{(\sqrt{s}u)^3} (\sqrt{s})^{n+5} \nonumber \\
	&\times \left\{ \left( n+4 + \frac{3 (u^0)^2}{(n+2)u^2} \right) 
	\left[\left(u^0 + u \right)^{n+3} - \left(u^0 - u\right)^{n+3} \right] 
	-\frac{3(n+3)u^0}{(n+2)u}
	\left[\left(u^0 + u \right)^{n+3} + \left(u^0 - u\right)^{n+3} \right] \right\} \; .
\end{align}	
Now, recalling Eqs.~\eqref{kkkk}, \eqref{kkk'k'} together with 
$k^{\langle\mu}k^{\nu\rangle} k_{\mu}k'_\nu = 2 k^3 k'x/3$, we replace
\begin{equation}
P^{\mu}_{T}P^{\nu}_{T} k_{\langle\mu}k_{\nu\rangle}
= \frac{k^2}{3}\left[2k^2 + 4kk'x +k'^2 \left( 3x^2 - 1\right) \right] \; .
\end{equation}
Hence, we can write the gain term as
\begin{align}
	\mathcal{\mathcal{G}}_{rn}^{(2)} 
	&=\frac{\sigma_T P_0^2\beta^{-2-r-n}}{15(n+1)(n+3) 2^{n+8}} 
	\int_0^\infty \d y \! \int_0^\infty \d y' e^{-y-y'} y^r (y+y')^{n+6} \nonumber \\
	&\times	\int_{\frac{|y-y'|}{y+y'}}^1 \d z \left(1 - z^2\right) \left[ 3 \left(y - y' \right)^2 
	+ 2 \left(y^2 - 3y'^2 \right)z^2 + 3\left(y + y'\right)^2 z^4 \right]\nonumber \\
	&\times \frac{1}{z^2}\left\{ \left( n+4 + \frac{3}{(n+2)z^2} \right) 
	\left[\left(1 + z \right)^{n+3} - \left(1 - z\right)^{n+3} \right] 
	-\frac{3(n+3)}{(n+2)z}
	\left[\left(1 + z \right)^{n+3} + \left(1 - z\right)^{n+3} \right] \right\} \; .
	\label{eq:gain_1_step2}
\end{align}
Solving the $y,y'$ integrals in a similar fashion as in the $\ell = 0$ and $\ell = 1$ cases, we find Eq. \eqref{eq:gain2_main},
\begin{align}
	\mathcal{G}_{rn}^{(2)} &= \frac{2 P_0^2\sigma_T \beta^{-2-r-n}}{15(1+n)(2+n)(3+n)(1+r)(2+r)(3+r)} \nonumber \\
	&\times \Big\{ \Gamma(8+r+n) \left[64-6(r+n)+2(r^2+n^2)-3rn +3(n^2 r+r^2 n)+r^2n^2 \right]\nonumber\\
	&- \Gamma(6+r)\Gamma(6+n) \left[22+4(r+n)+rn\right] \Big\} \; .
	\label{eq:gain2_final}  
\end{align}
Similar to the previous computations, this expression can be evaluated for $r\to -1$, yielding
\begin{equation}
 \mathcal{G}_{-1,n}^{(2)}= \frac{\sigma_T P_0^2 \beta^{-1-n}}{15  (n+1)(n+2)(n+3)} \Big\{(n+8)!
 -24(n+7)! + 120(n+5)! + 72(n+6)!\big[\psi(n+6) - \psi(2)\big]\Big\} \;.
 \label{gain_2_r-1}
\end{equation}

%%%
\section{Computation of the loss matrices}
\label{app:loss_mat}

Having computed the terms $\mathcal{L}_{rn}^{(\ell)}$ in Appendix \ref{app:loss}, we now have to perform the sums to obtain the loss part of the collision matrix $\mathcal{A}_{rn}^{(\ell),{\rm l}}$ defined in 
Eq.~\eqref{eq:A_expansion_loss}.
We start by substituting the explicit expression \eqref{eq:loss_aux_unit} into Eq.~\eqref{eq:A_expansion_loss} together with the expressions from Eqs.~\eqref{eq:Pm_coeffs} and \eqref{eq:W_coeffs}. After some straightforward algebra we obtain
\begin{align}
	\mathcal{A}_{rn}^{(\ell),{\rm l}} &= 
	\sigma_T P_0 \beta^{1 - r + n}  \frac{(-1)^{\ell+n} (2\ell+1)!!}{\ell! (n+2\ell+1)!}  \nonumber \\ 
	&\times
	 \sum_{m = n}^{N_\ell} \frac{(m+2\ell+1)!}{n! (m-n)!}
	\sum_{q = 0}^m (-1)^q \binom{m}{q} \left[\frac{(-1)^\ell \ell!}{(2\ell + 1)!!} 
	\frac{(r + q + 2\ell + 1)!}{(q + 2\ell + 1)!}   + \mathcal{B}^{(\ell)} 
	\frac{(r + \ell + 1)!  (q + \ell + 2)!}{(q + 2\ell + 1)!}\right] \; .
	\label{eq:loss_aux_polys}
\end{align}
The sum over $q$ can be evaluated with the help of the identity
\begin{align}
	 \sum_{q = 0}^m (-1)^q \binom{m}{q} \frac{(q + a + b)!}{(q+a)!} &=\frac{(a+b)!}{a!} {}_2F_1 (a+b+1,-m;a+1;1) \nonumber \\
	&= (-1)^m \frac{b! (a + b)!}{(b-m)!(m+a)!}\;,
	\label{eq:sum_binom}
\end{align}
where we used Eq.~(15.8.7) of Ref.~\cite{olver2010nist} to evaluate the Gauss hypergeometric function:
\begin{equation}
 {}_2F_1(a+b+1,-m;a+1;z) =  \frac{(-b)_m}{(a+1)_m} {}_2F_1(-m, a+b+1; b-m+1; 1-z)\;,
 \label{2F1_1}
\end{equation}
with the Pochhammer symbols evaluating to $(a+1)_m = (a + 1) \cdots (a + m) = (a + m)! / a!$ and $(-b)_m = (-b) (-b+1) \cdots (-b+m-1) = (-1)^m b! / (b-m)!$. Noting that ${}_2F_1(a,b;c;0) = 1$, we arrive at 
\begin{equation}
 {}_2F_1 (a+b+1,-m;a+1;1) = \frac{(-1)^m a! b!}{(a+m)! (b-m)!},
\end{equation}
in agreement with Eq.~\eqref{eq:sum_binom}.
Using Eq. \eqref{eq:sum_binom}, we arrive at
\begin{align}
	\mathcal{A}_{rn}^{(\ell),{\rm l}} &= 
	\sigma_T P_0 \beta^{1 - r + n} 
	\frac{(-1)^{\ell+n} (2\ell+1)!!}{\ell! (n+2\ell+1)!}(r+2\ell+1)! \nonumber\\
	&\times\sum_{m = n}^{N_\ell} (-1)^m \binom{m}{n} 
	\left[\frac{(-1)^\ell \ell!}{(2\ell + 1)!!} \binom{r}{m}   + \mathcal{B}^{(\ell)} 
	\frac{(1-\ell)!(\ell+2)!(r + \ell + 1)!}{m!(1-\ell-m)! (r+2\ell+1)!} \right] \;.
	\label{eq:loss_aux_sumq}
\end{align}
The binomial coefficient  $\binom{r}{m}$ vanishes when 
$m > r$, such that the first term in the sum over $m$ gives a nonvanishing contribution only when $n \le r$. In this case, the sum over $m$ runs between $n$ and $r$, yielding a Kronecker delta:
\begin{equation}
	\sum_{m = n}^{r} (-1)^m \binom{m}{n} \binom{r}{m} = (-1)^n \delta_{rn}\;.
\end{equation}
The term involving $\mathcal{B}^{(\ell)}$ vanishes for $\ell = 2$, since by definition $\mathcal{B}^{(2)} = 0$. For $\ell = 0$ and $\ell = 1$, the sum over $m$ terminates at $m = 1 - \ell$. Performing this sum separately for $\ell = 0$ and $\ell = 1$, we find
\begin{subequations}
\begin{alignat}
	\ell \ell &= 0:\qquad \quad 
	\sum_{m = n}^{1} \binom{m}{n} \frac{(-1)^m}{m!(1-m)!} &&= -\delta_{n1}\;, \\
	\ell &= 1:\qquad \quad  
	\sum_{m = n}^{0} \binom{m}{n} \frac{(-1)^m}{m!(-m)!} &&= \delta_{n0}\;.
\end{alignat}
\end{subequations}
With all these results, the contributions of the loss terms to the collision matrix are
\begin{subequations}
\begin{align}
	\mathcal{A}^{(0),{\rm l}}_{rn} &=  \sigma_T P_0 \beta \left[ \delta_{nr} + 
	\frac{(r+1)!}{2} \delta_{n1} \beta^{1-r}\right] \;,\\
	\mathcal{A}^{(1),{\rm l}}_{rn} &=  \sigma_T P_0 \beta \left[ \delta_{nr} - 
	\frac{(r+2)!}{6} \delta_{n0} \beta^{-r}\right]\; ,\\
	\mathcal{A}^{(2),{\rm l}}_{rn} &=  \sigma_T P_0 \beta \, \delta_{nr} \; ,
\end{align}
\end{subequations}
in agreement with Eqs. \eqref{eq:A_loss}.

%%%
\section{Computation of the gain matrices}
\label{app:gain_mat}

In this section we compute the gain part of the collision matrix defined in Eq.~\eqref{eq:A_expansion_gain}. 
As discussed in the main text, the matrices $\mathcal{A}^{(\ell),g}_{rn}$ will contain terms 
that diverge in the limit $N_\ell\to\infty$. These divergences will appear in the form of 
certain sums $S_n^{(\ell)}(N_\ell)$ defined in Eq. \eqref{eq:sum_aux}, that we list here again,
\begin{equation}
 S^{(\ell)}_n \left(N_\ell\right) \equiv \sum_{m = n}^{N_\ell} \binom{m}{n} \frac{1}{(m+\ell)(m+\ell+1)}\;.
 \label{eq:sum_aux_app}
 \end{equation}
These sums can be evaluated recursively using auxiliary sums
\begin{equation}
\widetilde{S}_n^{(\ell)}(N_\ell) \equiv \sum_{m=n}^{N_\ell}\binom{m}{n} \frac{1}{m+\ell}\;.
\label{eq:sum_aux_tilde}
\end{equation}
The explicit recursions will be listed at the end of the following subsections.

\subsection{Gain matrix for \boldmath\texorpdfstring{$\ell=0$}{l=0}}

Setting $\ell=0$ and inserting the results for $\mathcal{G}_{rn}^{(0)}$ from Eq.~\eqref{gain0_main} into 
Eq.~\eqref{eq:A_expansion_gain}, we find
\begin{equation}
 \mathcal{A}_{rn}^{(0),{\rm g}} = 
 \frac{2  \beta P_0 \sigma_T \beta^{n - r} (-1)^n}{r (n+1)!} 
 \sum_{m = n}^{N_\ell} \binom{m}{n} (m+1)!
 \sum_{q = 0}^m \frac{(-1)^q}{(q+1)!(m-q)!}
 \left[\frac{(q+r+2)!}{(q+1)!} - (q+2)(r+1)!\right]\;.
 \label{eq:gain0_aux}
\end{equation}
The above expression is indeterminate when $r = 0$. Let us first consider the case $r > 0$.
The sum over $q$ can be performed by shifting the summation index $q$ to $q+1$ and applying 
Eq.~\eqref{eq:sum_binom},
\begin{equation}
 \sum_{q = 0}^m \frac{(-1)^q}{(q+1)!(m-q)!}
 \frac{(q+k+2)!}{(q+1)!} = 
 \frac{(k+1)!}{(m+1)!}\left[1 + 
 \frac{(-1)^m (r+1)!}{(m+1)!(r-m)!}\right] \;.
\end{equation}
Applying the above formula with $k = r$ and $k = 0$, corresponding to the first and second term in square brackets in Eq.~\eqref{eq:gain0_aux}, respectively, we find
\begin{equation}
 \mathcal{A}_{r>0,n}^{(0),{\rm g}} = 
 \frac{2  (-1)^n (r+1)! \sigma_T P_0 \beta^{1+n-r}}{r (n+1)!} 
 \sum_{m = n}^{N_0} (-1)^m \binom{m}{n}
 \left[\frac{(r+1)!}{(r-m)!(m+1)!}- \delta_{m0}\right]\;.
 \label{eq:gain0_aux2}
\end{equation}

In order to perform the sum over $n$ in Eq.~\eqref{eq:gain0_aux2}, we introduce the function $S_{rn}(x)$ via
\begin{align}
 S_{rn}(x)  &\equiv  \sum_{m = 0}^{r-n}
 \binom{r-n}{m} (-1)^m x^{r-m} =  (-1)^{r-n} x^n (1 - x)^{r-n}\;.
 \label{eq:Srn}
\end{align}
Denoting the integral of order $q$ of $S_{rn}(x)$ by
\begin{align}
 S_{rn}^{(-q)}(x) &\equiv  \int_0^x \d x_1 \int_0^{x_1} \d x_2 \cdots 
 \int_0^{x_{q-1}} \d x_q S(x) \nonumber\\
 &=  \sum_{m = 0}^{r-n}
 \binom{r-n}{m} \frac{(r - m)!}{(r - m + q)!} (-1)^m x^{r-m+q}\;,
 \label{eq:Srn_int}
\end{align}
the sum in the first term in square brackets in Eq.~\eqref{eq:gain0_aux2} can be written as
\begin{equation}
 \sum_{m = n}^{N_0} (-1)^m \binom{m}{n}
 \frac{(r+1)!}{(r-m)!(m+1)!}=\frac{(-1)^n}{n!}\frac{(r+1)!}{(r-n)!}  S_{rn}^{(-1)}(1)\;.
\end{equation}
Then, we can reexpress Eq.~\eqref{eq:gain0_aux2} as
\begin{equation}
 \mathcal{A}_{r>0,n}^{(0),{\rm g}} = 
 \frac{2(r+1)! \sigma_T  P_0 \beta^{1-r}}{ r\,n! (n+1)!}
 \left[ \frac{(r+1)!}{(r-n)!}(-\beta)^{n}
 S_{rn}^{(-1)}(1)-\delta_{n0}\right]\;.
 \label{eq:gain0_aux3}
\end{equation}
 
Now, using the definition of the incomplete Euler Beta function \cite{olver2010nist},
\begin{equation}
 B_z(a,b) \equiv   \int_0^z \d t\, t^{a-1} (1-t)^{b-1}\;,
 \label{eq:Beta_incomplete}
\end{equation}
it can be seen that 
\begin{equation}
 S_{rn}^{(-1)}(z) = (-1)^{r-n} B_z(n+1,r-n+1) \;, \label{eq:Srn_m1}
\end{equation}
while $S_{rn}^{(-1)}(1) = (-1)^{r-n} B(n+1,r-n+1)$ can be written 
in terms of the complete Euler Beta function \cite{olver2010nist}, defined by
\begin{equation}
 B(n+1,r-n+1) = \frac{n!(r - n)!}{(r+1)!}\;.
 \label{eq:Beta_complete}
\end{equation}
Using these results, Eq.~\eqref{eq:gain0_aux3} reduces to
\begin{equation}
 \mathcal{A}_{r>0,n}^{(0),{\rm g}} = 
 \frac{2 (r+1)! \sigma_T P_0\beta^{1+n-r}}{ r (n+1)!}
 \left( 1 - \delta_{n0}\right)\;,
\end{equation}
recovering the first part of Eq. \eqref{eq:A_g_0}.

When $r = 0$, we find with the help of Eq.~\eqref{gain_0_r-1}
\begin{equation}
 \mathcal{A}_{0n}^{(0),{\rm g}} = 
 \frac{2  (-1)^n \sigma_T P_0  \beta^{n+1} }{(n+1)!} 
 \sum_{m = n}^{N_0} \binom{m}{n} (m+1)!
 \sum_{q = 0}^m \frac{(-1)^q(q+2)}{(q+1)!(m-q)!}
 \left[\psi(3+q) - \psi(2)\right]\;.
\end{equation}
The summation over $q$ gives
\begin{equation}
 \sum_{q = 0}^m \frac{(-1)^q(q+2)}{(q+1)!(m-q)!}
 \left[\psi(3+q) - \psi(2)\right] 
 = \begin{cases}
  -\frac{1}{m(m+1) (m+1)!}\;, & m>0\\
  1\;, & m=0
  \end{cases}\;.
\end{equation}
The $m = 0$ term contributes only when $n = 0$, in which case we have
\begin{equation}
 \mathcal{A}^{(0),{\rm g}}_{00} = 2 \sigma_T  P_0 \beta 
 \left[1 - \sum_{m = 1}^{N_0} \frac{1}{m(m+1)}\right]= \frac{2 \sigma_T P_0 \beta}{N_0 + 1}\;,
\end{equation}
approaching $\mathcal{A}^{(0),{\rm g}}_{00} \rightarrow  0 $ in the limit when $N_0 \rightarrow \infty$. 

When $n > 0$, we have
\begin{equation}
 \mathcal{A}_{0,n>0}^{(0),{\rm g}} = 
 \frac{2 (-1)^{n+1}  \sigma_T P_0 \beta^{n+1} }{(n+1)!} 
 S_n^{(0)}(N_0)\;,
\end{equation}
where we used the definition from Eq.~\eqref{eq:sum_aux_app}.
The sum $S_n^{(0)}(N_0)$ diverges as $\log N_0$ for $n = 1$. For small $n>1$, it diverges as $N_0^{n-1}$, while for large $n\leq N_0$ the divergence goes as
$N_0^{N_0-n-2}$, suggesting a maximum degree of divergence around $n\sim N_0/2$. 
In the case $n=1$, we have 
\begin{equation}
 S_1^{(0)}(N_0) = \psi(N_0 + 2) - \psi(2)\;,
\end{equation}
while for $n = 2$ it holds that
\begin{equation}
 S_2^{(0)}(N_0) = -\psi(N_0 + 2) + \psi(2) + \frac{N_0}{2}\;.
\end{equation}

Using the auxiliary sum $\widetilde{S}^{(0)}_n(N_0)$ defined in Eq.~\eqref{eq:sum_aux_tilde}, we can formulate a coupled recursion equation
\begin{subequations}
\begin{align}
	S^{(0)}_{n+1}(N_0) &= \frac{1}{n+1} \widetilde{S}^{(0)}_n(N_0)-S_n^{(0)}(N_0)\;,\\
    \widetilde{S}^{(0)}_{n+1}(N_0) &=  
 \frac{1}{n+1} \binom{N_0+ 1}{n+1}-\frac{n}{n+1}\widetilde{S}^{(0)}_n\;,\\
 \widetilde{S}^{(0)}_1(N_0) &= N_0\;,
\end{align}
\end{subequations}
while the recursion for $\widetilde{S}_1^{(0)}(N_0)$ can be solved exactly:
\begin{equation}
 \widetilde{S}^{(0)}_n(N_0) = \frac{1}{n} \binom{N_0}{n}\;.
\end{equation}

%%%
\subsection{Gain matrix for \boldmath\texorpdfstring{$\ell=1$}{l=1}}

We now compute the collision matrix $\mathcal{A}^{(1),{\rm g}}_{rn}$ defined in Eq.~\eqref{eq:A_expansion_gain}:
\begin{equation}
 \mathcal{A}_{rn}^{(1),{\rm g}} =
 \frac{6 (-1)^{n+1}}{n! (n+3)! P_0} 
 \sum_{m = n}^{N_1} \frac{m! (m+3)!}{(m-n)!} 
 \sum_{q = 0}^m \frac{(-1)^q\mathcal{G}^{(1)}_{r-1,q}}{q!(m-q)! (q+3)!} \;.
\end{equation}
Substituting Eq.~\eqref{eq:gain1_main} into the above leads to
\begin{align}
 \mathcal{A}_{rn}^{(1),{\rm g}} &= \frac{2(-1)^{n}  \sigma_T P_0\beta^{1+n-r}}{n! (n+3)! r(r+1) } 
 \sum_{m = n}^{N_1} \frac{m!(m+3)!}{(m-n)!} 
 \sum_{q = 0}^m \frac{(-1)^q}{(m-q)!(q+2)!(q+3)!}\nonumber\\
 &\qquad \qquad\times \Big\{r(q+5+r)! -(r+2)^2(q+4+r)! + (r+2)!\big[(2+r)(q+4)! - r(q+3)!\big]\Big\} \;.
 \label{eq:gain1_aux}
\end{align}
Similar to the $\ell=0$ case, special care must be taken when evaluating the expression above for 
$r = 0$, hence we start by assuming that $r > 0$.
The sum over $q$ is performed by first shifting $q$ upwards by two units, then extending the summation range from $(2,N_1+2)$ to $(0,N_1+2)$ and subtracting the $q = -1$ and $q = -2$ terms. Noting that these latter $q = -1$ and $q = -2$ contributions vanish identically, 
the sum can be evaluated using Eq.~\eqref{eq:sum_binom} as follows:
\begin{align}
&\sum_{q = 0}^{m+2} \frac{(-1)^q}{(m+2-q)!q!(q+1)!}\Big\{r(q+3+r)! 
  -(r+2)^2(q+4+r)! + (r+2)!\big[(2+r)(q+2)! - r(q+1)!\big]\Big\} \nonumber\\
 =&\, \frac{(-1)^m \left[(r + 2)!\right]^2 \left[r + m(r+2)\right]}{(m+2)!(m+3)!(r - m)!} \;. 
\end{align}
Finally, $\mathcal{A}_{rn}^{(1),{\rm g}}$ evaluates to
\begin{equation}
 \mathcal{A}_{rn}^{(1),{\rm g}} = \frac{2  [(r+2)!]^2 \sigma_T P_0 (-\beta)^{1+n-r} }{ n! (n+3)! (r-n)! r(r+1)}
 \left[ (r+4) S^{(-2)}_{rn}(1)-(r+2)S^{(-1)}_{rn}(1) \right] \;,
 \label{eq:gain1_aux2}
\end{equation}
where the notation $S^{(-q)}_{rn}$ was introduced in Eq.~\eqref{eq:Srn_int}. 
Using Eq.~\eqref{eq:Srn_m1}, the function $S^{(-2)}_{rn}(1)$
can be evaluated as
\begin{align}
 S^{(-2)}_{rn}(1) &\equiv (-1)^{r-n} \int_0^1 \d z\, B_z(n+1,r-n+1) \nonumber\\
 &= (-1)^{r - n} \frac{n! (r - n + 1)!}{(r+2)!}\; .
 \label{eq:Srn_m2}
\end{align}
Furthermore, using Eqs.~\eqref{eq:Srn_m1} and \eqref{eq:Beta_complete} to 
replace $S^{(-1)}_{rn}(1)$, Eq.~\eqref{eq:gain1_aux2} reduces to
\begin{equation}
 \mathcal{A}_{r>0,n \le r}^{(1),{\rm g}} = 
 \frac{2  (r+2)! \sigma_T P_0 \beta^{1+n-r}}{ (n+3)!}
 \frac{n(r+4)-r}{r(r+1)}\;,
\end{equation}
while $\mathcal{A}_{r>0,n>r}^{(1),{\rm g}} = 0$, agreeing to Eq. \eqref{eq:A_g_1}.

Considering now the case when $r = 0$, and using Eq.~\eqref{gain_1_r-1}, Eq.~\eqref{eq:gain1_aux} becomes 
\begin{equation}
 \mathcal{A}_{0n}^{(1),{\rm g}} = \frac{ 2(-1)^{n} \sigma_T P_0 \beta^{1+n}}{ n! (n+3)!} 
 \sum_{m = n}^{N_1} \frac{(m+3)!}{(m-n)!}  
 \sum_{q = 0}^m \frac{(-1)^q m!}{(m-q)!(q+2)!}
 \Big\{(q+4)(q+5) - 2 
 - 4(q+4) \big[\psi(q+5) - \psi(2)\big]\Big\}\;.
 \label{eq:gain1_r0_aux}
\end{equation}
The sum over $q$ can be performed for the terms not involving the digamma function $\psi(q+5)$ using the binomial expansion, as follows:
\begin{subequations}
\begin{align}
 \sum_{q = 0}^m \frac{(-1)^q m!}{(m-q)!q!} &= \delta_{m0}\;,\\
 \sum_{q = 0}^m \frac{(-1)^q m!}{(m-q)!(q+1)!} &= \frac{1}{m+1}\;,\\
 \sum_{q = 0}^m \frac{(-1)^q m!}{(m-q)!(q+2)!} &= \frac{1}{m+2}\;.
\end{align}
\end{subequations}
The sum over $q$ involving $\psi(q+5)$ can be performed
by noting that $\psi(q+5) = \psi(1) + \sum_{k = 1}^{q+4} \frac{1}{k}$, 
where $\psi(1) = -\gamma$ and $\gamma \simeq 0.577$ is the Euler-Mascheroni constant,
such that 
\begin{equation}
 \sum_{q = 0}^m \frac{(-1)^q m!}{(m-q)!(q+2)!} (q+4)\psi(q+5) 
 = \left(\frac{11}{6} - \gamma\right) \frac{3m+4}{(m+1)(m+2)} 
 + \frac{2 [3+ (m+1)(m+2)(m+3)]}{3(m+1)^2(m+2)^2(m+3)}\;.
\end{equation}
This leads to
\begin{equation}
 \mathcal{A}_{0n}^{(1),{\rm g}} =
 \frac{16(-1)^n  \sigma_T P_0 \beta^{1+n}}{(n+3)!}
 \Bigg[\frac{3}{4}\delta_{n0} -S_n^{(1)}(N_1)\Bigg]\;,
 \label{eq:gain1_r0_aux2}
\end{equation}
where we employed Eq.~\eqref{eq:sum_aux_app}.
Similar to the $\ell=0$ case, with the help of an auxiliary sum defined in Eq.~\eqref{eq:sum_aux_tilde}
we can write down a recursion relation
\begin{subequations}
\begin{align}
 S^{(1)}_{n+1}(N_1) &=  \frac{1}{n+1} \widetilde{S}^{(1)}_n(N_1)-\frac{n+2}{n+1} S^{(1)}_n\;,\\
 \widetilde{S}^{(1)}_{n+1}(N_1) &= \frac{1}{n+1}\binom{N_1 + 1}{n+1} -\widetilde{S}^{(1)}_n \;,\\
 S^{(1)}_0(N_1) &= \frac{N_1 + 1}{N_1 + 2}\;,\\
 \widetilde{S}^{(1)}_0(N_1) &= \psi(N_1 + 2) + \gamma\;.
 \label{eq:evil_S1_rec}
\end{align}
\end{subequations}
Note that now, Eq. \eqref{eq:gain1_r0_aux2}, can be evaluated explicitly in the case $n=0$:
\begin{equation}
\mathcal{A}_{00}^{(1),{\rm g}} = -\frac{2}{3}\sigma_T P_0 \beta \frac{N_1 -2}{N_1 + 2} \overset{N_1\to\infty}{\longrightarrow} -\frac{2}{3} \sigma_T P_0\beta\;.
\end{equation}

%%%
\subsection{Gain matrix for \boldmath\texorpdfstring{$\ell=2$}{l=2}}

Considering the case when $\ell=2$, we are inserting Eq. \eqref{eq:gain2_main} into Eq. \eqref{eq:A_expansion_gain}, which yields
\begin{align}
    \mathcal{A}^{(2),{\rm g}}_{rn} &\equiv \frac{15(-\beta)^{2+n}}{P_0 n!(n+5)!}
    \sum_{m = n}^{N_2} \frac{m!(m+5)!}{(m-n)!} 
    \sum_{q = 0}^m \frac{(-\beta)^q \mathcal{G}^{(2)}_{r-1,q}}
    {q!(m-q)!(q+5)!}\nonumber\\
    &= \frac{2(-1)^{n} \sigma_T P_0\beta^{1+n-r}}{ n! (n+5)! r(r+1)(r+2)}
 \sum_{m = n}^{N_2} \frac{m!(m+5)!}{(m-n)!} 
 \sum_{q = 0}^{m} \frac{(-1)^q f^{(2)}_{rq}}{(q+3)!(q+5)!(m-q)!}\;,\label{eq:A_gain_aux_1}
\end{align}
where we defined
\begin{align}
	f^{(2)}_{rq} &\equiv
	r(1+r)(q+r+8)! - 2r(3+r)(4+r)(q+r+7)!\nonumber\\
	& + (2+r)(3+r)^2 (4+r) (q+r+6)! - (r+4)!\left[(r+3)(q+6)!-2r(q+5)!\right]\;.
	\label{eq:gain2_frq_def}
\end{align}

As in the previous cases, Eq.~\eqref{eq:A_gain_aux_1} is also indeterminate when $r = 0$. 
For the time being, we focus on the case when $r > 0$. 
The sum over $q$ can be performed by shifting $q$ upwards by three units, hence the summation range can be extended downwards from $(3,m+3)$ to $(0,m+3)$, such that
\begin{equation}
 \sum_{q = 0}^{m} \frac{(-1)^q f^{(2)}_{rq}}{(q+3)!(q+5)! (m-q)!} 
 = -\sum_{q = 0}^{m+3} \frac{(-1)^q f^{(2)}_{r,q-3}}{q!(q+2)!(m+3-q)!}\;.
\end{equation}
With the above shift, the terms appearing inside the square brackets in the expression 
for $f^{(2)}_{rq}$ in Eq.~\eqref{eq:gain2_frq_def} lead to vanishing contributions:
\begin{equation}
 \sum_{q = 0}^{m+3} \frac{(-1)^q}{q!(m+3-q)!} =
 \sum_{q = 0}^{m+3} \frac{(-1)^q (q+3)}{q!(m+3-q)!} = 0\;.
\end{equation}
The other terms can be summed using the binomial theorem, as indicated 
in Eq.~\eqref{eq:sum_binom}, by setting $a = 2$ (for all terms) and 
$b = r+3$, $r +2$ and $r+1$.
The final result is
\begin{equation}
 \mathcal{A}^{(2),{\rm g}}_{r>0,n} \equiv \frac{2(-1)^{n}   (r+3)!(r+4)! \sigma_T P_0\beta^{1+n-r}}{ n! (n+5)! r(r+1)(r+2)}
 \sum_{m = n}^{N_2} \frac{(-1)^m[(3+r)(m+2)! - 6(m+1)!]}
 {(m-n)! (r-m)!(m+3)!}\;.
\end{equation}
The term $(r-m)!$ appearing in the denominator on the second line 
is indicative that $\mathcal{A}^{(2),{\rm g}}_{r>0,n}$ vanishes 
when $n > r$ due to the fact that $\Gamma(n)$ diverges for integer $n\leq 0$. Performing the sum over $m$ yields the first part of Eq. \eqref{eq:A_g_2},
\begin{equation}
\mathcal{A}^{(2),{\rm g}}_{r>0,n\le r} = \frac{2\sigma_T P_0\beta^{1+n-r} (r+4)! (n+1)(9n+nr-4r)}{(n+5)! r(r+1)(r+2)}\;,
\end{equation}
while $\mathcal{A}^{(2),{\rm g}}_{r>0,n > r} = 0$.

We now focus on the $r = 0$ case, which can be evaluated using Eq. \eqref{gain_2_r-1}.
Performing the summation gives
\begin{equation}
  \mathcal{A}^{(2),{\rm g}}_{0n} = \frac{432 (-1)^n\sigma_T P_0 \beta^{1+n}}{ (n+5)!} \left[\frac{5}{18}\delta_{n0}-S^{(2)}_n(N_2)\right] \;,
\end{equation}
where we used Eq. \eqref{eq:sum_aux_app}.

With the help of the auxiliary sum defined in Eq. \eqref{eq:sum_aux_tilde}, we arrive at the following recursions
\begin{subequations}
\begin{align}
 S^{(2)}_{n+1}(N_2) &=\frac{1}{n+1} \widetilde{S}^{(2)}_n(N_2) -\frac{n+3}{n+1} S^{(2)}_n(N_2) \;,\\
 \widetilde{S}^{(2)}_{n+1}(N_2) &=\frac{1}{n+1} 
 \binom{N_2 + 1}{n+1} -\frac{n+2}{n+1} \widetilde{S}^{(2)}_n(N_2) \;,\\
 S^{(2)}_0(N_2) &= \frac{N_2 + 1}{2(N_2 + 3)}\;,\\
 \widetilde{S}^{(2)}_0(N_2) &= \psi(N_2 + 3) -\psi(2)\;.
\end{align}
\end{subequations}
\end{widetext}

%%%
%%%
\section{Calculations for the bulk viscous pressure}
\label{app:bulk_calc}
In this Appendix, we provide some intermediate calculations needed to arrive at the exact results for the inverse collision matrix $\tau^{(0)}$, the bulk viscosity $\zeta$ and the relaxation time of the bulk viscous pressure $\tau_\Pi$. Furthermore, we show how to compute the correction to the local-equilibrium distribution function proportional to the bulk viscous pressure.

\subsection{Inverse matrix}
\label{subsec:app_inv_matrix}

The matrix in the scalar case has the following structure:
\begin{equation}
 \mathcal{A}^{(0)}_{rn} = 
 \begin{pmatrix}
  \mathcal{A}^{(0)}_{00} & \mathcal{A}^{(0)}_{0,n>2} \\
  0 & \mathcal{A}^{(0)}_{r>2,n>2}
 \end{pmatrix},
\end{equation}
where $\mathcal{A}^{(0)}_{r > 2, n>r} = 0$, i.e. the matrix appearing in the bottom-right corner of the above expression is lower-triangular. The inverse matrix $\tau^{(0)}_{rn}$ inherits the same form,
\begin{equation}
 \tau^{(0)}_{rn} = \begin{pmatrix}
  \tau^{(0)}_{00} & \tau^{(0)}_{0,n>2} \\
  0 & \tau^{(0)}_{r>2,n>2}
 \end{pmatrix},
\end{equation}
where $\tau^{(0)}_{r>2,n>2}$ is also lower-triangular. It is easy to see that 
\begin{equation}
 \tau^{(0)}_{00} = \frac{1}{\mathcal{A}^{(0)}_{00}} = \lambda_{\rm mfp} \frac{N_0 + 1}{N_0 - 1} \;.
 \label{eq:tau000}
\end{equation}

For future convenience, we parametrize $\tau^{(0)}_{rn}$ for $3 \le n \le r \le N_0$ as
\begin{equation}
 \tau^{(0)}_{rn} = \frac{\lambda_{\rm mfp} (r+1)!}{\beta^{r-n}(n+1)!} \left(\delta_{rn} + \tilde{\tau}^{(0)}_{rn}\right) \;.
\end{equation}
Imposing $\sum_{m = 0,\neq 1,2}^{N_0} \tau^{(0)}_{rm} \mathcal{A}^{(0)}_{mn} = \delta_{rn}$ 
gives for $r,n > 2$:
\begin{equation}
 \tilde{\tau}^{(0)}_{rm} - \sum_{n = m}^r \frac{2}{n} \tilde{\tau}^{(0)}_{rn} = \frac{2}{r} \;. 
\end{equation}
The above relation can be arranged into a simple recurrsion,
\begin{equation}
 \tilde{\tau}^{(0)}_{rm} = \frac{m}{m-2} \tilde{\tau}^{(0)}_{r,m+1} \;.
\end{equation}
Noting that $\tau^{(0)}_{rr} = 1 / \mathcal{A}^{(0)}_{rr}$, we have $\tilde{\tau}^{(0)}_{rr} = 2 / (r - 2)$, such that 
\begin{equation}
 \tilde{\tau}^{(0)}_{rm} = \frac{2(r-1)}{(m-1)(m-2)},
\end{equation}
leading to
\begin{equation}
 \tau^{(0)}_{r>2,2< n \le r} = \frac{\lambda_{\rm mfp} (r+1)!}{\beta^{r-n} (n+1)!} \left[\delta_{rn} + \frac{2(r-1)}{(m-1)(m-2)}\right] \;.
 \label{tau_0_rn}
\end{equation}

The elements on the zeroth line can be found by imposing $\sum_{r = 0}^{N_0} \mathcal{A}^{(0)}_{0r} \tau^{(0)}_{rn} = 0$ for $n > 2$:
\begin{equation}
 \mathcal{A}^{(0)}_{00} \tau^{(0)}_{0,n>2} = -\sum_{r = n}^{N_0} \mathcal{A}^{(0)}_{0r} \tau^{(0)}_{rn} \;.
\end{equation}

Using Eqs.~\eqref{A_0}, \eqref{eq:tau000}, and \eqref{tau_0_rn}, we get
\begin{multline}
	\frac{\tau_{0,n>0}^{(0)}}{\tau^{(0)}_{00}} = 
	-\frac{2 \beta^n}{(n-1)(n-2)(n+1)!} \\\times 
	\sum_{r = n}^{N_0} (-1)^r (r-1) S_r^{(0)}(N_0) [2 + (r-2)\delta_{rn}]\;.
\end{multline}
Using the explicit expression \eqref{eq:sum_aux} for $S^{(0)}_n$, the summation over $r$ can be performed, leading to:
\begin{align}
	\frac{\tau_{0,n>0}^{(0)}}{\tau^{(0)}_{00}} &= -\frac{2 (-\beta)^n}{(n-1)(n-2)(n+1)!} 
	\nonumber\\
	&\times 
	\sum_{m = n}^{N_0} \binom{m}{n} \frac{2n + m(n-2)(nm - m + n + 1)}{m^2(m^2-1)}\nonumber\\
	&=-\frac{2 (-\beta)^n}{(n-1)(n-2)(n+1)!}\nonumber\\
	&\times\binom{1+N_0}{n} \frac{(1+N_0-n)[N_0(n-2)+n]}{N_0(N_0+1)^2}\;.
	\label{eq:tau00n}
\end{align}
Collecting the above results, we find Eqs.~\eqref{eq:tau0_exact}.

\subsection{Bulk viscosity}
\label{subsec:app_bulk_viscosity}
We compute $\zeta/m^4\equiv \zeta_0 / m^4 $ by substituting Eqs.~\eqref{eq:tau000} and \eqref{eq:tau00n} in Eq.~\eqref{eq:zeta_def}:\begin{multline}
	\frac{1}{m_0^4} \zeta = \frac{P_0 \beta^4 \lambda_{\rm mfp} (N_0 + 1)}{54  (N_0 - 1)} 
	\Bigg[1 - \sum_{n= 3}^{N_0} \frac{(-1)^n}{n+1} \\\times 
	\sum_{m = n}^{N_0} \binom{m}{n} 
	\frac{n + m(n-2)(nm-m+1)}{m^2 (m^2 - 1)}\Bigg]\;.
	\label{eq:zeta_aux}
\end{multline}
Swapping the summation with respect to $n$ with that with respect to $m$ and using the properties
\begin{subequations}
	\label{eq:sum_ids}
	\begin{align}
		\sum_{n = 3}^m \binom{m}{n} \frac{(-1)^n}{n+1} &= -\frac{m(m-1)(m-2)}{6(m+1)}\;,\\
		\sum_{n = 3}^m \binom{m}{n} (-1)^n &= -\frac{1}{2}(m-1)(m-2)\;,\\
		\sum_{n = 3}^m \binom{m}{n} (-1)^n n &= -m(m-2)\;.
	\end{align}
\end{subequations}
Eq.~\eqref{eq:zeta_aux} can be reduced to
\begin{multline}
	\frac{1}{m_0^4} \zeta = \frac{P_0 \beta^4 \lambda_{\rm mfp} (N_0 + 1)}{54  (N_0 - 1)} \\\times 
	\left[1 + \sum_{m = 3}^{N_0} \frac{(m-2)(11m^2 + 4m - 3)}{3m^2 (m-1) (m+1)^2}\right]\;.
\end{multline}
The sum over $m$ appearing above represents a correction to the 14-moment approximation, represented by the prefactor of the square brackets. After preforming this sum, we arrive at Eq.~\eqref{zeta_0_exact_tau}.

\subsection{IReD relaxation time}
\label{subsec:app_tau_IReD}
We begin with $\tau_\Pi \equiv  \tau_{\Pi;0}$ and use Eqs.~\eqref{eq:zeta_r} and~\eqref{zeta_0_exact_tau},
\begin{align}
	\tau_\Pi &= \tau^{(0)}_{00} + \sum_{r = 3}^{N_0} \tau^{(0)}_{0r} \frac{\zeta_r}{\zeta} \nonumber\\
	&= \tau^{(0)}_{00} \Bigg\{1 - \frac{6N_0(N_0^2-1)}{6+7N_0+11N_0^3} \sum_{m = 3}^{N_0} \frac{1}{m(m+1)} \nonumber\\
	& \times \sum_{r = 3}^{m} \binom{m}{r} (-1)^r
	\left(2H_r - \frac{1}{r+1} - \frac{8}{3}\right) \nonumber\\ 
	& \times\left[r - 1 + \frac{2r}{m-1} + \frac{2r}{m(m-1)(r-2)} \right]\Bigg\}\;.
\end{align}
In order to evaluate the sums not containing harmonic numbers, we need the identities \eqref{eq:sum_ids}, as well as
\begin{equation}
	\sum_{r = 3}^m \binom{m}{r} \frac{(-1)^r}{r-2} = \frac{1}{4} m(m-1)(3 - 2H_m)\;. \label{eq:s_rm2}
\end{equation}
In order to perform the summation over $r$ for the terms involving $H_r$, we employ its integral representation,
\begin{equation}
	H_r = \int_0^1 \d t\, \frac{1 - t^r}{1 - t}\;,
\end{equation}
together with the relations
\begin{align}
	\sum_{r = 3}^m \binom{m}{r} (-t)^r &= -1 + (1-t)^m + \frac{m t}{2} (2 + t - mt)\;,\nonumber\\
	\sum_{r = 3}^m \binom{m}{r} (-1)^r  r &= m t [1 - (1 - t)^{m-1} - t(m - 1)]\;,\nonumber\\
	\sum_{r = 3}^m \binom{m}{r} \frac{(-1)^r}{r-2} &= -\frac{m t^3}{6} (m - 2)(m-1) \nonumber\\
	& \times {}_3F_2(1,1,3-m;2,4; t)\;. \label{eq:sH_rm2}
\end{align}
Interchanging the summation with respect to $r$ with the integration with respect to $t$, we arrive at
\begin{subequations}
	\begin{align}
		\sum_{r = 3}^m \binom{m}{r} H_r &= -\frac{3m^3 - 7m^2 + 4}{4m}\;,\label{eq:iH1}\\
		\sum_{r = 3}^m \binom{m}{r} (-1)^r r H_r  &= \frac{m}{2}(5 - 3m) + \frac{1}{m-1}\;,
		\label{eq:iHr} \\
		\sum_{r = 3}^m \binom{m}{r} \frac{(-1)^r H_r}{r-2} &= \frac{m}{4}(m - 1)(7 - 3 H_m - 2H_{m,2})\;, \label{eq:iH_rm2}
	\end{align}
\end{subequations}
where $H_{m,n} = \sum_{r = 1}^m r^{-n}$ is the generalized Harmonic number, with $H_m \equiv H_{m,1}$. Adding everything up, we find
\begin{multline}
	\tau_\Pi = \tau^{(0)}_{00} \Bigg\{1 + \frac{4N_0(N_0^2-1)}{6+7N_0+11N_0^3} \sum_{m = 3}^{N_0}
	\frac{1}{m(m+1)}\\\times \left[\frac{11}{6} - \frac{2m}{m^2-1} - \frac{6 + m}{m^2} - \frac{6}{(m-1)^2} + 6H_{m,2}\right]\Bigg\}\;.
\end{multline}
The summation over $m$ can be performed, yielding Eq.~\eqref{eq:tau_Pi_N0}.
For the relaxation times of the higher-order moments we have
\begin{multline}
	\tau_{\Pi;r}=\sum_{n=0,\neq 1,2}^{N_0}\tau^{(0)}_{rn}\frac{\zeta_n}{\zeta_r}\\
	= \frac{m^4}{\zeta_r} \frac{\lambda_{\rm mfp} P \beta^{4-r}}{108r} (r-2)(r-1)(r+1)! 
	\Bigg[2H_r - \frac{1}{r+1} \\
	- \frac{8}{3} + 2 \sum_{n = 3}^r \frac{1}{n-2} \left(2H_n - \frac{1}{n+1} - \frac{8}{3} \right)\Bigg]\;.
\end{multline}
Performing the summation then gives Eq.~\eqref{eq:tau_Pi_r_N0}.

\subsection{Correction to the distribution function}
\label{subsec:app_corr_dist}
We start from Eq. \eqref{eq:delta_f_0_start} and use
Eqs.~\eqref{eq:zeta_r} and \eqref{zeta_0_exact_tau} for $\zeta_n$ and $\zeta_0$, as well as Eq.~\eqref{eq:UR_H} for $\mathcal{H}^{(0)}_{\bk n}$, arriving at
\begin{multline}
 \frac{\delta f^{(0)}_\bk}{f_{0\bk}} = -\frac{6 \Pi}{m_0^2 \beta^2 P} \left[
 \sum_{m = 0}^{N_0} L^{(1)}_m(\beta E_\bk) + 
 \frac{3 N_0(N_0^2 -1)}{6 + 7N_0 + 11 N_0^3} \right.\\
 \times \sum_{n = 3}^{N_0} (-1)^n(n-1) \left(2H_n - \frac{1}{n+1} -\frac{8}{3}\right) \\
 \left. \times \sum_{m =n}^{N_0} \frac{m!}{n!(m-n)!} L^{(1)}_m(\beta E_\bk) \right]\;.
 \label{eq:df0_aux}
\end{multline}
In the second term, the sums over $n$ and $m$ can be swapped, while the sum over $n$ can be evaluated as follows:
\begin{multline}
 \sum_{n = 3}^m \frac{m! (-1)^n (n-1)}{n!(m-n)!}\left(2H_n - \frac{1}{n+1} - \frac{8}{3}\right)\\ = -\frac{11}{3} + 2\left(\frac{1}{m - 1} + \frac{1}{m} + \frac{1}{m+1}\right) \;.
\end{multline}
Plugging the above into Eq.~\eqref{eq:df0_aux} leads to
\begin{multline}
 \frac{\delta f^{(0)}_\bk}{f_{0\bk}} = -\frac{6 \Pi}{m_0^2 \beta^2 P} \left[\sum_{m = 0}^{N_0} L^{(1)}_m(\beta E_\bk) + \frac{3 N_0(N_0^2 - 1)}{6 + 7N_0 + 11N_0^3}\right.\\
 \left. \times \sum_{m = 3}^{N_0} L^{(1)}_m(\beta E_\bk) \left(\frac{2}{m-1} + \frac{2}{m} + \frac{2}{m+1} - \frac{11}{3}\right)\right].
\end{multline}

We now consider the limit $N_0 \rightarrow \infty$, when $3N_0(N_0^2 - 1) / (6 + 7N_0 + 11 N_0^3) \rightarrow 3/11$. This leads to the expression 
\begin{multline}
 \frac{\delta f^{(0)}_\bk}{f_{0\bk}} = -\frac{6\Pi}{m_0^2 \beta^2 P}
 \Bigg[1 + L^{(1)}_1(\beta E_\bk) + L^{(1)}_2(\beta E_\bk) \\
 + \frac{6}{11} \sum_{m = 3}^{\infty} L^{(1)}_m(\beta E_\bk) \left(\frac{1}{m-1} + \frac{1}{m} + \frac{1}{m+1}\right)\Bigg]\;.
 \label{eq:df0_aux2}
\end{multline}
The summation over $m$ can be performed by introducing a fictitious parameter $0 < t < 1$ and employing the generating function 
\begin{equation}
 \sum_{m = 0}^\infty t^m L^{(\alpha)}_m(x) = \frac{1}{(1 - t)^{\alpha + 1}} e^{-tx / (1 - t)}\;.
\end{equation}
In our case, we must evaluate 
\begin{multline}
 \sum_{m = 3}^{\infty} L^{(1)}_m(x) \left(\frac{1}{m-1} + \frac{1}{m} + \frac{1}{m+1}\right) \\
 = \int_0^1 \d t \left(1 + \frac{1}{t} + \frac{1}{t^2}\right) \left[\frac{e^{-xt / (1 - t)}}{(1 - t)^2} - \sum_{m=0}^2 t^m L^{(1)}_m(x)\right]\;.
\end{multline}
It can be checked that the integrand behaves like $O(t)$ around $t = 0$ and thus the integral converges. The result is
\begin{multline}
 \sum_{m = 3}^{\infty} L^{(1)}_m(x) \left(\frac{1}{m-1} + \frac{1}{m} + \frac{1}{m+1}\right)
 = \frac{3}{x} \\
 - (3 - x) \ln x -\frac{19}{2} + \gamma(x - 3) + 6x - \frac{11x^2}{12}\;.
\end{multline}
Plugging the above into Eq.~\eqref{eq:df0_aux2} then gives Eq. \eqref{eq:delta_f_0_final}.
\vspace{-10pt}

\bibliographystyle{unsrtnat}
\bibliography{bib_URHS}

\end{document}